\documentclass[useAMS,usenatbib,twocolumn]{mnras}
\pdfoutput=1
\usepackage{amssymb}
\usepackage{amsmath}
\usepackage{natbib}
\usepackage[pdftex]{graphicx}
\usepackage{subfigure}
\usepackage{booktabs}
\usepackage{xcolor}
\usepackage{textcase}

\begin{document}

\title[GSF instability]
{Angular momentum transport, layering, and zonal jet formation by the GSF instability: nonlinear simulations at a general latitude}
  \author[A. J. Barker, C. A. Jones \& S. M. Tobias]{A. J. Barker\thanks{Email address: A.J.Barker@leeds.ac.uk}, C. A. Jones and S. M. Tobias \\ Department of Applied Mathematics, School of Mathematics, University of Leeds, Leeds, LS2 9JT, UK
}

\pagerange{\pageref{firstpage}--\pageref{lastpage}} \pubyear{2020}
\maketitle
\label{firstpage}

\begin{abstract}
We continue our investigation into the nonlinear evolution of the Goldreich-Schubert-Fricke (GSF) instability in differentially rotating radiation zones. This instability may be a key player in transporting angular momentum in stars and giant planets, but its nonlinear evolution remains mostly unexplored. In a previous paper we considered the equatorial instability, whereas here we simulate the instability at a general latitude for the first time. We adopt a local Cartesian Boussinesq model in a modified shearing box for most of our simulations, but we also perform some simulations with stress-free, impenetrable, radial boundaries. We first revisit the linear instability and derive some new results, before studying its nonlinear evolution. The instability is found to behave very differently compared with its behaviour at the equator. In particular, here we observe the development of strong zonal jets (``layering" in the angular momentum), which can considerably enhance angular momentum transport, particularly in axisymmetric simulations. The jets are, in general, tilted with respect to the local gravity by an angle that corresponds initially with that of the linear modes, but which evolves with time and depends on the strength of the flow. The instability transports angular momentum much more efficiently (by several orders of magnitude) than it does at the equator, and we estimate that the GSF instability could contribute to the missing angular momentum transport required in both red giant and subgiant stars. It could also play a role in the long-term evolution of the solar tachocline and the atmospheric dynamics of hot Jupiters.
\end{abstract}

\begin{keywords}
Sun: rotation -- stars: rotation -- hydrodynamics -- waves -- instabilities
\end{keywords}

\section{Introduction}

Stably-stratified radiation zones are unlikely to be quiescent, and are potentially subject to a number of (magneto-) hydrodynamic instabilities that can drive turbulence or wave activity. The resulting mixing and angular momentum transport produced by these instabilities is important for the evolution of the global properties and the internal structures of rotating stars (e.g.~\citealt{Maeder2009,Maederetal2013,Meynetetal2013,ARM2018}). Radiation zones also couple with neighbouring convection zones through the excitation, propagation and dissipation of waves \citep[e.g.][]{Rogers2006,Lecoanet2013,Rogers2013,Couston2018,Auguston2019,Korre2019} and via magnetic fields \citep[e.g.][]{Spruit1999,Zahn2007,Garaud2008,Strugarek2011,Wood2011,Fuller2019}. Despite much research, the mechanisms responsible for mixing and for transporting angular momentum in radiation zones remain poorly understood. 

Observational advances in helio- and astero-seismology have shown that our current understanding of transport processes in radiation zones is inadequate. Unsolved problems include the inferred internal rotation rates of red giant and sub-giant stars \citep{Beck2012,Mosser2012,Cantielloetal2014,Spada2016,Eggenberger2016,Eggenberger2017}, whose cores rotate slower than expected, and the formation and maintenance of the solar tachocline \citep{Thompson2003,Tobias2005,Garaud2008,Wood2011,gilman2017,gilman2018}. A separate problem is the atmospheric dynamics of hot Jupiters, particularly regarding whether the jets that advect heat from dayside to nightside are subject to small-scale hydrodynamic instabilities that are currently unresolved in global simulations (e.g.~\citealt{Goodman2009,Showman2009,DobbsDixon2010,LiGoodman2010,Fromang2016,Mayne2017,Menou2019}).

The Goldreich-Schubert-Fricke (GSF) instability \citep{GS1967,Fricke1968} has long been considered as a possible mechanism for angular momentum transport in the radiation zones of stars (or planets). It is essentially an axisymmetric centrifugal instability that is facilitated by the action of thermal diffusion, which neutralises the otherwise stabilising effects of buoyancy. The instability grows if the differential rotation is sufficiently strong (e.g.~\citealt{Acheson1978,KnoblochSpruit1982,Caleo2016a,Caleo2016b}). In the simplest case in which the thermal Prandtl number (the ratio of viscosity to thermal diffusivity) is strictly zero, the instability occurs if the angular momentum per unit mass decreases outwards from the rotation axis, or if there is any nonzero gradient of the angular velocity along the rotation axis. The latter is generally much easier to satisfy. Until recently (\citealt{BJT2019}; hereafter paper I), the nonlinear development of this instability in stellar interiors had only been studied in axisymmetric (two-dimensional) simulations by \cite{Kory1991} and briefly in small domains by \cite{Rashid2010}. In paper I, we presented a comprehensive study into the nonlinear evolution of the equatorial GSF instability using both axisymmetric and three-dimensional simulations. We demonstrated that the linear and nonlinear equations governing the axisymmetric equatorial instability are equivalent to those of salt fingering (for a certain diffusivity ratio), where the angular momentum field plays the role of salinity (see also \citealt{Knobloch1982}). This analogy was found to be helpful to interpret our results in light of much recent work on the salt fingering problem (e.g.~\citealt{Traxler2011,Brownetal2013,GaraudBrummell2015,Garaud2018,Xie2019}). However, the three-dimensional nonlinear evolution is strictly not equivalent, even if it bears some similarities with salt fingering.

In paper I, the equatorial GSF was typically observed to produce homogeneous turbulence with enhanced transport properties. The instability did not generally form large-scale structures such as layering or strong zonal jets, and the properties of the instability were found to be well explained by a simple single-mode theory. This theory can in principle be applied straightforwardly to predict the resulting angular momentum and heat transport in stars when the equatorial instability produces homogeneous turbulence. Meridional jets were observed in simulations with shearing-periodic boundaries in small azimuthal domains, which acted as barriers to transport. However, these jets were not typically observed with stress-free conditions or in simulations with wider azimuthal domains, so we speculate that they are unimportant for stars.

The nonlinear evolution of the GSF instability at a general latitude has not yet been explored. There are several reasons why the non-equatorial instability could differ in interesting ways from the equatorial case. Firstly, the differential rotation required for the non-equatorial instability to onset is generally much weaker. The criterion at the equator is particularly restrictive and requires the presence of centrifugally unstable flows that violate Rayleigh's criterion. This corresponds to a very strong radial differential rotation. On the other hand, at a general latitude, the instability occurs if the variation in the angular velocity along the rotation axis is sufficiently strong, which is usually a much easier criterion to satisfy.

The GSF instability is related to the ``secular" shear instabilities that have been proposed to contribute to the missing mixing in stellar radiation zones (e.g.~\citealt{Zahn1974,Zahn1992}). Standard shear instabilities, in which perturbations are assumed to be adiabatic, are not usually expected to develop in stellar radiation zones owing to the strong stabilising effect of the stratification. However ``secular" shear instabilities, which require finite-amplitude perturbations, are believed to be important by producing thermally-diffusive shear-induced turbulence when the Richardson number Ri (which measures the ratio of the strength of the stratification to the shear) of the flow is large, provided the P\'{e}clet number Pe (which measures the ratio of thermal diffusion to advection timescales) is sufficiently small. Simulations of these instabilities indicate that this is a promising mechanism of angular momentum transport and mixing in radiation zones (e.g.~\citealt{Prat2013,Prat2014,Prat2016,Garaud2017,Gagnier2018,Kulenthirarajah2018,Mathis2018}), which can be expected when RiPe or RiPr (where Pr is the Prandtl number, the ratio of viscosity to thermal diffusivity) is sufficiently small. The GSF instability is, on the other hand, a linear instability, but we will show in \S~\ref{lineargrowth} that it also onsets when RiPr is sufficiently small ($<1/4$). The effect of rotation on secular shear instabilities remains to be explored, and we expect that the resulting flows will interact with those generated by the GSF instability.

The GSF instability may also occur in astrophysical discs, where it has been referred to as the Vertical Shear Instability or VSI (e.g.~\citealt{Urpin1998,Nelsonetal2013,StollKley2014,BarkerLatter2015,LinYoudin2015,LatterPap2018}). This may drive weakly turbulent motions and stir solid material in regions of protoplanetary discs that are not subject to the magneto-rotational instability. Indeed, simulations using a local model like the ones that we will present in this paper but for parameters relevant for astrophysical discs, may shed some light on the nonlinear evolution of the VSI. This topic is left for future work.

Our primary goal is to understand the nonlinear evolution of the GSF instability at a general latitude and to derive physically-motivated prescriptions for the transport of angular momentum, as well as other quantities such as heat or heavy elements, that can be implemented in stellar evolution codes. As we will demonstrate, the instability behaves very differently from the equatorial case, making it difficult to propose a simple prescription for the transport that adequately describes all of our simulation results. This is because the instability generates strong zonal jets (``layering" in the angular momentum) and these significantly enhance the momentum transport (particularly in axisymmetric cases -- the effect is weaker in 3D). We speculate that the interaction of the strong jets with the turbulent transport may better be parameterised via a quasilinear turbulence/mean flow interaction theory \citep{Diamond2005,mct2016}. Our paper is structured as follows: in \S~\ref{model} we describe our model and numerical approach. In \S~\ref{lineargrowth}, we revisit the axisymmetric linear instability and derive some new results, including a simple criterion for the onset of instability, and analyse its properties. We then turn to describe the results of a set of axisymmetric and three-dimensional simulations of the instability in \S~\ref{results30}. We compare our results with a generalisation of the theory presented in Paper I in \S~\ref{theorycomparison}, and discuss the astrophysical implications of our work in \S~\ref{astrophysical}. Finally, we conclude in \S~\ref{Conclusions}.

\section{Local Cartesian model: small patch of a radiation zone}
\label{model}

\begin{figure}
  \begin{center}
    \subfigure{\includegraphics[trim=0cm 0cm 0cm 0cm, clip=true,width=0.45\textwidth]{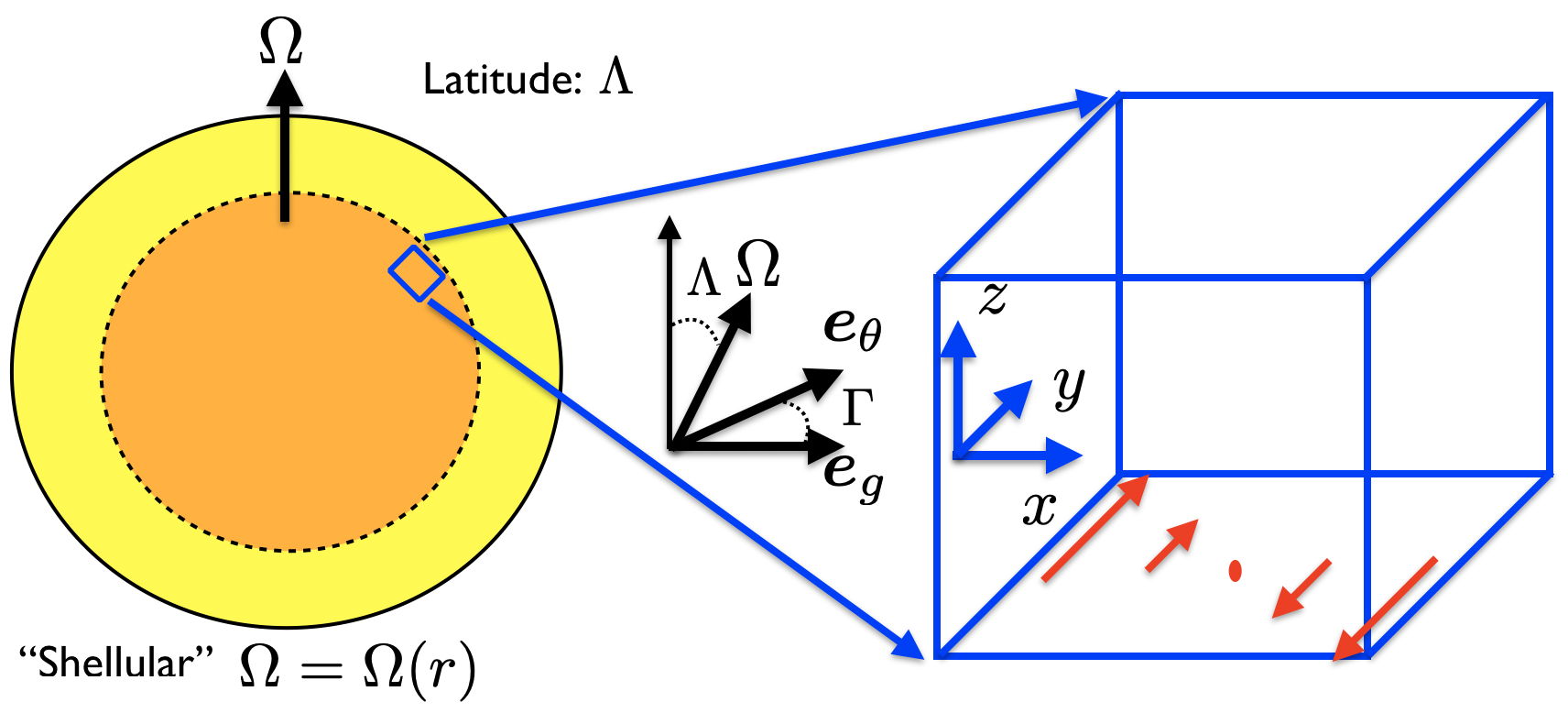}}
    \end{center}
  \caption{Local Cartesian model to study the GSF instability at a general latitude. For illustration, the dark orange region may represent a radiation zone and the yellow region an overlying convection zone, so that the Cartesian domain represents a small patch in the solar tachocline. The rotation vector is inclined by an angle $\Lambda$ from $z$, or by $90^{\circ}-\Lambda$ from the local radial direction ($x$). In general, when we are not at the equator, the normal to the stratification surfaces (i.e.~along the temperature gradient) $\boldsymbol{e}_\theta$ is inclined relative to the local gravity vector $\boldsymbol{e}_g=\boldsymbol{e}_x$ by an angle $\Gamma$ that is determined by the thermal wind equation.}
  \label{Model}
\end{figure}

We consider a local Cartesian representation of a small patch of a stably-stratified radiation zone of a differentially rotating star (or planet). Our coordinate axes $(x,y,z)$ are defined such that $x$ is the local radial, $y$ is the local azimuthal, and $z$ is the other meridional direction (see Fig.~\ref{Model}), and the box has size $L_x\times L_y \times L_z$. The star is assumed to possess a ``shellular" differential rotation, such that the angular velocity $\Omega(r)$ depends only on spherical radius $r$ (e.g.~\citealt{Zahn1992}), though our model can be readily extended to consider more general profiles. The differential rotation can be locally decomposed into a uniform rotation $\boldsymbol{\Omega}=\Omega \hat{\boldsymbol{\Omega}}$ and a linear (radial) shear flow $\boldsymbol{U}_0=-\mathcal{S}x\boldsymbol{e}_y$, where $\mathcal{S}$ is the local value of $-\varpi \frac{\mathrm{d}\Omega}{\mathrm{d}r}$, and $\varpi$ is the cylindrical radius. At a general latitude $\Lambda$, $\hat{\boldsymbol{\Omega}}=(\sin\Lambda,0,\cos\Lambda)$, where $\Lambda=0^{\circ}$ at the equator and $90^{\circ}$ at the pole.

We adopt the Boussinesq approximation \citep{SpiegelVeronis1960}, which is valid for subsonic flows with length-scales that are much shorter than a density or pressure scale height, both of which are expected to be appropriate for the GSF instability. We also assume $\Omega^2 \varpi \ll g$ so gravity is in the radial direction\footnote{This assumption is straightforward to relax \citep[e.g.][]{KnoblochSpruit1982}.}, so $\boldsymbol{e}_g=\boldsymbol{e}_x$. Perturbations to the shear flow $\boldsymbol{U}_{0}$, are governed by the dimensional governing equations
\begin{eqnarray}
\label{EQ1}
&&D\boldsymbol{u} + 2\boldsymbol{\Omega}\times \boldsymbol{u} +\boldsymbol{u}\cdot \nabla \boldsymbol{U}_0 = -\nabla p +  \theta\boldsymbol{e}_{x}+ \nu\nabla^{2}\boldsymbol{u}, \\
&& D\theta +\mathcal{N}^2 \boldsymbol{u}\cdot\boldsymbol{e}_{\theta}=  \kappa\nabla^{2}\theta, \\
&& \nabla \cdot \boldsymbol{u} = 0, \\
&& D \equiv \partial_{t} + \boldsymbol{u}\cdot \nabla + \boldsymbol{U}_0 \cdot \nabla,
\label{EQ4}
\end{eqnarray}
where $\boldsymbol{u}$ is the velocity perturbation and $p$ is a pressure variable. We use $\theta$ as our ``temperature perturbation", which has the units of an acceleration and is related to the usual temperature perturbation $T$ by $\theta=\alpha g T$, where $\alpha$ is the thermal expansion coefficient and $g$ is the acceleration due to gravity. The background reference density has been set to unity. We adopt a background temperature (entropy) profile $T(\boldsymbol{x})$, with uniform gradient $\alpha g\nabla T = \mathcal{N}^2 \boldsymbol{e}_{\theta}$, where $\boldsymbol{e}_\theta = (\cos\Gamma,0,\sin\Gamma)$, and $\mathcal{N}^2>0$ in a radiation zone. We also adopt a constant kinematic viscosity $\nu$ and thermal diffusivity $\kappa$.

At the equator ($\Lambda=0$), the rotation is constant on cylinders and surfaces of constant density and pressure are aligned. This is 
equivalent to the shearing box model of an astrophysical disc with radial stratification and shear. Studying this case was the focus of paper I. Here we instead focus on cases with $\Lambda\ne 0$, in which surfaces of constant density and pressure are misaligned ($\Gamma\ne 0$) in general. We assume that the degree of misalignment is determined by the ``thermal wind equation"
\begin{eqnarray}
\label{TWE}
2 \Omega \mathcal{S} \sin \Lambda &=& \mathcal{N}^2 \sin \Gamma,
\end{eqnarray}
which follows from the azimuthal component of the vorticity equation for the basic flow, rather than by any external forcing or transient phenomena. Our approach will be to choose values of $\Omega, \mathcal{S}, \mathcal{N}^2$ and $\Lambda$, so that Eq.~\ref{TWE} determines $\Gamma$ i.e. the degree of ``baroclinicity". An alternative viewpoint (taken by \citealt{Rashid2008}) is to consider the temperature gradient to be imposed, then the thermal wind equation determines the corresponding differential rotation (i.e. the ``baroclinic shear"). Note that the thermal wind equation does not constrain the ``barotropic shear". For example, at the equator the thermal wind equation is trivially satisfied and arbitrary profiles of $\Omega(\varpi)$ are permitted.

As in paper I, we adopt $\Omega^{-1}$ as our unit of time and take the lengthscale $d$ to define our unit of length, where 
\begin{eqnarray}
d=\left(\frac{\nu\kappa}{\mathcal{N}^2}\right)^{\frac{1}{4}}.
\label{d_def_eqn}
\end{eqnarray}
This lengthscale was chosen because the fastest growing modes typically have wavelengths $O(d)$, just like in other related double-diffusive problems \citep[e.g.][]{Garaud2018}. We also define $N=\mathcal{N}/\Omega$ to be our dimensionless buoyancy frequency and $S=\mathcal{S}/\Omega$ to denote our dimensionless shear rate, which can be thought of as a Rossby number. We also define the Prandtl number
\begin{eqnarray}
\mathrm{Pr} = \frac{\nu}{\kappa}.
\end{eqnarray}

This problem has 4 remaining independent physical parameters: $S, \mathrm{Pr}, N^2$, and $\Lambda$, in addition to the dimensions of the box, $L_x$, $L_y$ and $L_z$ in units of $d$. We also define the derived non-dimensional parameters, including the Ekman number
\begin{eqnarray}
\mathrm{E} = \frac{\nu}{\Omega d^2} = \mathrm{Pr}^{1/2} N,
\end{eqnarray}
and the Richardson number
\begin{eqnarray}
\mathrm{Ri} = \frac{\mathcal{N}^2}{\mathcal{S}^2}=\mathrm{E}^2\mathrm{Pr}^{-1}S^{-2}.
\label{Ridef}
\end{eqnarray}
The non-dimensional momentum and heat equations can then be written in the form
\begin{eqnarray}
\label{Eq1ND}
&&D\boldsymbol{u} + 2\hat{\boldsymbol{\Omega}}\times \boldsymbol{u} -Su_x\boldsymbol{e}_y = -\nabla p +  \theta\boldsymbol{e}_{g}+ \mathrm{E}\nabla^{2}\boldsymbol{u}, \\
\label{Eq2ND}
&& D\theta +N^2\boldsymbol{u}\cdot\boldsymbol{e}_\theta=  \frac{\mathrm{E}}{\mathrm{Pr}}\nabla^{2}\theta,
\end{eqnarray}
where we have scaled the time by $\Omega^{-1}$, lengths by $d$, velocities by $\Omega d$ and the temperature $
T=\theta/g \alpha$ by $\Omega^2 d / g \alpha$. We have not added hats to denote non-dimensional quantities (i.e.~$u_x, u_y, u_z$ and $\theta$) to simplify the presentation. We use these dimensionless variables when discussing our simulations results in \S~\ref{results30}. 

Most of our simulations use a modified version of the Cartesian pseudo-spectral code SNOOPY \citep{Lesur2005}. This uses a basis of shearing waves, which is equivalent to using shearing-periodic boundary conditions in $x$. In real space, using un-sheared coordinates, these would specify that
\begin{eqnarray}
u_x\left(-\frac{L_x}{2},y,z,t\right)=u_x\left(\frac{L_x}{2},(y-S L_{x} t)\textrm{mod}(L_y),z,t\right),
\end{eqnarray}
and similarly for the other variables. We adopt periodic boundary conditions in $y$ and $z$. The code uses a 3rd order Runga-Kutta method for time-stepping, and the diffusion terms are accounted for using an integrating factor. We have tested our modifications to the code to ensure that it correctly captures the linear growth of the GSF instability. We also ensure that each simulation is adequately resolved by either running selected simulations at higher resolution to ensure convergence of the bulk statistics, or by ensuring that the relative spectral kinetic energy in the modes at the de-aliasing wavenumber is smaller than $10^{-3}$ of the maximum. As in paper I, we found it necessary to enforce the box-averaged velocity components (i.e.~the zero wavenumber mode) to be zero periodically (with a typical period of between 1 and 20 timesteps) to avoid unphysical growth of these quantities. This is explained in paper I, and is particularly important when the flow is centrifugally unstable, since this component can grow owing to small numerical errors.

We have performed a suite of both axisymmetric ($y$-invariant) and three-dimensional simulations. Our typical simulation domain has $L_x = L_z = 100 d$, unless otherwise specified, which was found to be sufficiently large to contain several wavelengths of the fastest growing linear mode. $L_y$ is varied separately in 3D simulations to explore the importance of 3D effects. We initialise the flow using solenoidal random noise of amplitude $10^{-3}$ for all wavenumbers in the range $\hat{i},\hat{j},\hat{k}\in [1,21]$, where $k_x=\frac{2\pi}{L_x}\hat{i}$, $k_y=\frac{2\pi}{L_y}\hat{j}$ and $k_z=\frac{2\pi}{L_z}\hat{k}$.

We also present the results of several three-dimensional simulations using the spectral element code Nek5000 \citep{Nek5000}, which allows us to consider different boundary conditions to shearing-periodic conditions in $x$. These simulations solve Eqs.~\ref{EQ1}--\ref{EQ4} for the same linear shear flow and temperature gradient, but we adopt impenetrable, stress-free, fixed temperature conditions at the boundaries in $x$ for these simulations. These specify that
\begin{eqnarray}
\theta=u_x=\partial_x u_y=\partial_x u_z=0 \;\;\;\; \text{on} \;\;\;\; x=\pm\frac{L_x}{2}.
\end{eqnarray}
Nek5000 adopts $\mathcal{E}$ elements and within each element the velocity components and the pressure are represented as tensor product Legendre polynomials of order $\mathcal{N}_p$ and $\mathcal{N}_p-2$, respectively.  The total number of grid points is therefore $\mathcal{E}\mathcal{N}_p^3$. We also use a 3rd order mixed implicit-explicit scheme with a variable time-step.

\section{Axisymmetric linear instability at a general latitude}
\label{lineargrowth}

\begin{figure}
  \begin{center}
    \subfigure{\includegraphics[trim=0cm 0cm 0cm 0cm, clip=true,width=0.33\textwidth]{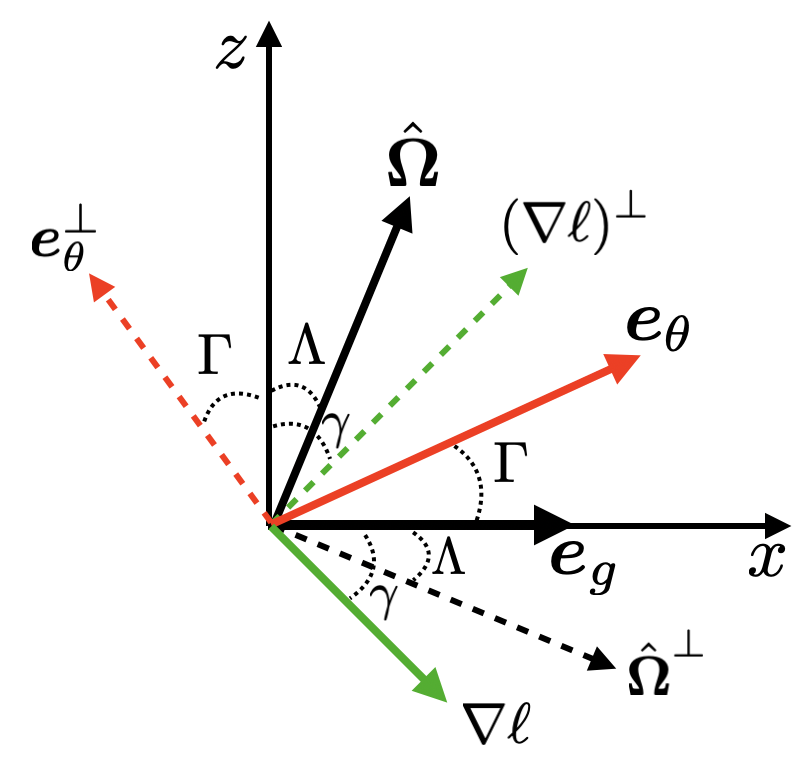}}
    \end{center}
  \caption{Illustration of the various vectors and corresponding ang
  les in the $(x,z)$-plane as defined in the text. The angles $\Lambda$, $\Gamma$ and $\gamma$ are all positive in the
  northern hemisphere in the case $\mathcal{S}>0$, that is $\mathrm{d} \Omega /\mathrm{d}r <0.$}
  \label{Arrows}
\end{figure}

In this linear stability section we use dimensional quantities throughout. 
We consider axisymmetric modes which have an azimuthal wavenumber $k_y=0$, 
as these are known to be important for GSF instability, and we may consider quantities to vary locally as $\exp (\mathrm{i} k_x x + \mathrm{i} k_z z + st)$, where $k_x$ and $k_z$ are the wavevector components along the radial and the other meridional direction. The growth rate $s$ can be shown to satisfy \citep[e.g.][]{GS1967,Acheson1978,KnoblochSpruit1982}
\begin{eqnarray}
\label{cubic}
s_\nu^2 s_\kappa + a s_\kappa + b s_\nu=0,
\end{eqnarray}
where $s_\nu=s+\nu k^2$, $s_\kappa=s+\kappa k^2$, and
\begin{eqnarray}
a &=& \frac{2}{\varpi} \left(\hat{\boldsymbol{k}}\cdot\boldsymbol{\Omega}\right)\left(\hat{\boldsymbol{k}}\cdot(\nabla \boldsymbol{\ell})^{\perp}\right), \\
b &=& \mathcal{N}^2 \left(\hat{\boldsymbol{k}}\cdot\boldsymbol{e}_\theta^{\perp}\right)\left(\hat{\boldsymbol{k}}\cdot\boldsymbol{e}_g^{\perp}\right),
\end{eqnarray}
where $\hat{\boldsymbol{k}}$ is the unit vector in the direction of the wavevector $\boldsymbol{k}=(k_x,0,k_z)$, and $k=\sqrt{k_x^2+k_z^2}$ is the wavenumber. We define several vectors in the $(x,z)$-plane, starting with the local specific angular momentum gradient $\nabla \ell = \nabla (\varpi^2 \Omega)$,
\begin{eqnarray}
\nonumber
\nabla \ell &=& \varpi(2\Omega c_\Lambda -\mathcal{S}, 0, -2\Omega s_ \Lambda), \\
&=& |\nabla \ell|(c_{\gamma}, 0, -s_{\gamma}),
\label{gradldef}
\end{eqnarray}
and its normal, 
\begin{eqnarray}
\nonumber
(\nabla \ell)^{\perp} &=& \varpi(2\Omega s_\Lambda, 0,  2\Omega c_ \Lambda-\mathcal{S}), \\
&=& |\nabla \ell|(s_{\gamma}, 0, c_{\gamma}),
\end{eqnarray}
where the squared magnitude of the local angular momentum is defined by
\begin{eqnarray}
|\nabla \ell|^2 = \varpi^2 \mathcal{S}^2+4\varpi^2\Omega(\Omega-\mathcal{S}c_\Lambda).
\label{gradlsq}
\end{eqnarray}
We have also introduced an additional angle $\gamma$, which defines the direction of the local angular momentum gradient relative to $x$. Furthermore, we have denoted $\cos\Lambda$ and $\sin\Lambda$ by $c_\Lambda$ and $s_\Lambda$, respectively, and similarly for other angles, to simplify the presentation.
We further define the local vector parallel to stratification surfaces (normal to $\boldsymbol{e}_\theta$), 
\begin{eqnarray}
\boldsymbol{e}_\theta^{\perp} &=& (-s_\Gamma, 0, c_\Gamma),
\end{eqnarray}
and the vector perpendicular to gravity 
\begin{eqnarray}
\boldsymbol{e}_g^{\perp} &=& (0, 0, 1).
\end{eqnarray}
Finally, we define the vector perpendicular to the rotation axis, i.e.~the local cylindrical radial direction:
\begin{eqnarray}
\hat{\boldsymbol{\Omega}}^\perp &=& (c_\Lambda, 0, -s_\Lambda).
\end{eqnarray}
Note that the ``baroclinic shear" is given by
\begin{eqnarray}
\hat{\boldsymbol{\Omega}}\cdot (\nabla \ell)=-\mathcal{S} \varpi s_\Lambda,
\label{bseqn}
\end{eqnarray}
and hence the angle between the rotation axis and the angular momentum gradient is $\mathrm{cos}^{-1}\left(-\mathcal{S}s_\Lambda/|\nabla \ell|\right)$. It is helpful to also define a modified Richardson number
\begin{eqnarray}
\mathrm{R}=\frac{\mathcal{N}^2 \varpi}{2\Omega |\nabla\ell|},
\label{modifiedRi}
\end{eqnarray}
which is one possible measure of the ratio of the stabilising effects of stratification to the destabilising effects of the angular momentum gradient \citep[e.g.][]{KnoblochSpruit1982}. 
We can also derive an alternative form of the thermal wind equation, by using
Eq.~\ref{bseqn} to eliminate
$\mathcal{S}$ from the thermal wind equation Eq.~\ref{TWE}. Then equation Eq.~\ref{modifiedRi}
gives 
\begin{eqnarray}
s_{\gamma-\Lambda} = \mathrm{R} s_{\Gamma}.
\label{TWE2}
\end{eqnarray}

We show all of the vectors and corresponding angles on the $(x,z)$-plane in Fig.~\ref{Arrows}.
In  the case $\mathcal{S}>0$, corresponding to $\mathrm{d} \Omega /\mathrm{d}r <0$ as expected in stars, 
Eq.~\ref{gradldef} implies that $\gamma>\Lambda$ in the northern hemisphere, so Eq.~\ref{TWE2}
gives $\Gamma>0$. In the southern hemisphere, the signs of all the angles in Fig.~\ref{Arrows}
are reversed.

\subsection{Nondiffusive stability}

We first consider nondiffusive (adiabatic) stability, meaning the case with $\nu=\kappa=0$. The growth rate is determined by
\begin{eqnarray}
s^2 = -(a+b),
\end{eqnarray}
and hence we have stability when 
\begin{eqnarray}
\label{adcritineq}
a+b>0.
\end{eqnarray}
As it stands, this expression involves the wavevector orientation, and so must be manipulated to derive an expression that is independent of $\boldsymbol{k}$. This is best done by defining $p=k_x/k_z$, then Eq.~\ref{adcritineq} can be written as a quadratic for $p$:
\begin{eqnarray}
\label{adquadratic}
p^2 s_{\gamma}s_\Lambda+p (s_{\gamma+\Lambda}-\mathrm{R}s_\Gamma)+(\mathrm{R} c_\Gamma+c_\Lambda c_{\gamma})>0.
\end{eqnarray}
This is always satisfied if the left hand side has no real roots, i.e. if
\begin{eqnarray}
(s_{\gamma+\Lambda}-\mathrm{R} s_\Gamma)^2-4 s_{\gamma}s_\Lambda (\mathrm{R} c_\Gamma+c_\Lambda c_{\gamma})<0,
\label{rootcondition}
\end{eqnarray}
and we have $s_{\gamma} s_{\Lambda} > 0$. This latter condition is always satisfied in the northern hemisphere, since then Eq.~\ref{gradldef} implies $s_{\gamma}>0$, and in the southern hemisphere
both $s_{\gamma}$ and $s_{\Lambda}$ reverse signs, so it holds there too. 
Using Eq.~\ref{TWE2} to eliminate R from Eq.~\ref{rootcondition}, sufficient conditions for stability reduce to
\begin{eqnarray}
s_{\Lambda} s_{\gamma+\Gamma}>0. 
\label{dynamstab1}
\end{eqnarray}
This is equivalent to the Solberg-H\o iland criterion \citep{Solberg1936,Hoiland1941}: that the angular momentum must increase outwards on surfaces of constant entropy for adiabatic dynamical stability, i.e.~we require
\begin{eqnarray}
(\nabla \ell)\cdot \boldsymbol{e}_\theta^{\perp}<0,
\end{eqnarray}
when $\Lambda>0$ (and the opposite inequality when $\Lambda<0$). 
Using Eq.~\ref{gradldef}, Eq.~\ref{TWE}, and noting that in a radiative zone
$c_{\Gamma} = \sqrt{1 - s^2_{\Gamma}}>0$, the criterion Eq.~\ref{dynamstab1} can also be written as
\begin{eqnarray}
\left( 1 - \frac{4 \Omega^2 S^2 s^2_{\Lambda}}{\mathcal{N}^4} \right)^{1/2} > \ \ 
\frac{\mathcal{S}(\mathcal{S}-2 \Omega c_{\Lambda})}{\mathcal{N}^2}.
\label{dynamstab2}
\end{eqnarray}
In the case when $\mathcal{S}>0$ and the radial component of the angular momentum points 
outward, $\gamma < \pi/2$ in Fig.~\ref{Arrows}, and then Eq.~\ref{gradldef} shows $\mathcal{S} < 2 \Omega c_{\Lambda}$,
so Eq.~\ref{dynamstab2} shows there is always dynamical stability. In the opposite 
case, $\mathcal{S} > 2 \Omega c_{\Lambda}$, $\gamma > \pi/2$, we can square the inequality
to get (using Eq.~\ref{gradlsq})
\begin{eqnarray}
\label{adcrit2}
\varpi^2 \mathcal{N}^4>\mathcal{S}^2|\nabla \ell|^2.
\end{eqnarray}
The physical significance of Eq.~\ref{adcrit2} is that if the radial component of
the angular momentum gradient is inward, we need a sufficiently strong stable entropy gradient $\mathcal{N}^2$
to ensure dynamical stability. In this paper, we will primarily consider cases that are adiabatically stable according to Eq.~\ref{dynamstab1} but for which thermal diffusion enables the GSF instability.

We can also show that Eq.~\ref{dynamstab1} is equivalent to Eq. 31 in \cite{KnoblochSpruit1982}. The angles in their figure 4 correspond (if positive) 
to the case $\mathcal{S}<0$, so angular velocity increasing outward. To recover their
result we must set take our $\Gamma <0$, in which case our $\Lambda > \gamma$. Then if the various angles interchanged according to their$\rightarrow$our: $\Lambda\rightarrow\Lambda$,
$\theta\rightarrow\Gamma+\Lambda$, $\Gamma\rightarrow \Lambda-\gamma$. 

In the absence of stable stratification, i.e. if $\mathcal{N}^2=0$, the thermal wind equation Eq.~\ref{TWE2} means that either $\gamma=\Lambda$, in which case the angular momentum
increases in the $ \hat{\boldsymbol{\Omega}}^\perp$ direction,  
\begin{eqnarray}
\nabla \ell \cdot \hat{\boldsymbol{\Omega}}^{\perp} >0,
\label{Rayleigh}
\end{eqnarray}
and so is stable by the Rayleigh criterion, or $\gamma=\Lambda + \pi$, in which case angular momentum decreases outward, which is the Rayleigh unstable case. At the equator, the GSF instability occurs only if this criterion is not satisfied. It is one of our primary goals to explore the efficiency of the non-equatorial GSF instability in the regime of weaker differential rotation in which this criterion (and Eq.~\ref{adcrit2}) is satisfied, but the system is nonetheless unstable to the (diffusive) GSF instability.

Finally, we consider the case where the radial entropy gradients are much larger than the latitudinal gradients, i.e.~$\mathrm{R} \gg1$, as is frequently the case in stars. In this limit, Eq.~\ref{TWE2} implies $\Gamma$ is small, so in Eq.~\ref{adquadratic} the
$s_{\Gamma}$ term is negligible and $c_{\Gamma} \approx 1$, so the nondiffusive stability criterion at large $R$ is
\begin{eqnarray}
\label{adcrit3}
\mathrm{R} > \frac{s_{\gamma-\Lambda}^2}{4 s_\Lambda s_{\gamma}}=\frac{\mathcal{S}^2 \varpi}{8 |\nabla \ell|\Omega}, \ \ 
\textrm{or} \ \ \mathrm{Ri} > \frac{1}{4},
\end{eqnarray}
using Eq. \ref{modifiedRi} and Eq. \ref{Ridef}.

\subsection{Diffusive (GSF) instability}

Thermal diffusion enables instability even if the differential rotation is adiabatically stable. This is referred to as the GSF instability, and is the primary focus of this paper. We can derive a criterion for the onset of steady modes (which are the relevant ones e.g.~\citealt{Knobloch1982}) by considering when the constant term in Eq.~\ref{cubic} becomes negative, i.e.~when
\begin{eqnarray}
\label{consttermcubic}
a+\mathrm{Pr}b + \nu^2k^4<0.
\end{eqnarray}
If the stratification is stabilising $b>0$, so for diffusive instability $a$ must be negative. Note that even though Pr is small, Ri might be large, so the term $\mathrm{Pr} b$ is not necessarily small. Following a similar approach to Eq.~\ref{adcrit3}, we obtain the following criterion for instability in the strongly stratified limit (so that $\Gamma\approx 0$):
\begin{eqnarray}
\label{GSFstrongstrat}
\mathrm{R Pr}<\frac{s_{\gamma-\Lambda}^2}{4s_\Lambda s_{\gamma}}
=\frac{\mathcal{S}^2 \varpi}{8 |\nabla \ell|\Omega}, \ \ 
\textrm{or} \ \ \mathrm{Ri Pr}  < \frac{1}{4}. \label{RiPrcrit}
\end{eqnarray}
This is equivalent to \cite{KnoblochSpruit1982} Eq.~34, and it 
must be satisfied for the occurrence of the GSF instability at a general latitude. Given that $\mathrm{Pr}\ll 1$ in stars, this criterion can easily be  satisfied even when the nondiffusive stability criterion Eq.~\ref{adcrit3}
is satisfied. This criterion was derived by \cite{Rashid2008} at the poles ($\Lambda=90^\circ$), but we have just demonstrated that this result holds for any latitude $\Lambda\ne 0$ if we adopt a shellular profile of differential rotation. At the equator, instability occurs if
\begin{eqnarray}
\kappa_{ep}^2=2\Omega(2\Omega-\mathcal{S})<0,
\end{eqnarray}
which implies that much stronger differential rotation is required there.

\subsubsection{Limit as $\mathrm{Pr}\rightarrow 0$, with $\mathrm{RiPr}\rightarrow 0$}
\label{limprto0}

Since $\mathrm{Pr}$ is very small in stellar interiors, we now consider the properties of the instability in the limit $\mathrm{Pr}\rightarrow 0$, with $\mathrm{RiPr}\rightarrow 0$. This may be relevant for rapidly 
rotating stars, since then $\mathrm{Ri}$ is not so large, allowing 
$\mathrm{RiPr}$ to be small. 
In this limit, taking $\mathcal{S}\sim O(\Omega)$, $a$ and $b$ are $O(\Omega^2)$,
$s \sim O(\Omega)$, and $k^2 \sim O(\Omega/\sqrt{\kappa \nu})$.
Then Eq.~\ref{cubic} reduces to 
\begin{eqnarray}
s^2=-a=-\frac{2 \Omega |\nabla \ell|}{\varpi}\left(\hat{\boldsymbol{k}}\cdot\hat{\boldsymbol{\Omega}}\right)\left(\hat{\boldsymbol{k}}\cdot\hat{(\nabla \boldsymbol{\ell})}^{\perp}\right),
\label{smallRiPrdisp_eqn}
\end{eqnarray}
which indicates that stability is determined by the sign of $a$. Marginal stability ($s=0$) occurs when the wavevector is either perpendicular to the rotation axis, meaning that $\hat{\boldsymbol{k}}\cdot\boldsymbol{\Omega}=0$ (with motions that are parallel to the rotation axis), or when the wavevector is parallel to the angular momentum gradient $(\nabla \ell)$, meaning that $\hat{\boldsymbol{k}}\cdot (\nabla \ell)^{\perp}=0$ (with corresponding motions that are perpendicular to the angular momentum gradient, or along surfaces of constant angular momentum). We will show below that in this scaling the fastest growing modes have a wavevector a
ngle that is half-way between the two unit vectors $\hat{\boldsymbol{\Omega}}^\perp$ and $\hat{(\nabla \ell)}$, i.e. between the rotation axis and a surface of constant angular momentum (see also \citealt{KnoblochSpruit1982}).

In the limit of small Pr, the stabilising effects of the stratification have been eliminated and the growth rate is independent of $\mathrm{Pr}$, $N^2$ and $\Gamma$. The fastest growing mode can be determined by maximising $a$ with respect to the wavevector orientation (or w.r.t. both $k_x$ and $k_z$). We find
\begin{eqnarray}
\label{FGMdirection}
\frac{k_z}{k_x}
&=& -\mathrm{tan}\left(\frac{1}{2}\left(\gamma+\Lambda\right)\right) \quad\left(\text{or} \;\; \mathrm{cot}\left(\frac{1}{2}\left(\gamma+\Lambda\right)\right)\right) \\
&=&\frac{\mathcal{S}c_\Lambda-2\Omega c_{2\Lambda}+|\nabla \ell|/\varpi}{(\mathcal{S}-4\Omega c_\Lambda)s_\Lambda}.
\label{FGMdirection1}
\end{eqnarray}
This implies that the wavevector of the fastest growing mode  in this limit lies half-way between $\boldsymbol{\Omega}^{\perp}$ and $\nabla \ell$. Note that at the pole, $k_z/k_x\approx 4\Omega/\mathcal{S}$ in the limit $\Omega^2\gg \mathcal{S}^2$, which agrees with \cite{Rashid2008} Eq.~35 (noting that our $k_z/k_x\rightarrow -k_y/k_z$ in their notation). At the equator, $k_z/k_x\rightarrow \infty$, indicating that the instability preferentially excites elevator modes with $k_x=0$, as shown in paper I.

The growth rate of the fastest growing mode satisfying Eq.~\ref{FGMdirection} is then
\begin{eqnarray}
\label{svalue_1}
s^2=\frac{2\Omega|\nabla \ell|}{\varpi} \sin^2 \left(\frac{1}{2}(\gamma-\Lambda)\right),
\end{eqnarray}
and this is maximal when the unstable wedge is as wide as possible. 
This can be re-written as
\begin{eqnarray}
\label{FGMgrowthlowPr}
s^2=\Omega (|\nabla\ell|/\varpi +\mathcal{S}c_\Lambda-2\Omega).
\end{eqnarray}
At the equator, $\Lambda=0$, therefore $|\nabla \ell|/\varpi=2\Omega- \cal{S}$ so that the above expression reduces to $s^2=-\kappa_{ep}^2= 2\Omega(\mathcal{S}-2\Omega)$, which agrees with the result derived in paper I. 

By maximising Eq.~\ref{cubic} with respect to $k^2$ in this limit (noting that $a$ and $b$ only depend on the wavevector orientation and not its magnitude), we may show that
\begin{eqnarray}
\label{kvalue_1}
k^4  = \frac{1}{2 d^4}\sin^2 \left( \frac{\gamma + \Lambda}{2} \right) ,
\end{eqnarray}
independently of $\mathrm{Pr}$. This is consistent with the results of paper I at the equator, where $k\rightarrow 2^{-1/4}d^{-1}$. We have therefore obtained asymptotic expressions for the fastest growing wavenumber $k$, the wavevector orientation $k_z/k_x$, and the corresponding growth rate $s$ for the case of small $\mathrm{Pr}$ and finite $\mathrm{Ri}$. We will later use these results.

In Appendix~\ref{RiPrO1}, we present a complementary asymptotic analysis to explore the limit as $\mathrm{Pr}\to 0$ with $\mathrm{RiPr}=O(1)$. This analysis extends \cite{Rashid2008} to a general latitude.

\subsubsection{Properties of the instability; an illustrative case}

\begin{figure}
  \begin{center}
    \subfigure{\includegraphics[trim=4cm 0cm 4.5cm 0cm, clip=true,width=0.4\textwidth]{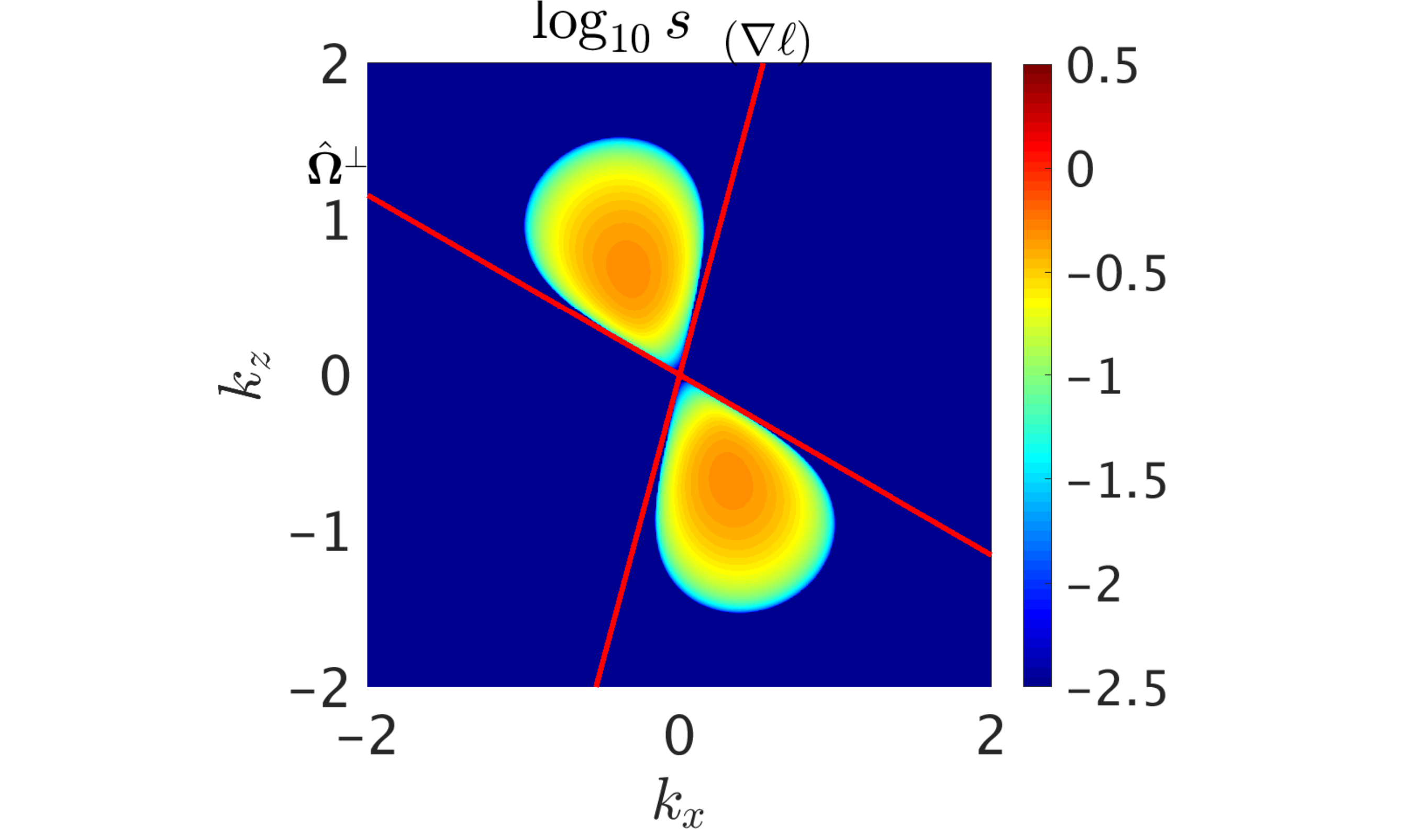}}
    \end{center}
  \caption{Logarithm of the linear growth rate ($\log_{10}s/\Omega$) for the axisymmetric GSF instability on the $(k_x,k_z)$-plane with $S=2, \Lambda=30^{\circ}, N^2=10, \mathrm{Pr}=10^{-2}$. The solid red lines demarcate the region of linear instability, and are parallel to the vectors $\boldsymbol{\Omega}^{\perp}$ and $\nabla \ell$).
   Unstable modes are contained within the wedge bounded by these two vectors.}
  \label{growthrate}
\end{figure}

Fig.~\ref{growthrate} shows the base 10 logarithm of the growth rate from solving Eq.~\ref{cubic} on the $(k_x,k_z)$-plane with $S=\mathcal{S}/\Omega=2 , \Lambda=30^{\circ}, N^2={\mathcal{N}}^2/\Omega^2=10$ and $\mathrm{Pr}=10^{-2}$. For these parameters, $\mathrm{Ri}=2.5$ and $\mathrm{E}=10^{-0.5}$. The red solid lines are parallel to the vectors $\hat{\boldsymbol{\Omega}}^{\perp}$ and $(\nabla \ell)$, which represent the boundaries of the unstable region, in accordance with our above discussion. The fastest growing modes with growthrates $O(1)$ are observed to lie along the line that is approximately half-way between these two vectors, as expected. The corresponding velocity perturbation for the fastest growing mode in the $(x,z)$-plane is perpendicular to this, since $\boldsymbol{k}\cdot\boldsymbol{u}=0$. 
This figure also shows that the wavelength of the fastest growing modes in this case are $O(d)$. Note that this value of $S$ would be marginally stable at the equator even if $N^2=0$. The presence of instability here illustrates that weaker shears are required to excite the GSF instability at non-equatorial latitudes.

Non-axisymmetric disturbances tend to orient themselves along the gradient of $\boldsymbol{\Omega}$, and therefore these modes becomes stable after some point in their evolution, and hence ultimately decay (e.g.~\citealt{LatterPap2018}). Hence, we have focussed on axisymmetric disturbances in this section, since they are likely to be the most important linear modes. Non-axisymmetric modes are likely to be essential for the nonlinear evolution however.

One might suppose that the GSF instability will saturate by transporting angular momentum to modify the mean flow, to the extent that the boundary conditions allow this, such that $(\nabla \ell)^{\perp}$ coincides with $\hat{\boldsymbol{\Omega}}$ i.e. by eliminating the unstable wedge, driving the system towards marginal stability. We will later show that our simulations provide some support for this hypothesis.

\section{Illustrative nonlinear results with $\Gamma=30^{\circ}$}
\label{results30}
\begin{table}
\begin{tabular}{cccccccc}
\hline
$S$ & Ri & $\Gamma$ & $\gamma$ & $\gamma-\Lambda$ &  $s_\mathrm{max}$ & $\theta_k$ & $k$ \\
\hline 
1 & 10 & $5.74^{\circ}$ & $53.8^{\circ}$ & $23.8^{\circ}$ & 0.065 & $40.0^{\circ}$ & 0.55  \\
1.5 & 4.44 & $8.63^{\circ}$ & $76.9^{\circ}$ & $46.9^{\circ}$ & 0.24 & $49.9^{\circ}$ & 0.67  \\
2 & 2.5 &  $11.54^{\circ}$ & $105^{\circ}$ & $75^{\circ}$ & 0.49 & $64.2^{\circ}$ & 0.74 \\
2.5 & 1.6 & $14.5^{\circ}$ & $127.5^{\circ}$ & $97.5^{\circ}$ & 0.78 & $77.8^{\circ}$ & 0.77 \\
3 & 1.11 & $17.5^{\circ}$ & $141.7^{\circ}$ & $111.7^{\circ}$ & 1.08 & $87.0^{\circ}$ & 0.78 \\
\hline
\end{tabular}
\caption{Table of the various angles and parameters for all simulations performed with $\Lambda=30^{\circ}$, $\mathrm{Pr}=10^{-2}$, $N^2=10$. The latter three columns give the growth rate (units of $\Omega$) and the angle and the magnitude (units of $d^{-1}$) of the wavenumber of the fastest growing mode.}
\label{TableAngles}
\end{table}

Our primary aim is to understand the nonlinear evolution of the non-equatorial GSF instability, and to quantify its angular momentum transport. In this section we present some illustrative nonlinear axisymmetric and 3D simulations with $\Lambda=30^{\circ}$, using dimensionless quantities throughout. We will assume $\mathrm{Pr}=10^{-2}$, $N^2=10$ and consider a range of values of $S$, noting that we are once again using the non-dimensional quantities specified in \S~\ref{model}. We will also vary $L_y$ to probe the importance of 3D effects, and we will take $L_x=L_z=100$ except where specified otherwise. With these parameters, the critical values of $S$ delineating the various regimes are: Solberg-H\o iland stability (Eq.~\ref{adcrit2}) if $S<4.01$ ($\mathrm{Ri}>0.622$) and GSF instability (Eq.~\ref{RiPrcrit}) if $S>0.633$ ($\mathrm{Ri}<25$). In the absence of stable stratification, we would also have Rayleigh stability (Eq.~\ref{Rayleigh}) if $S<2.31$ ($\mathrm{Ri}>1.87$). We consider the evolution for a number of cases in the various regimes. In the GSF-unstable cases with weak shears (that would be Rayleigh-stable), we have $S=1$, $1.5$ and $2$ ($\mathrm{Ri}=10$, $4.4$ and $2.5$). In the GSF-unstable regime with stronger shears (that would be Rayleigh-unstable) we have $S=2.5$ and $3$ ($\mathrm{Ri}=1.6$ and $1.11$). Note that, $S>2$ would be required for instability at the equator ($\Lambda=0$). Table~\ref{TableAngles} lists the various angles from linear theory for these simulations, as well as predictions for the maximum growth rate and corresponding wavenumber. Table~\ref{Table} lists the simulation parameters.

\subsection{$S=2$ with shearing-periodic BCs: axisymmetric case}
\label{Seq2shearingperiodicAxi}

\begin{figure}
  \begin{center}
      \subfigure{\includegraphics[trim=1cm 0cm 3cm 0cm, clip=true,width=0.45\textwidth]{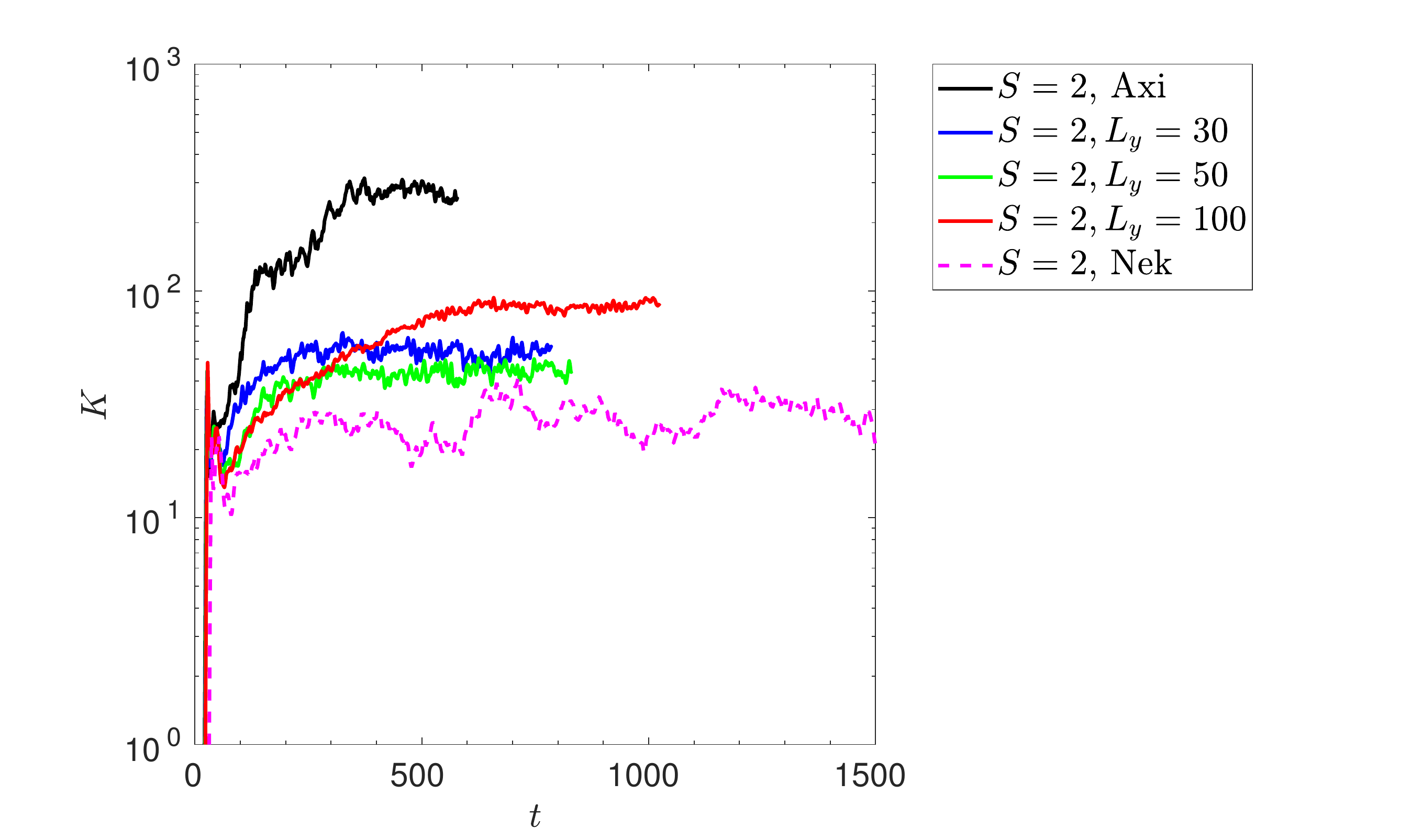}}
       \subfigure{\includegraphics[trim=1cm 0cm 3cm 0cm, clip=true,width=0.45\textwidth]{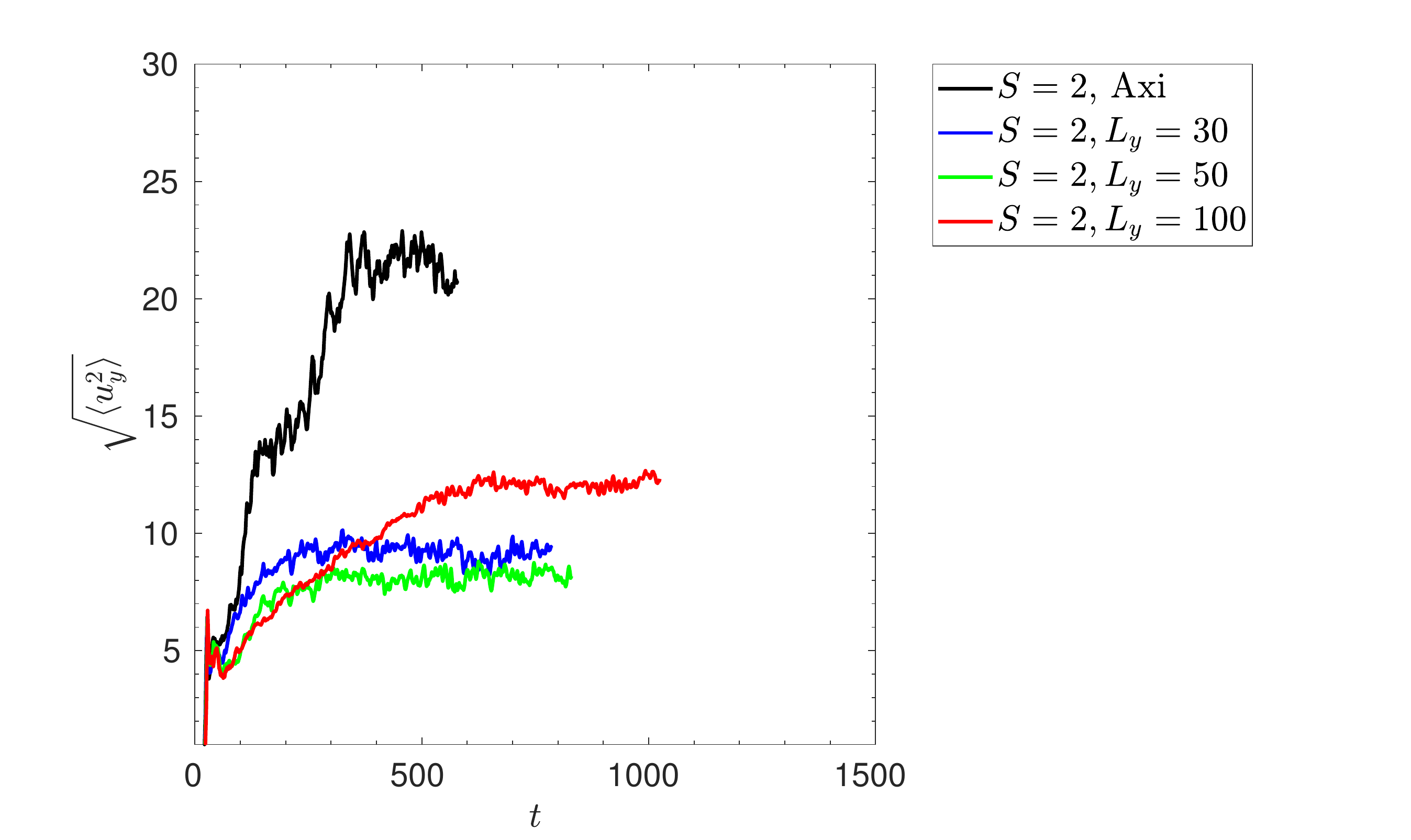}}
      \subfigure{\includegraphics[trim=1cm 0cm 3cm 0cm, clip=true,width=0.45\textwidth]{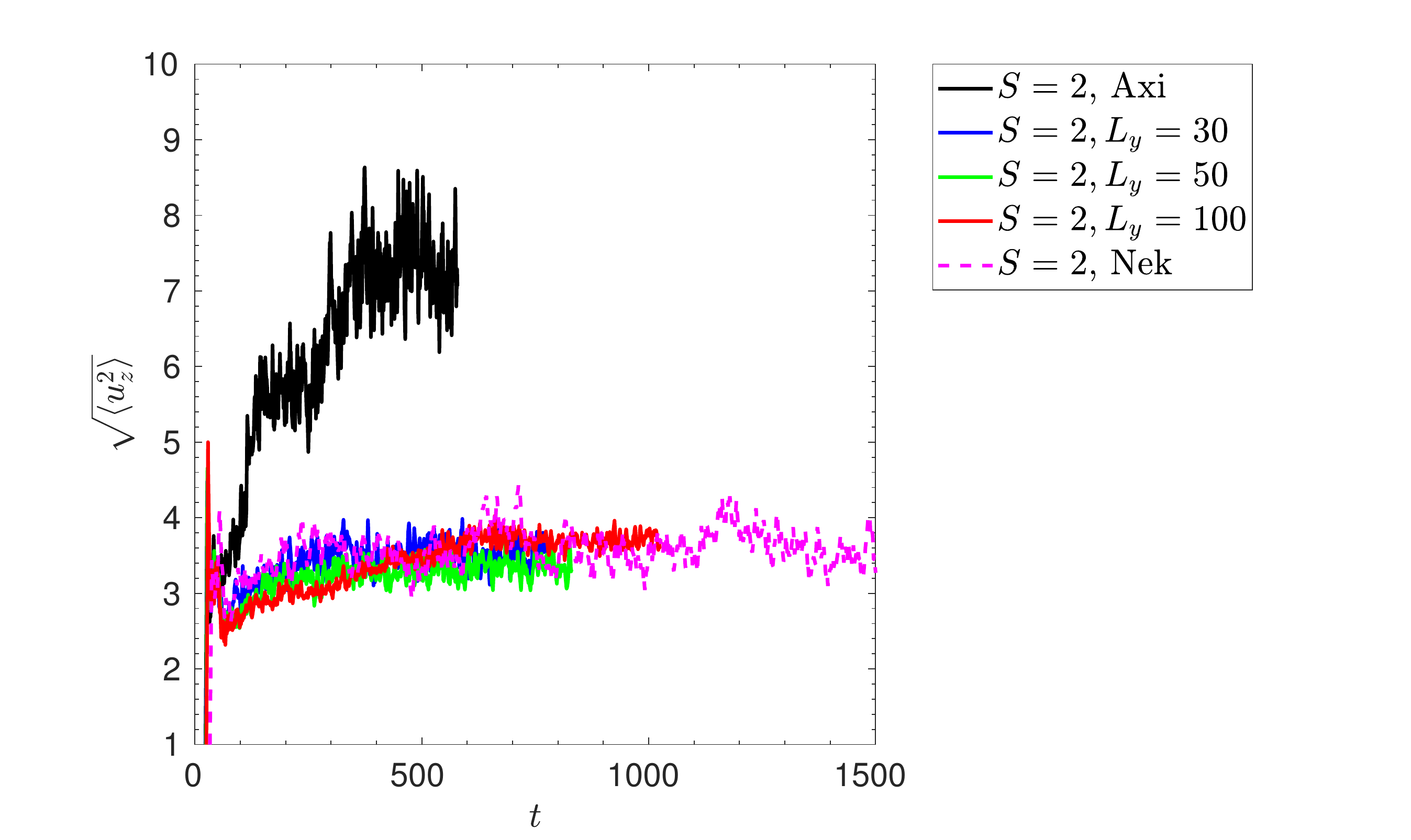}}
    \end{center}
  \caption{Temporal evolution of $K$, $v_y$ and $v_z$ in a set of simulations with S = 2, $\Lambda=30^{\circ}$, $N^2 = 10$, and $\mathrm{Pr} = 10^{-2}$, with various different $L_y$. The axisymmetric simulation exhibits much stronger flows than the 3D simulations, but there is only a weak dependence on $L_y\ne 0$. We have also plotted a simulation performed with stress-free radial boundaries in the top and bottom panels (labelled `Nek'), which will be discussed in \S~\ref{Seq2stressfree}.}
  \label{Seq2}
\end{figure}

\begin{figure}
  \begin{center}
      \subfigure{\includegraphics[trim=1cm 0cm 3cm 0cm, clip=true,width=0.45\textwidth]{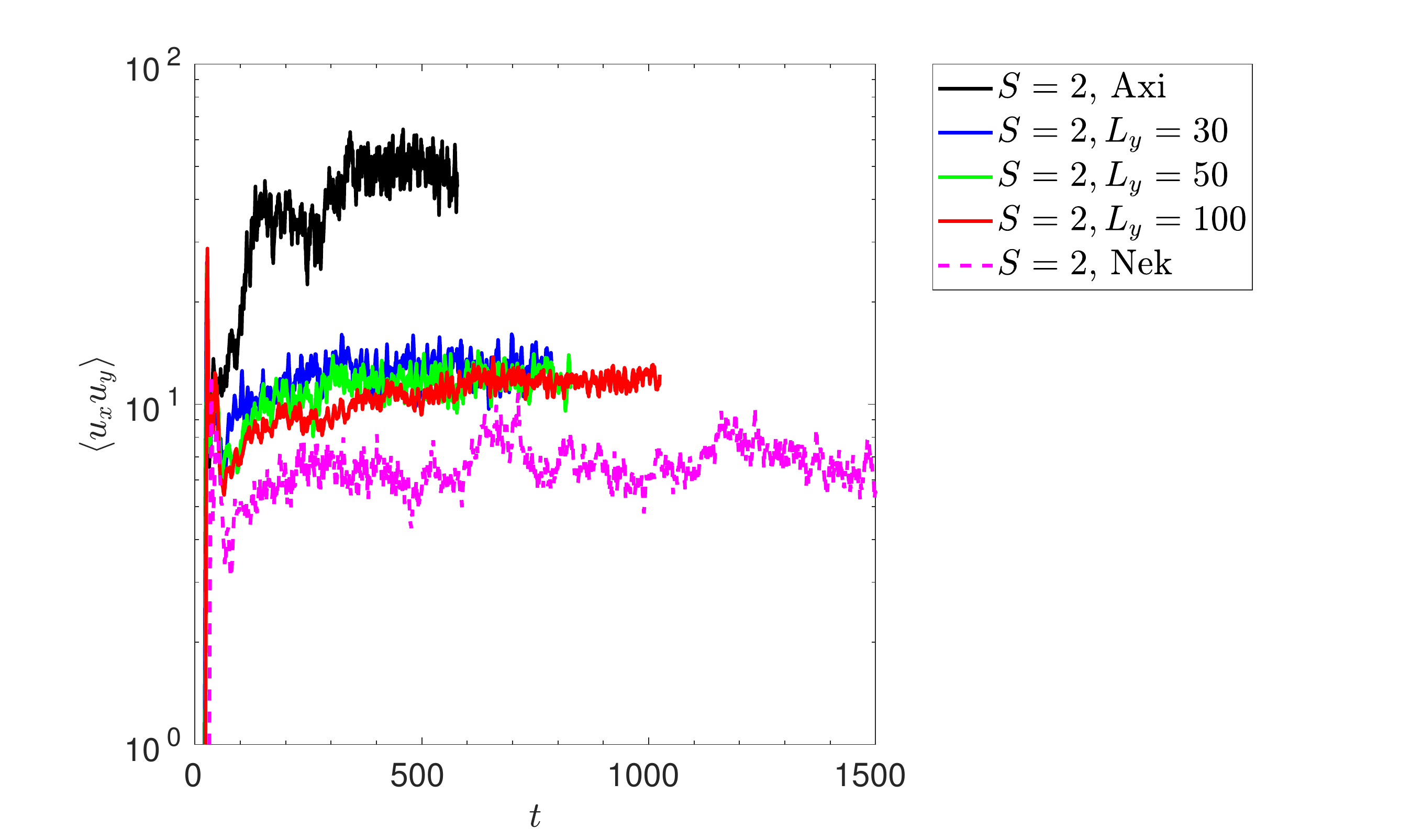}}
      \subfigure{\includegraphics[trim=1cm 0cm 3cm 0cm, clip=true,width=0.45\textwidth]{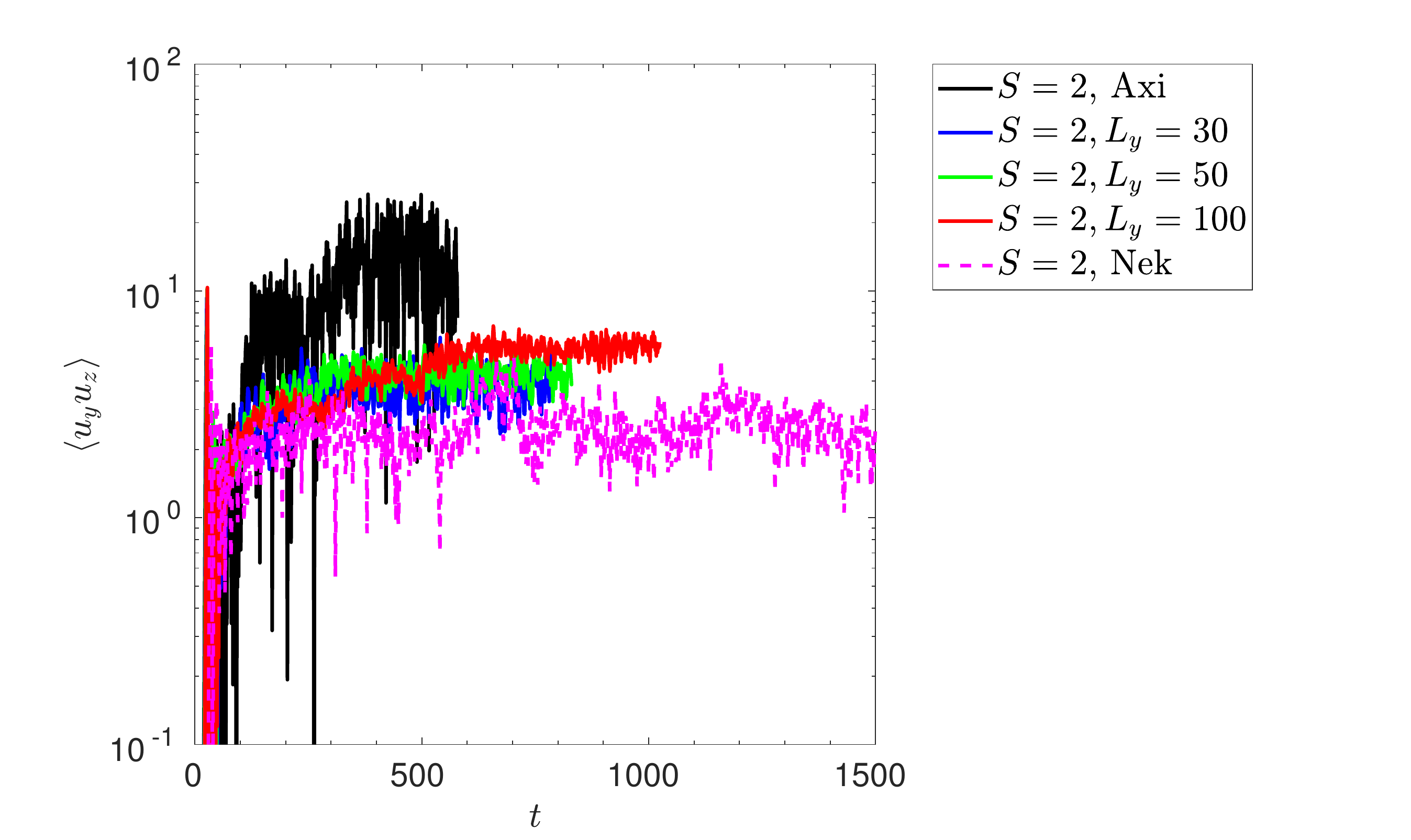}}
      \subfigure{\includegraphics[trim=1cm 0cm 3cm 0cm, clip=true,width=0.45\textwidth]{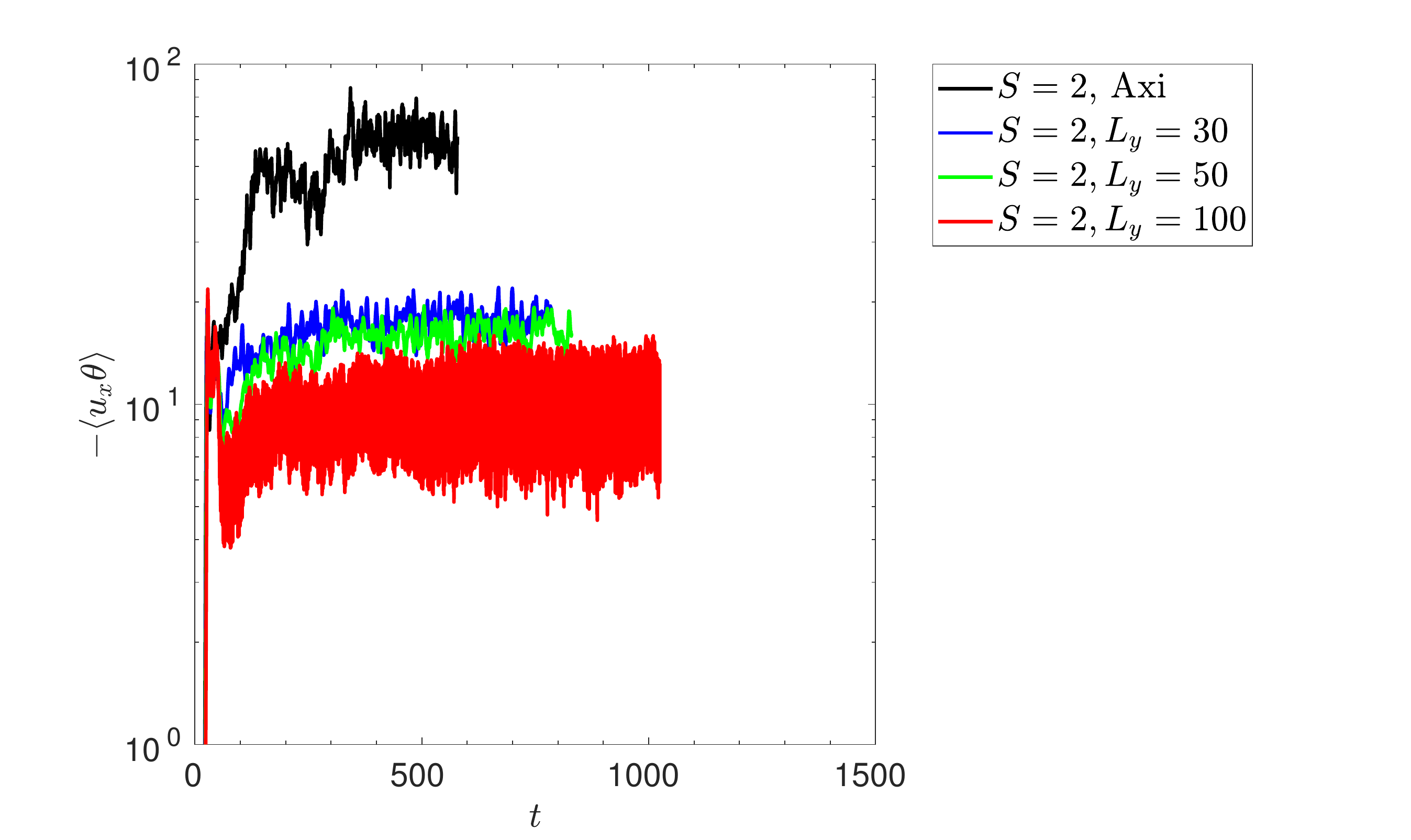}}
    \end{center}
  \caption{Temporal evolution of $\langle u_xu_y\rangle$, $\langle u_yu_z\rangle$ and $-\langle u_x\theta\rangle$ in a set of simulations with $S = 2$, $\Lambda=30^{\circ}, N^2 = 10$, and $\mathrm{Pr} = 10^{-2}$,
  with various different $L_y$. The axisymmetric simulation transports momentum and heat much more efficiently than the 3D simulations, but there is only a weak dependence on $L_y\ne 0$. We have also plotted a simulation performed with stress-free radial boundaries in the top and middle panels (labelled `Nek'), which will be discussed in \S~\ref{Seq2stressfree}.}
  \label{Seq2a}
\end{figure}

\begin{figure*}
  \begin{center}
     \subfigure[$t=22$]{\includegraphics[trim=4cm 0cm 5cm 0cm, clip=true,width=0.4\textwidth]{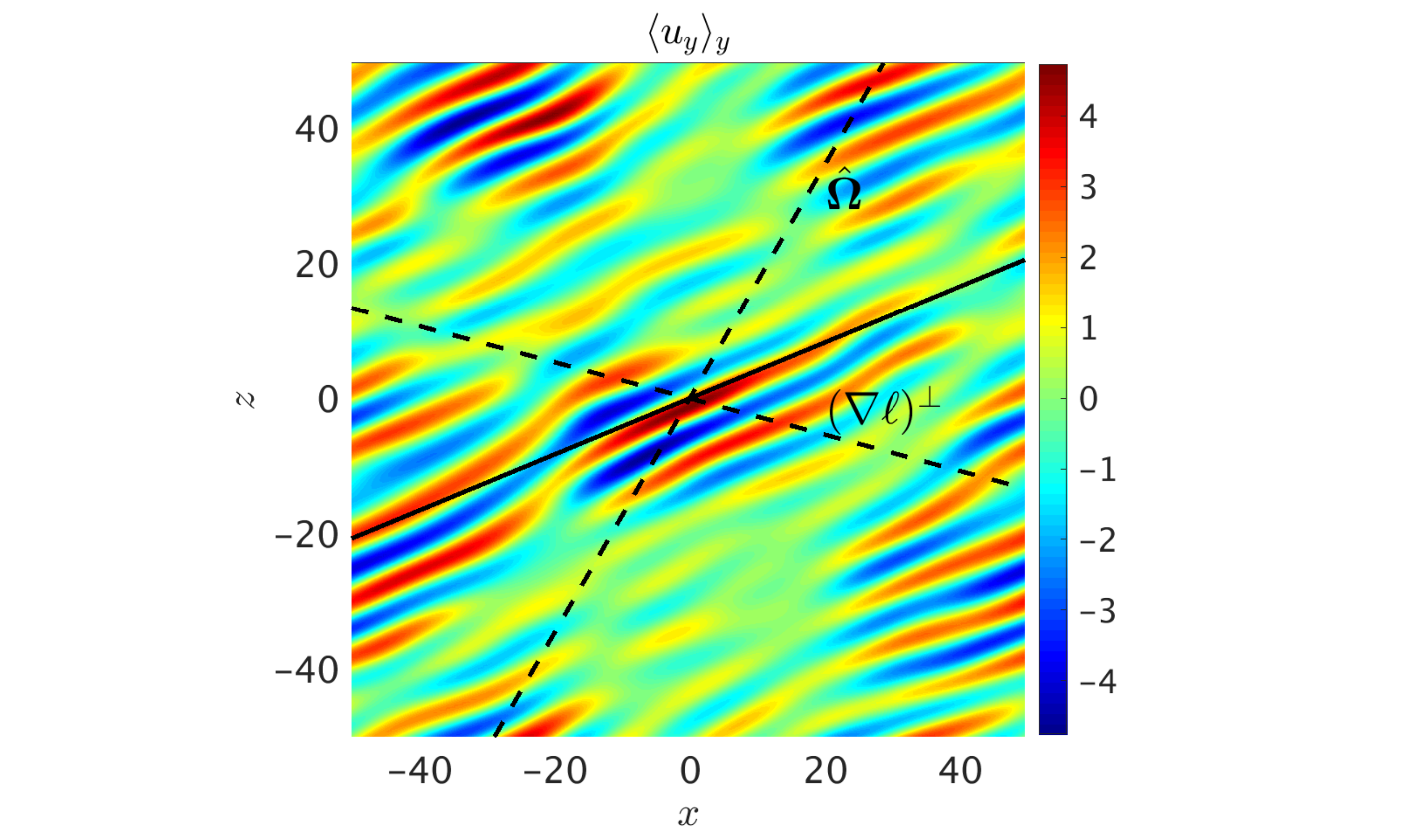}}
    \subfigure[$t=50$]{\includegraphics[trim=4cm 0cm 5cm 0cm, clip=true,width=0.4\textwidth]{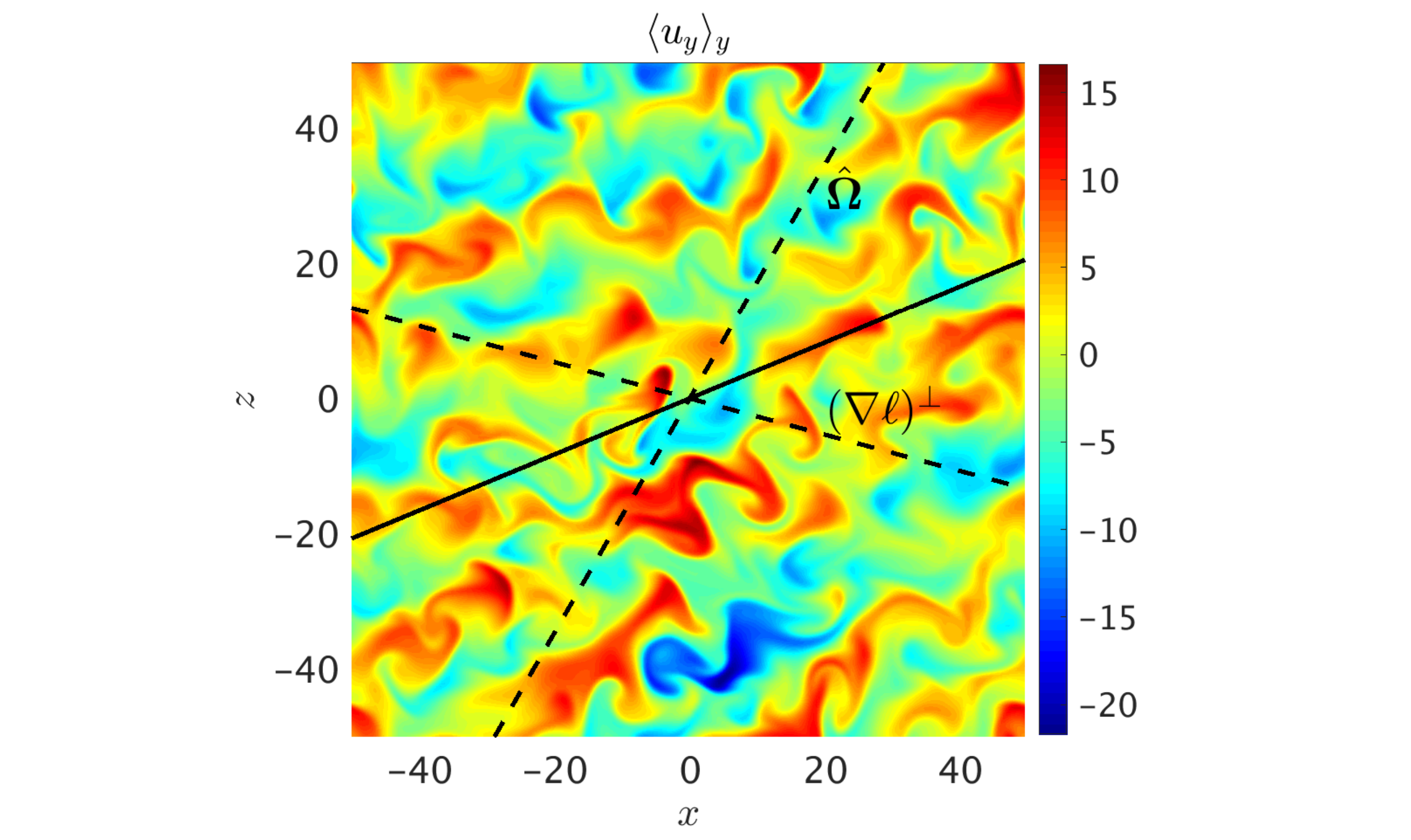}}
    \subfigure[$t=100$]{\includegraphics[trim=4cm 0cm 5cm 0cm, clip=true,width=0.4\textwidth]{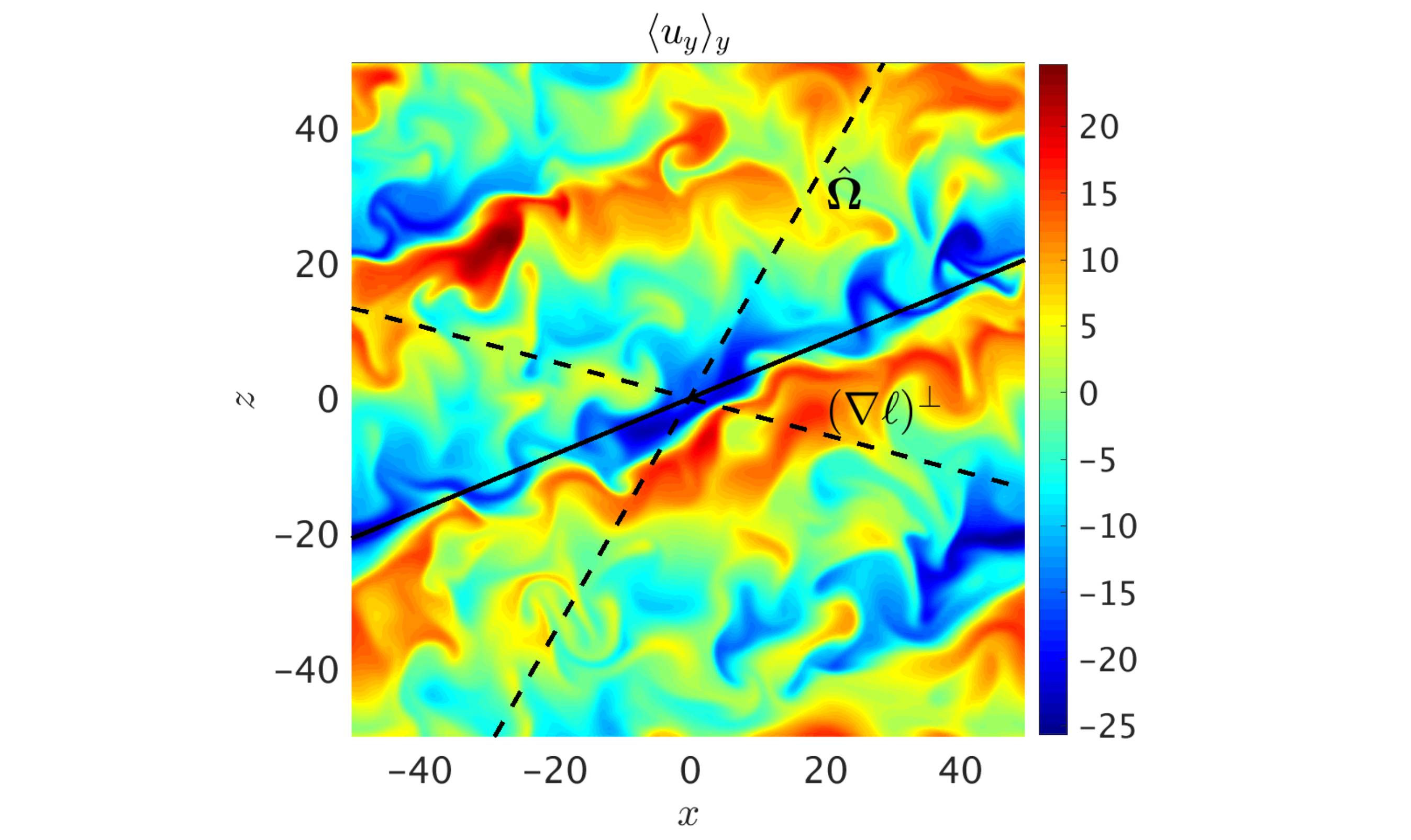}}
    \subfigure[$t=300$]{\includegraphics[trim=4cm 0cm 5cm 0cm, clip=true,width=0.4\textwidth]{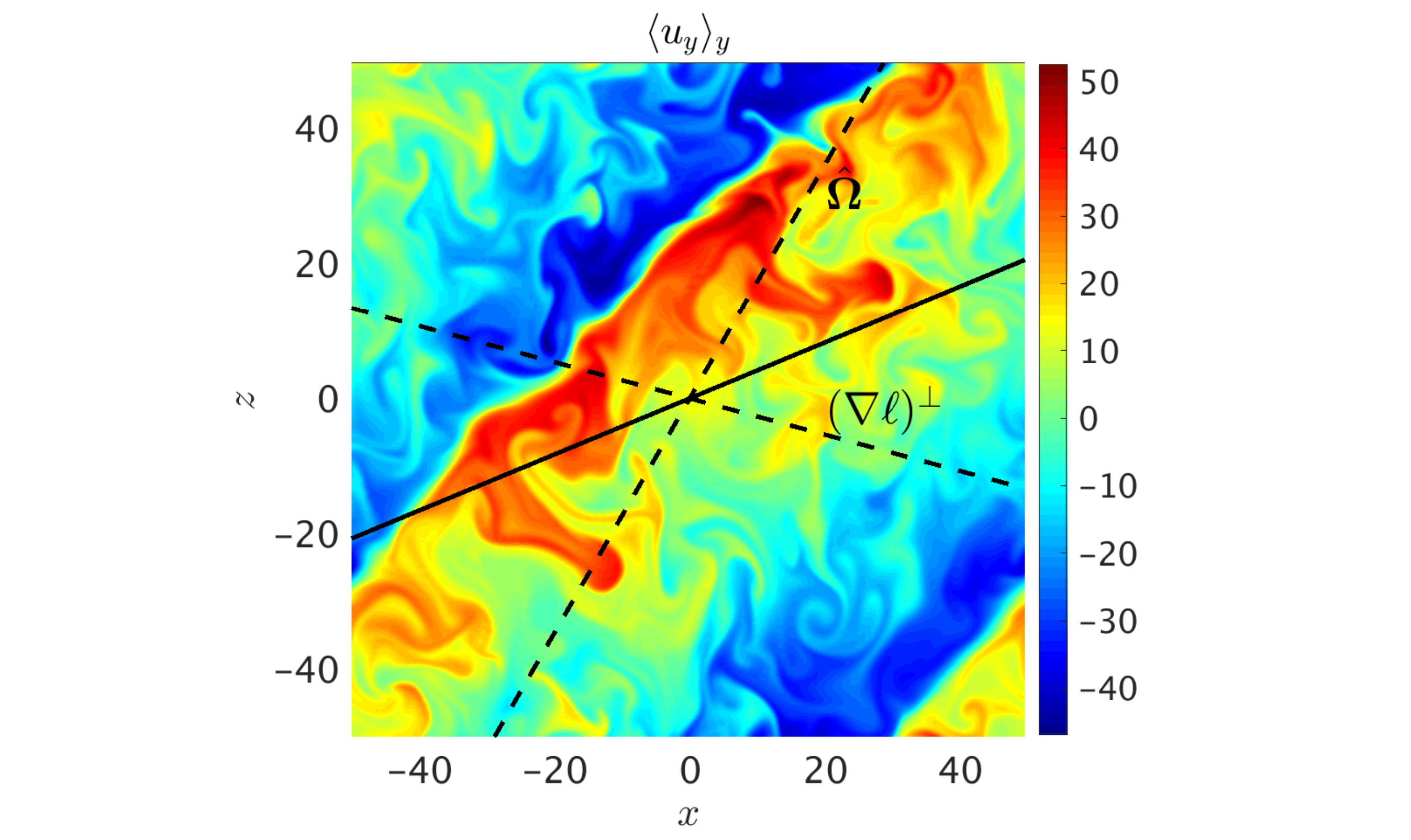}}
    \end{center}
  \caption{Snapshots of $u_y$ in the $(x, z)$-plane for an axisymmetric simulation with $S = 2$, $\Lambda=30^{\circ}$, $N^2 = 10$, and $\mathrm{Pr} = 10^{-2}$, at various times. The top panel shows the linear growing modes, which are slanted along the black solid line, which is half-way between the rotation axis and a surface of constant angular momentum (shown as black dashed lines). The remaining panels show the formation of zonal jets that merge and strengthen until they occupy the full extent of the box.}
  \label{Seq2_2D}
\end{figure*}

\begin{figure}
  \begin{center}
    \subfigure{\includegraphics[trim=4cm 0cm 5cm 0cm, clip=true,width=0.4\textwidth]{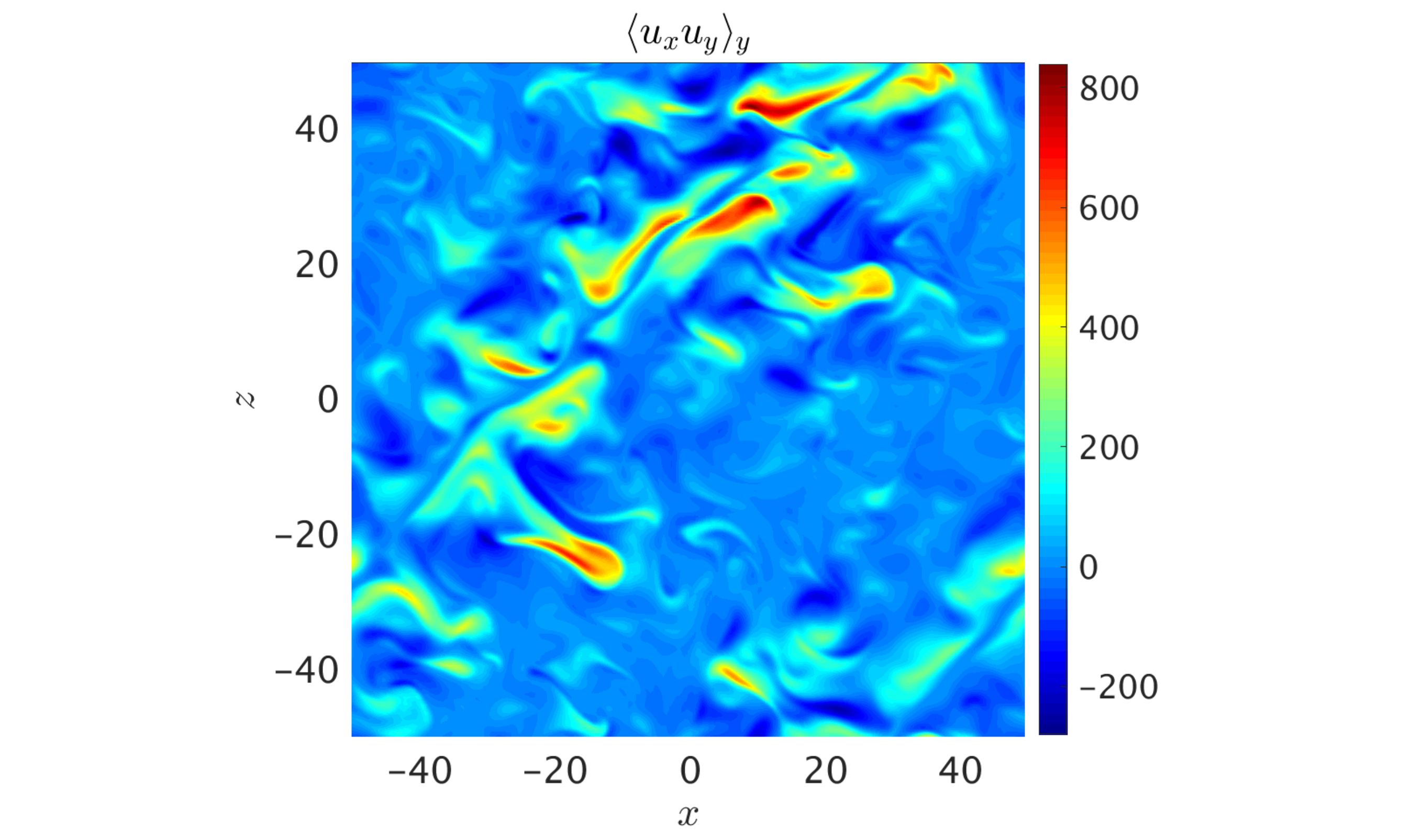}}
    \end{center}
  \caption{Snapshot of $u_x u_y$ in the $(x, z)$-plane at $t=300$ in the axisymmetric simulation with $S = 2$, $\Lambda=30^{\circ}$, $N^2 = 10$, and $\mathrm{Pr} = 10^{-2}$. Comparing this with Fig.~\ref{Seq2_2D} shows that momentum transport is dominated by the interfaces between layers.}
  \label{Seq2_2Duxuy}
\end{figure}

\begin{figure*}
  \begin{center}
     \subfigure[$t=22$]{\includegraphics[trim=4cm 0cm 5cm 0cm, clip=true,width=0.35\textwidth]{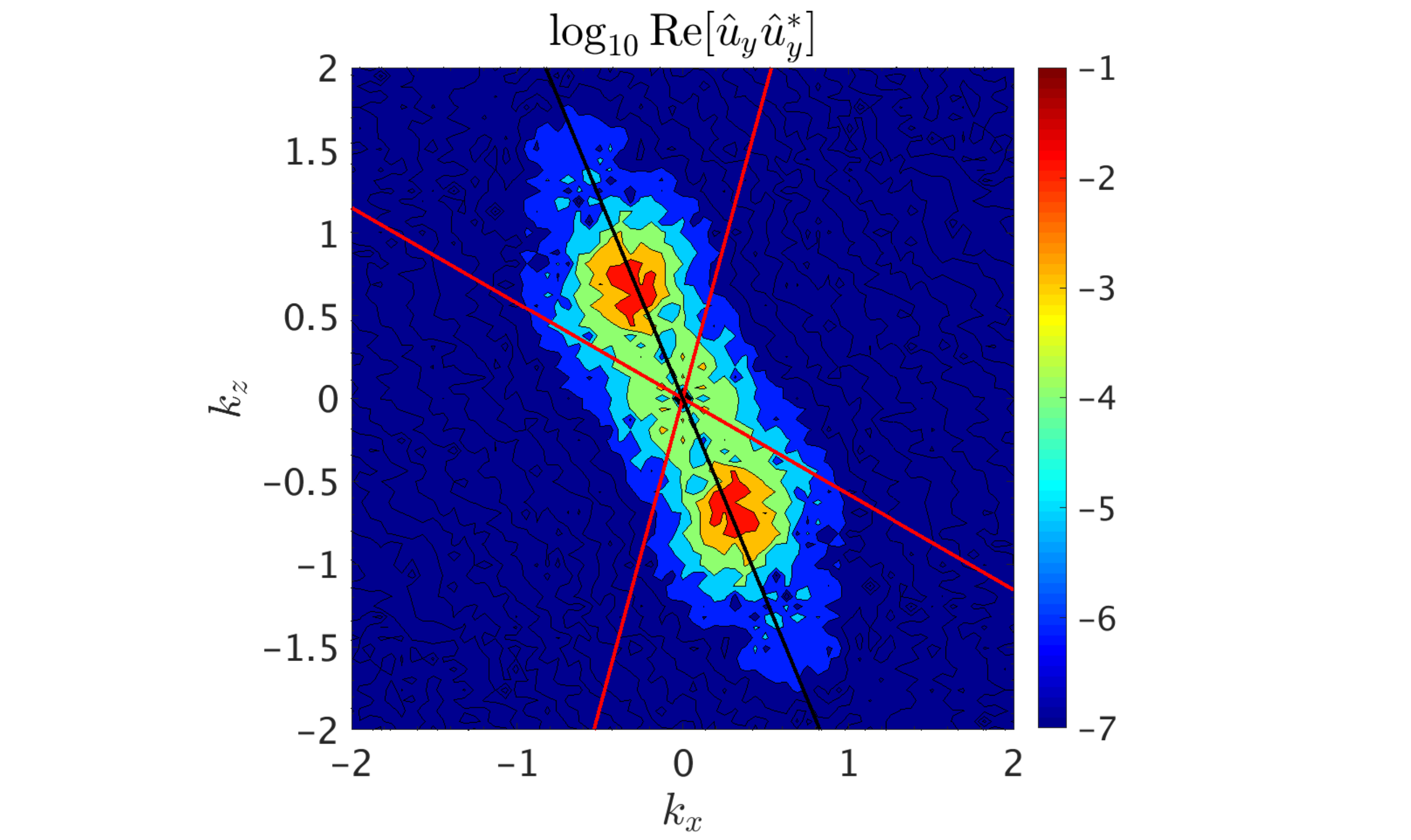}}
     \subfigure[$t=40$]{\includegraphics[trim=4cm 0cm 5cm 0cm, clip=true,width=0.35\textwidth]{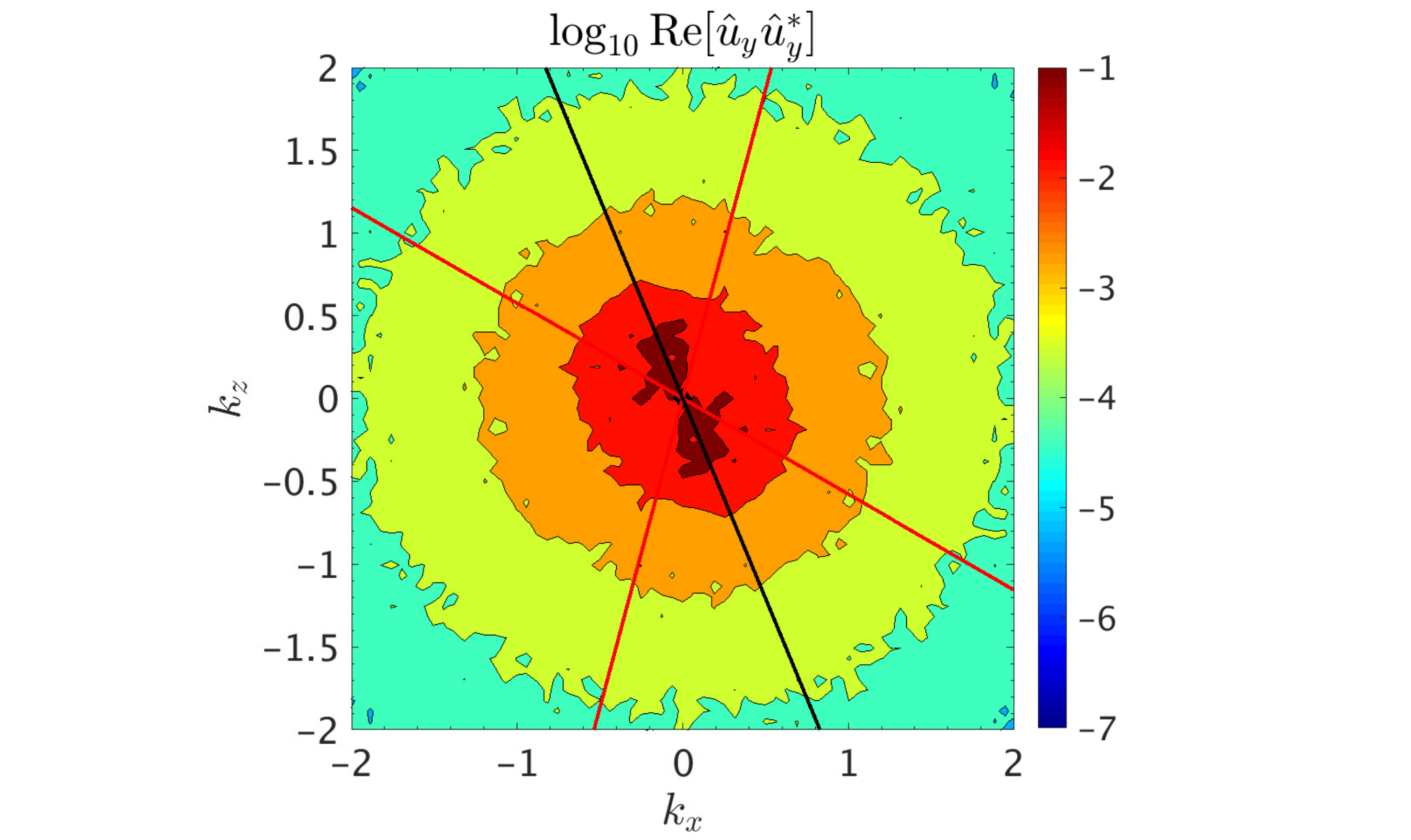}}
     \subfigure[$t=100$]{\includegraphics[trim=4cm 0cm 5cm 0cm, clip=true,width=0.35\textwidth]{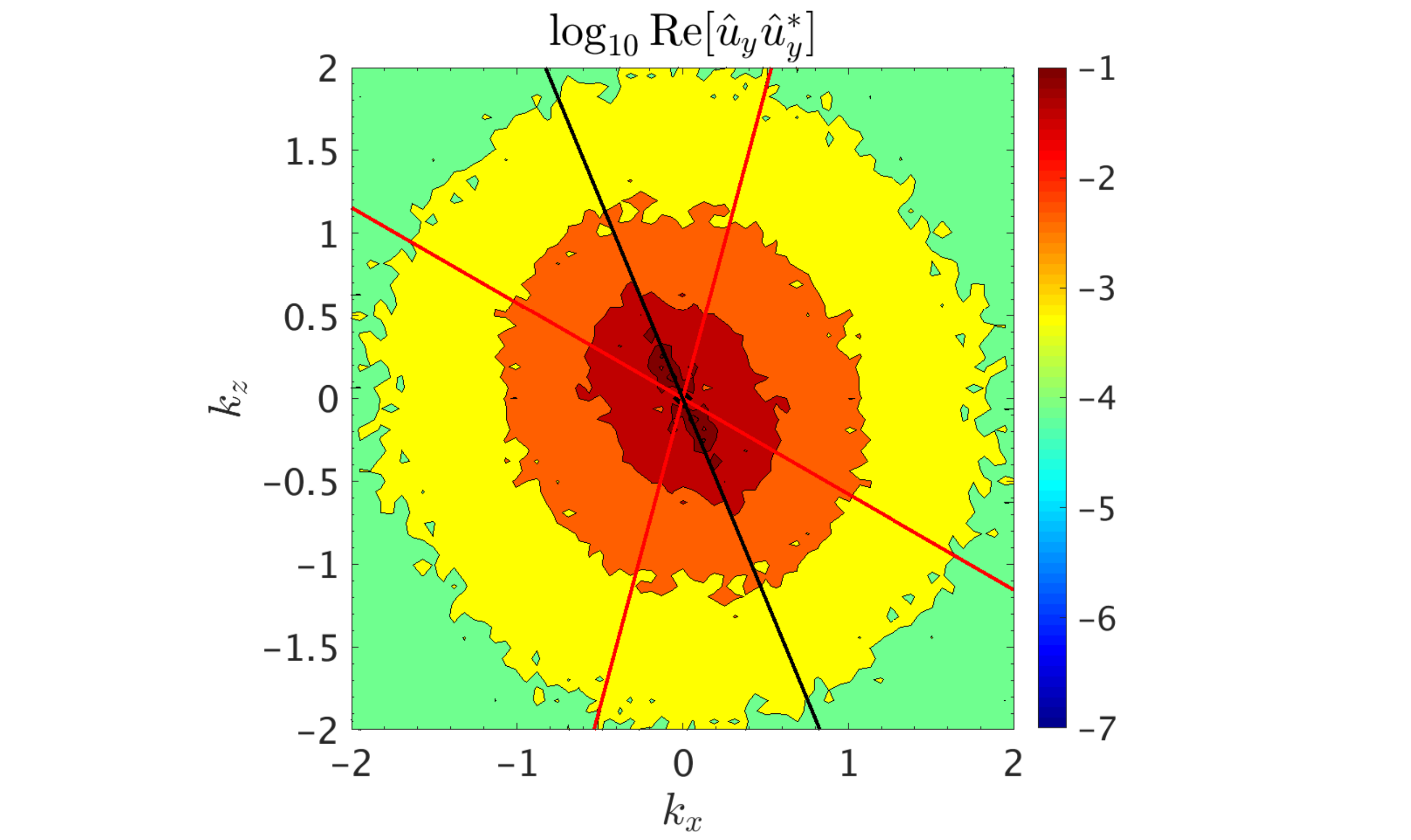}}
     \subfigure[$t=300$]{\includegraphics[trim=4cm 0cm 5cm 0cm, clip=true,width=0.35\textwidth]{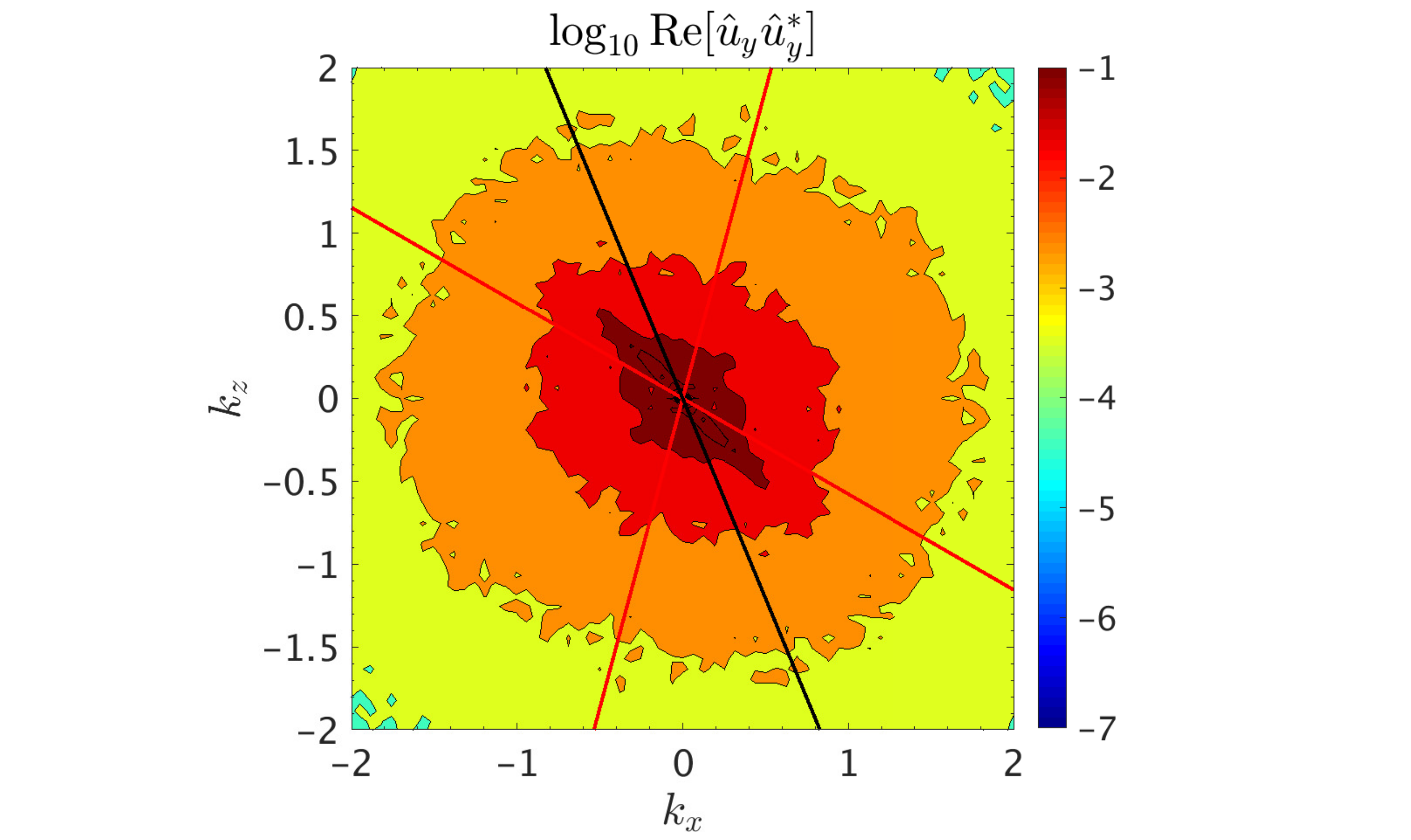}}
    \end{center}
  \caption{Fourier spectrum of $\log_{10} \mathrm{Re}[\hat{u}_y \hat{u}_y^*]$ on the ($k_x,k_z$)-plane for an axisymmetric simulation with $S = 2$, $\Lambda=30^{\circ}$, $N^2 = 10$, and $\mathrm{Pr} = 10^{-2}$, at various times. The black line indicates the direction of the fastest growing mode, and the red lines indicate $\hat{\boldsymbol{\Omega}}^\perp$ and $\nabla \ell$, and demarcate the boundaries of the linear GSF-unstable region (see Fig.~\ref{growthrate}). This shows that the modes are preferentially oriented along the linearly unstable direction until the later nonlinear phases (by $t\sim 300$), when the total flow is significantly modified.}
  \label{S2_spec}
\end{figure*}

We begin by presenting an axisymmetric simulation with $S=2$. Note that this case is Solberg-H\o iland stable, and would also be marginally Rayleigh-stable in the absence of stratification, but here it is GSF-unstable due to the presence of thermal diffusion. Figs.~\ref{Seq2} and \ref{Seq2a} show the temporal evolution of various volume-averaged quantities in these simulations, along with results from several 3D simulations with $L_y=30, 50$ and 100, which will be discussed further in the next section. Fig.~\ref{Seq2} shows the kinetic energy $K=\frac{1}{2}\langle |\boldsymbol{u}|^2\rangle$, where $\langle \cdot\rangle$ denotes a volume average, and the RMS velocity components $v_y=\langle u_y^2\rangle^{1/2}$ and $v_z=\langle u_z^2\rangle^{1/2}$. We have found $v_x=\langle u_x^2\rangle^{1/2}$ to be slightly larger, though comparable, with $v_z$, so we have omitted showing this. Fig.~\ref{Seq2a} shows the momentum flux components (Reynolds stresses) $\langle u_xu_y\rangle$ and $\langle u_yu_z\rangle$, as well as the radial buoyancy flux $-\langle u_x\theta\rangle$. Note that for our purposes we consider any systematic mean flows, such as azimuthal jets, to contribute to the Reynolds stress i.e.~we do not decompose the flow into a mean flow plus turbulent fluctuations to define the Reynolds stress. The corresponding azimuthal flow $u_y$ is shown on the $(x,z)$-plane in Fig.~\ref{Seq2_2D} at several different times in the axisymmetric simulation: during the linear growth phase at $t=22$, the initial nonlinear saturation at $t=50$, and finally at two later stages in the nonlinear evolution at $t=100$ and $t=300$.

The linear growth phase is dominated by modes that have a slanted structure, as we show in the top panel of Fig.~\ref{Seq2_2D}, consisting of finger-like motions along a direction (indicated by the solid black line) that lies approximately halfway between the rotation axis and a surface of constant angular momentum (parallel with $(\nabla \ell)^\perp$) -- both of these directions are indicated by black dashed lines -- as explained in \S \ref{lineargrowth}. At $t\sim 50$, the linear growth has saturated, and the initial finger-like motions have begun to merge into a number of zonal ($u_y$) jets that extend across the box. At this stage, these jets possess a similar orientation to the linear modes.

At later times, the jets undergo further mergers, which strengthens them and enhances the momentum transport. By $t\sim 100$, there are two jets along $z$ (or $x$), but by $t\sim 300$ the jets have merged until there is only one wavelength along $z$ (or $x$), after which this state is observed to persist. The strengthening of zonal jets as they merge can be clearly observed in the rapid transitions in the kinetic energy in the top panel of Fig.~\ref{Seq2}. As we show in the top two panels of Fig.~\ref{Seq2a}, the momentum transport is enhanced with each successive merger, such that $\langle u_x u_y\rangle$ has grown to be approximately 5 times larger than in the initial nonlinear phases. We also observe non-negligible $\langle u_y u_z\rangle$, though this is somewhat smaller than $\langle u_x u_y\rangle$. In Fig.~\ref{Seq2_2Duxuy} we show a snapshot of $u_x u_y$ on the $(x,z)$-plane at $t=300$, which shows that the interfaces between steps with oppositely-signed zonal flows dominantly contribute to $\langle u_xu_y\rangle$.

At $t=300$, the bottom right panel of Fig.~\ref{Seq2_2D} shows that the jet is no longer aligned with the linear modes. The maximum $|u_y|\sim 50$, which is comparable in strength with the background flow ($|\boldsymbol{U}_0|\leq 100$), indicating that the instability has significantly modified the (total) flow. It is interesting to note that the angle of the jets (measured from the $x$-axis) increases towards the rotation axis, as we might expect if the instability modifies the flow by ``shrinking the wedge" in Fig.~\ref{growthrate}. In other words, the instability appears to drive the flow towards marginal stability so that the surfaces of constant angular momentum (for the total flow) have a tendency to  coincide with the rotation axis. However, the boundary conditions in our setup do not allow the flow to be modified at the boundaries, so by this final stage, the boundaries are certainly constraining the flow. In \S~\ref{Seq2stressfree}, we will describe a complementary simulation with stress-free boundaries, and in \S~\ref{LargerBoxes} we will describe how the box size and aspect ratio affect the transport and the flow.

We can further analyse the flow by computing the Fourier spectrum of the velocity field. In Fig.~\ref{S2_spec}, we show $\log_{10} \mathrm{Re}[\hat{u}_y \hat{u}_y^*]$ on the $(k_x,k_z)$-plane, where hats denote quantities in spectral space, during the same times in the simulation as Fig.~\ref{Seq2_2D}. The first panel is at $t=22$, and the remaining three panels are averaged over 15 snapshots (spaced every time unit) starting at $t=40, 100$ and 300, respectively. Note that the modes with non-negligible energies at $t=22$ are those within the unstable wedge shown in Fig.~\ref{growthrate}, where the solid red lines indicate the directions of $\hat{\boldsymbol{\Omega}}^\perp$ and $\nabla \ell$. We have also found $\log_{10} \mathrm{Re}[\hat{u}_x \hat{u}_y^*]$ (the spectrum of the Reynolds stress) to exhibit similar features, indicating that the strong zonal jets are primarily responsible for the momentum transport. The peak of the spectrum at $t=300$ has shifted towards the left red solid line, indicating again that the instability acts to drive the system towards marginality.

\subsection{$S=2$ with shearing-periodic BCs: 3D cases}
\label{Seq2shearingperiodic3D}

\begin{figure}
  \begin{center}
     \subfigure[$t=100$]{\includegraphics[trim=4cm 0cm 5cm 0cm, clip=true,width=0.4\textwidth]{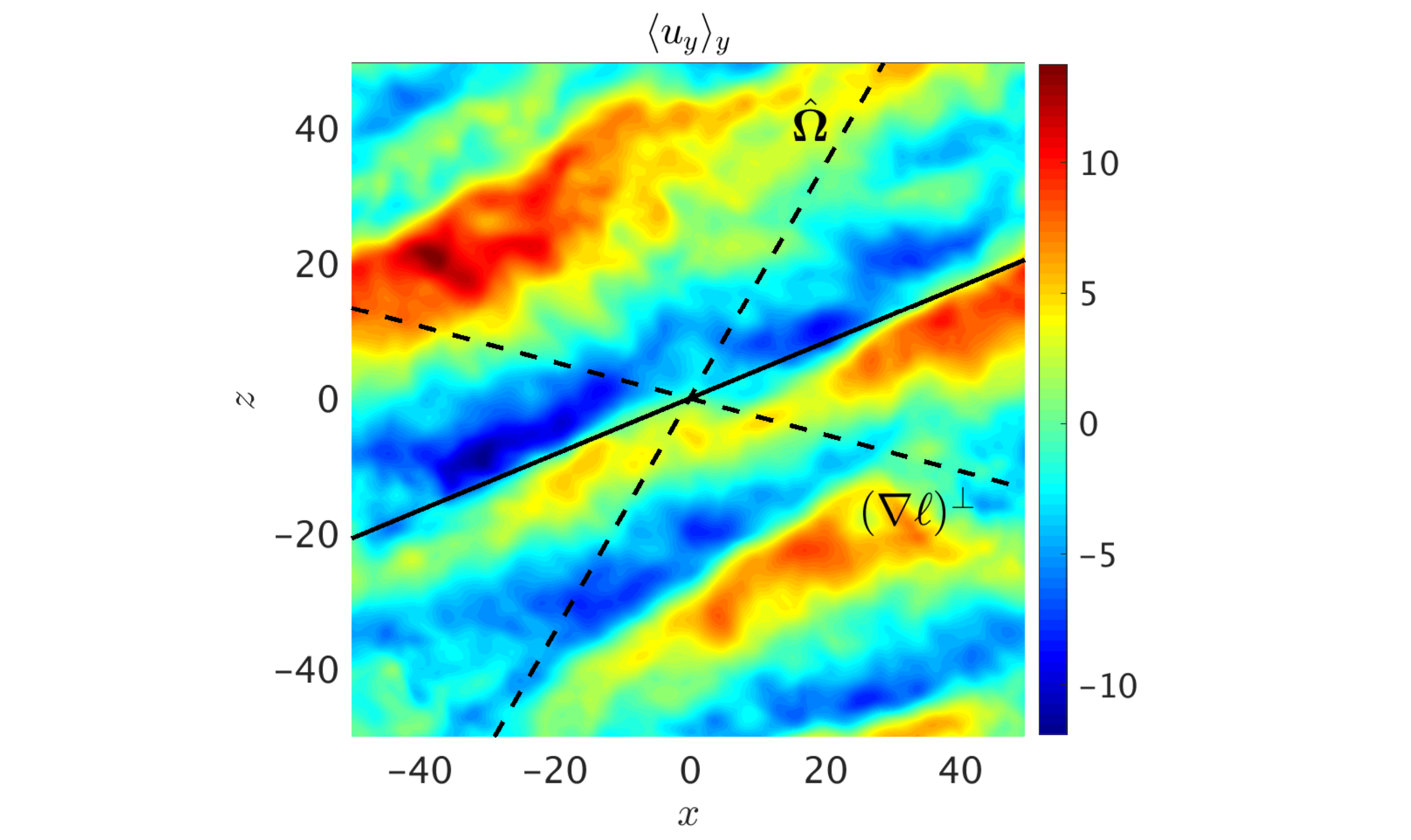}}
     \subfigure[$t=1000$]{\includegraphics[trim=4cm 0cm 5cm 0cm, clip=true,width=0.4\textwidth]{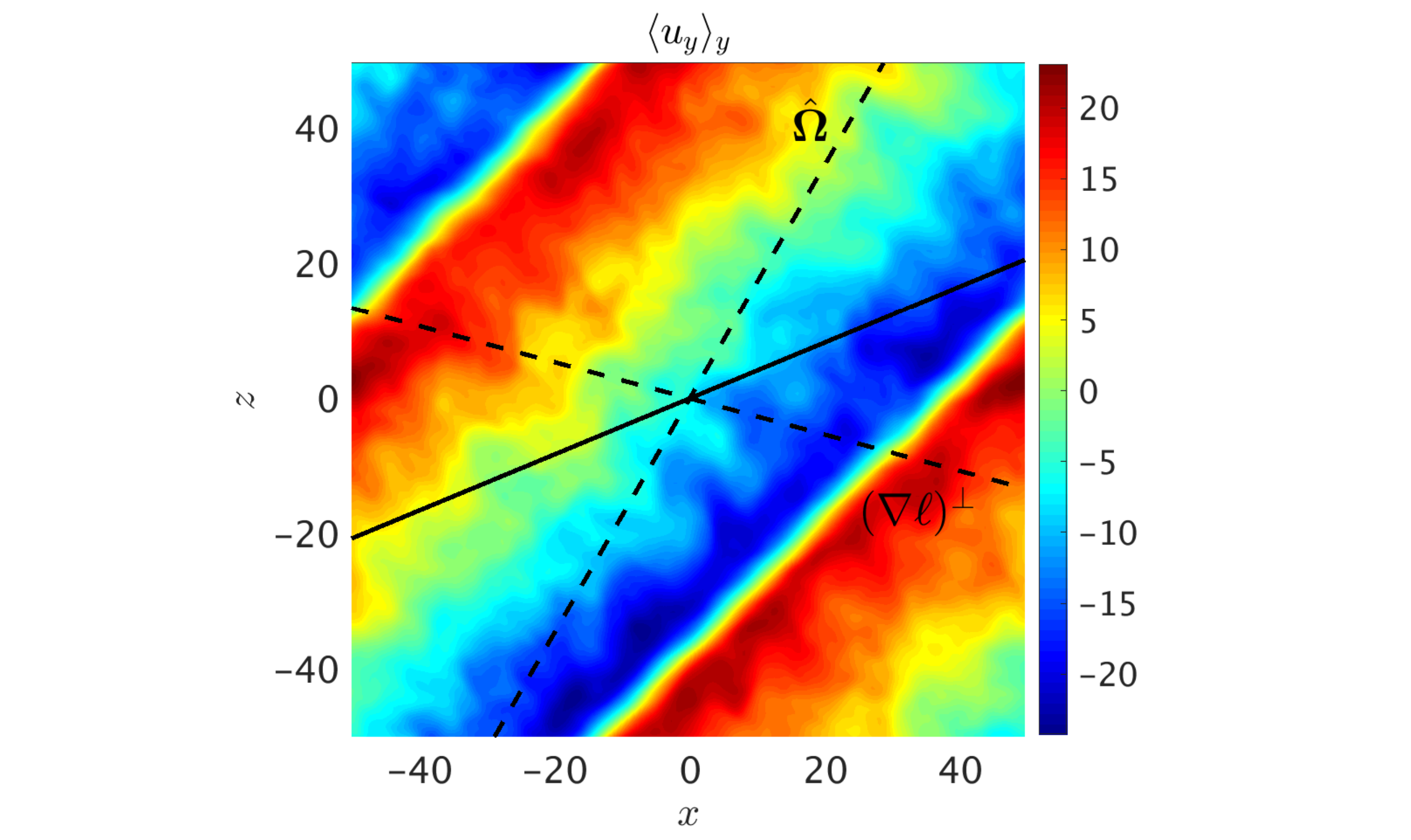}}
    \end{center}
  \caption{Snapshots of $y$-averaged $u_y$ in the $(x, z)$-plane for the 3D simulation with $L_y=100$, and $S = 2$, $\Lambda=30^{\circ}$, $N^2 = 10$, $\mathrm{Pr} = 10^{-2}$, at two different times. This illustrates that qualitatively similar nonlinear behaviour is obtained in 3D.}
  \label{Seq2_3D}
\end{figure}

Three-dimensional effects play a key role in the equatorial GSF instability (paper I), so we now turn to explore whether they are also important for the non-equatorial instability with $S=2$. The time-evolution of volume-averaged flow quantities for several 3D simulations is presented in Figs.~\ref{Seq2} and \ref{Seq2a} for cases with $L_y=30, 50$ and 100. All of the 3D simulations develop much weaker flows, having approximately one quarter of the energy of the axisymmetric case in the nonlinear state, but there is only weak dependence on $L_y$ between these different 3D simulations.

Fig.~\ref{Seq2_3D} shows the spatial structure of the $y$-averaged zonal flow ($u_y$) in a 3D simulation with $L_y=100$ at $t=100$ and $1000$, which can be compared with Fig.~\ref{Seq2_2D}. This demonstrates that strong zonal jets are also produced in three dimensions, but that the velocity magnitude of the jets (and of the GSF-driven turbulent flows, according to Fig.~\ref{Seq2}) is somewhat weaker than in the axisymmetric case by approximately a factor of 2. These jets merge and strengthen just as in the axisymmetric case. The corresponding momentum transport, shown in Fig.~\ref{Seq2a}, does not appear to be enhanced as significantly by the jet mergers in 3D however, and increases by less than a factor of 2 from $t\sim 100$ to $t\sim 1000$. Indeed, contrary to the axisymmetric case, $\langle u_x u_y\rangle$ in the 3D simulations remains at a level similar to its value at the initial saturation, even once the jets have merged to fill the box. This may be related to the weaker zonal flows here compared with those presented in \S~\ref{Seq2shearingperiodicAxi}. As a result, the transport is smaller by approximately a factor of 5 compared with the axisymmetric case once jets have merged by $t\sim300$. The strength of the jets doesn't depend strongly on $L_y\ne 0$, as is shown in the middle panel of Fig.~\ref{Seq2}. The radial buoyancy flux is also larger in the axisymmetric simulation (bottom panel of Fig.~\ref{Seq2a}), further indicating that the zonal jets do not enhance transport as efficiently in 3D. Note that the jets appear to enhance the buoyancy flux here, which is the opposite behaviour to the meridional jets produced by the equatorial instability presented in paper I. This difference is presumably due to their different orientation with respect to $x$.

In summary, this illustrative set of simulations highlights that the non-equatorial GSF instability produces strong zonal jets, which can be thought of as ``layering" or ``staircasing" of the angular momentum. The mechanism for the formation of these jets is complicated, as for other systems where layering occurs. Physically it is plausible that the instability saturates by a combination of modifying the large-scale state of the system (both in terms of temperature and angular momentum) and increasing the dissipation (via the presence of turbulent interactions). Because the overall gradients remain fixed, the system may only mix locally saturating with layers where the shear profile has been mixed, interleaved with layers where the overall shear is stronger; this leads to the formation of jets. However the turbulence also modifies the underlying temperature field, which is not aligned with that of angular momentum and so the saturation is complicated. The jets transport angular momentum and appear to drive the system towards marginal stability, as far as this is allowed by the boundary conditions. The jets are observed to merge until they grow to the box size, superficially similar to the behaviour of layers in salt fingering (e.g.~\citealt{Garaud2018}). In axisymmetric cases, the momentum transport is significantly enhanced by these strong jets, though their effects are somewhat weaker in 3D. This suggests that 3D simulations are probably required for evaluating the astrophysical importance of the instability. Since the momentum and heat transport in 3D simulations remains similar to the initial saturated value, this suggests that a simple single-mode mode theory for homogeneous GSF-driven turbulence may approximately explain the transport in our 3D simulations. We will turn to make this comparison in \S~\ref{theorycomparison}. However, whenever these jets form, they could play an important role in enhancing angular momentum transport in stellar interiors. How are these results affected by the shearing-periodic boundaries? In \S~\ref{Seq2stressfree}, we turn to analyse a complementary simulation performed with stress-free, impenetrable, radial boundaries to answer this question.

\subsection{$S=2$: 3D case with stress-free BCs}
\label{Seq2stressfree}

Here we present a 3D simulation performed using Nek5000 with stress-free, impenetrable, fixed temperature boundaries in $x$, with $L_y=30$ (using $\mathcal{E}=20\times3\times20$ elements and $\mathcal{N}_p=10$ and 15 for nonlinear terms -- a simulation with 6 elements in $y$ was also performed that gave essentially the same results). Our smallest 3D domain in $y$ was chosen for computational efficiency, and was motivated by the weak dependence of our 3D simulations on $L_y$. The time-evolution of volume-averaged flow quantities for this simulation is also presented in Figs.~\ref{Seq2} and \ref{Seq2a} as the magenta dashed lines. We observe that the kinetic energy is approximately a factor of 2 smaller than in the corresponding simulation with shearing-periodic boundaries, though the $x$-velocity magnitude is similar. As shown in the top panel of Fig.~\ref{Seq2a}, $\langle u_x u_y\rangle$ is approximately $20\%$ smaller than in the corresponding case with shearing-periodic boundaries by a similar factor, while $\langle u_y u_z\rangle$ is similar. 

\begin{figure}
  \begin{center}
     \subfigure[$t=100$]{\includegraphics[trim=4cm 0cm 5cm 0cm, clip=true,width=0.4\textwidth]{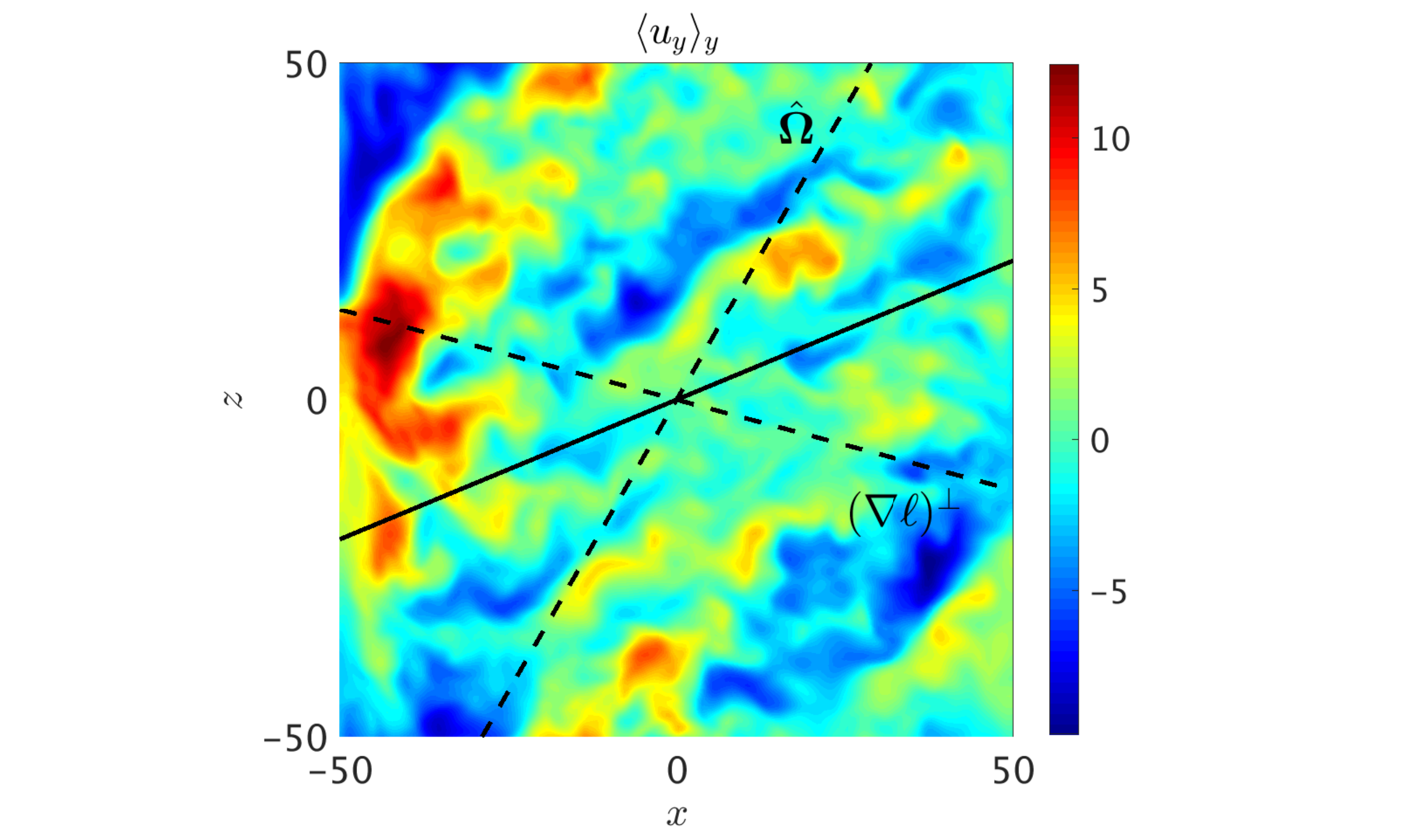}}
     \subfigure[$t=1000$]{\includegraphics[trim=4cm 0cm 5cm 0cm, clip=true,width=0.4\textwidth]{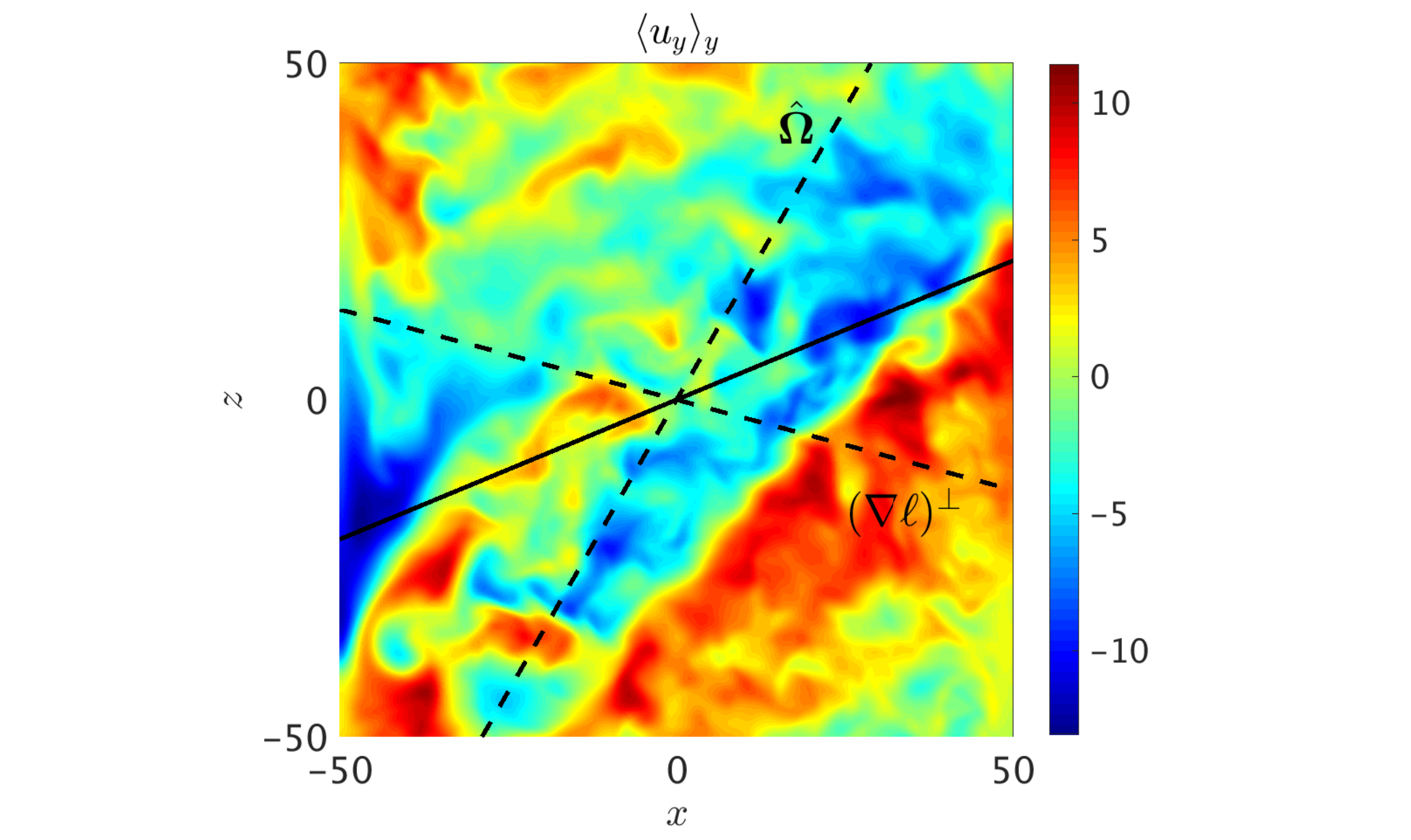}}
    \end{center}
  \caption{Snapshots $y$-averaged $u_y$ in the $(x, z)$-plane at $y=0$ for the 3D simulation with stress-free, impenetrable radial boundaries with $L_y=30$, $S = 2$, $\Lambda=30^{\circ}$, $N^2 = 10$ and $\mathrm{Pr} = 10^{-2}$, at two different times. This illustrates that qualitatively similar behaviour is observed using stress-free and shearing-periodic boundary conditions.}
  \label{Seq2_3DNek}
\end{figure}

Fig.~\ref{Seq2_3DNek} presents snapshots of the $y$-averaged $u_y$ flow component in the $(x,z)$-plane, which can be compared with Fig.~\ref{Seq2_3D}. The flow is broadly similar to the shearing-periodic case plotted in Fig.~\ref{Seq2_3D}, though it is approximately half the strength. Throughout the bulk of the flow, the tilt angle of the zonal flows is similar at $t=100$, with the flows being primarily aligned along the direction of the fastest growing mode, but by $t=1000$ they exhibit a steeper tilt angle. The flow does differ near the inner boundary however, and exhibits a much steeper tilt angle than for shearing box calculations even at $t=100$, as we show in Fig.~\ref{Seq2_3DNek}. This is presumably because the boundary conditions prevent radial flow. They also allow the basic flow to be modified by the instability at the boundaries.

This example illustrates that the nonlinear evolution of the GSF instability is not strongly affected by modifying the radial boundary conditions from shearing-periodic to stress-free and impenetrable. The main difference observed is that the flow near the boundaries is modified with stress-free conditions, which results in a slightly weaker turbulent energy and transport because the flow can evolve to better match the marginal state in this case.

\subsection{Two further weak shear cases ($S=1, 1.5$)}
\label{Gam30RayleighStable}

\begin{figure}
  \begin{center}
   \subfigure{\includegraphics[trim=9cm 1cm 9cm 1cm, clip=true,width=0.48\textwidth]{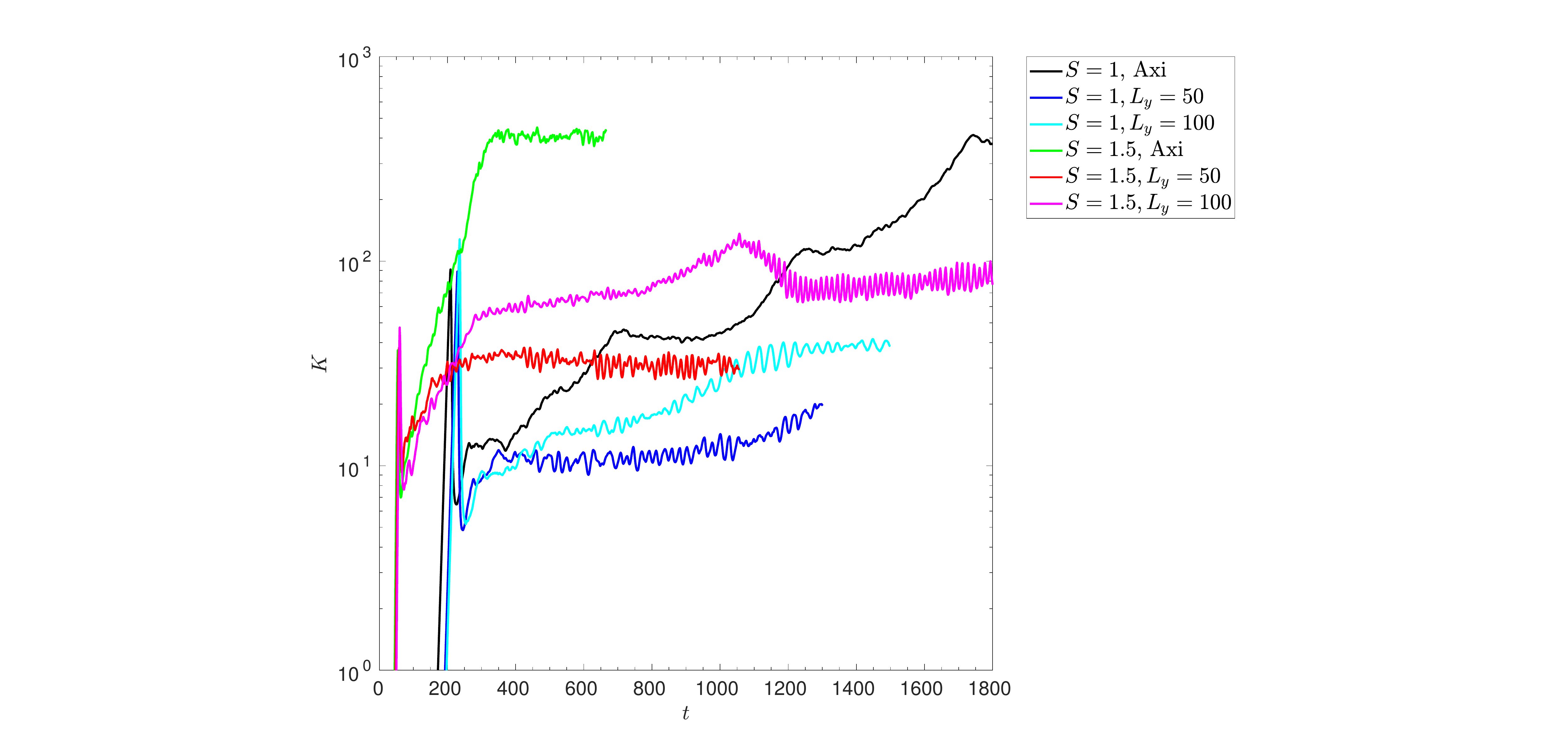}}
   \subfigure{\includegraphics[trim=9cm 1cm 9cm 1cm, clip=true,width=0.48\textwidth]{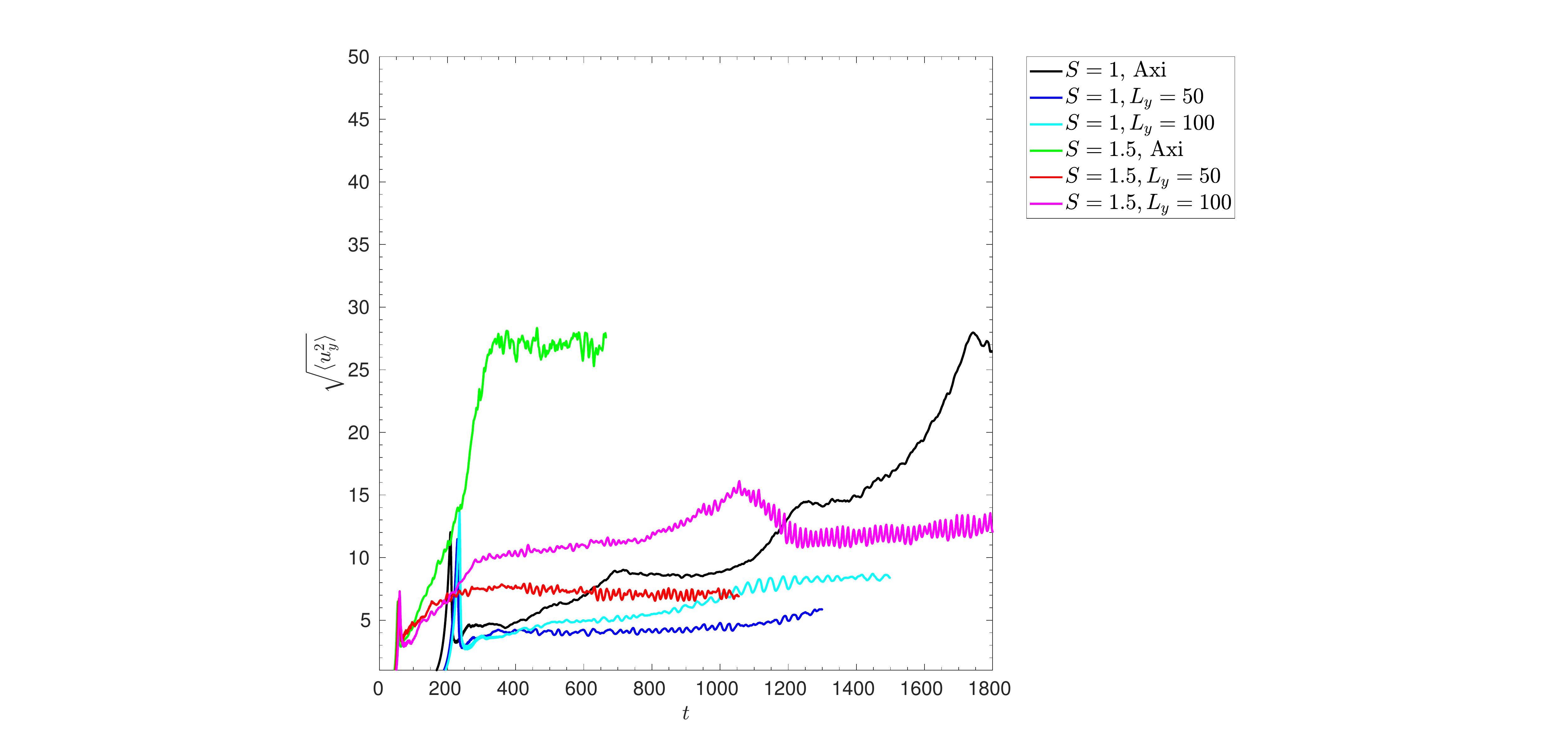}}
   \subfigure{\includegraphics[trim=9cm 1cm 9cm 1cm, clip=true,width=0.48\textwidth]{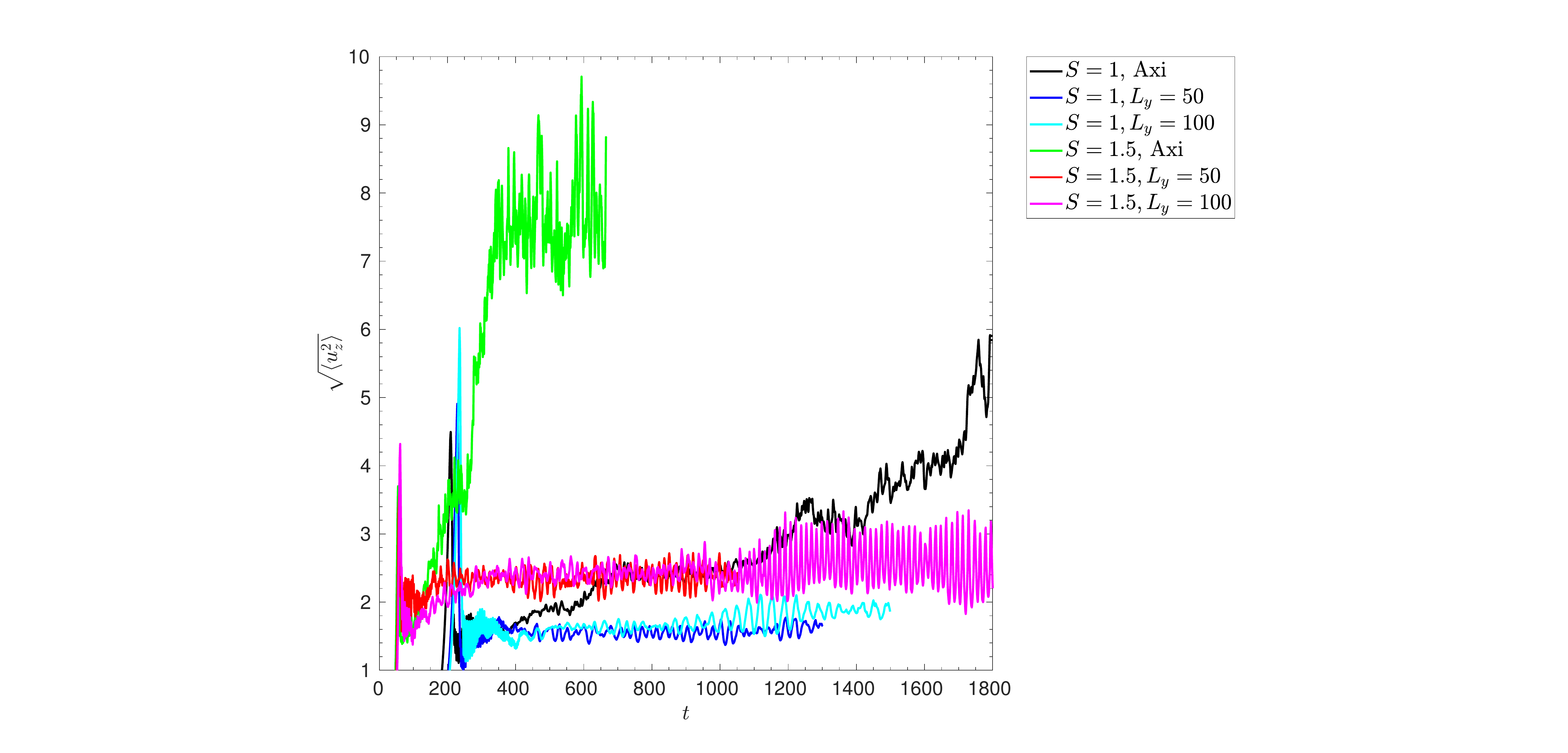}}
    \end{center}
  \caption{Same as Fig.~\ref{Seq2} but for simulations with $S=1$ and $1.5$.}
  \label{Seq1}
\end{figure}

\begin{figure}
  \begin{center}
  \subfigure{\includegraphics[trim=9cm 1cm 9cm 1cm, clip=true,width=0.48\textwidth]{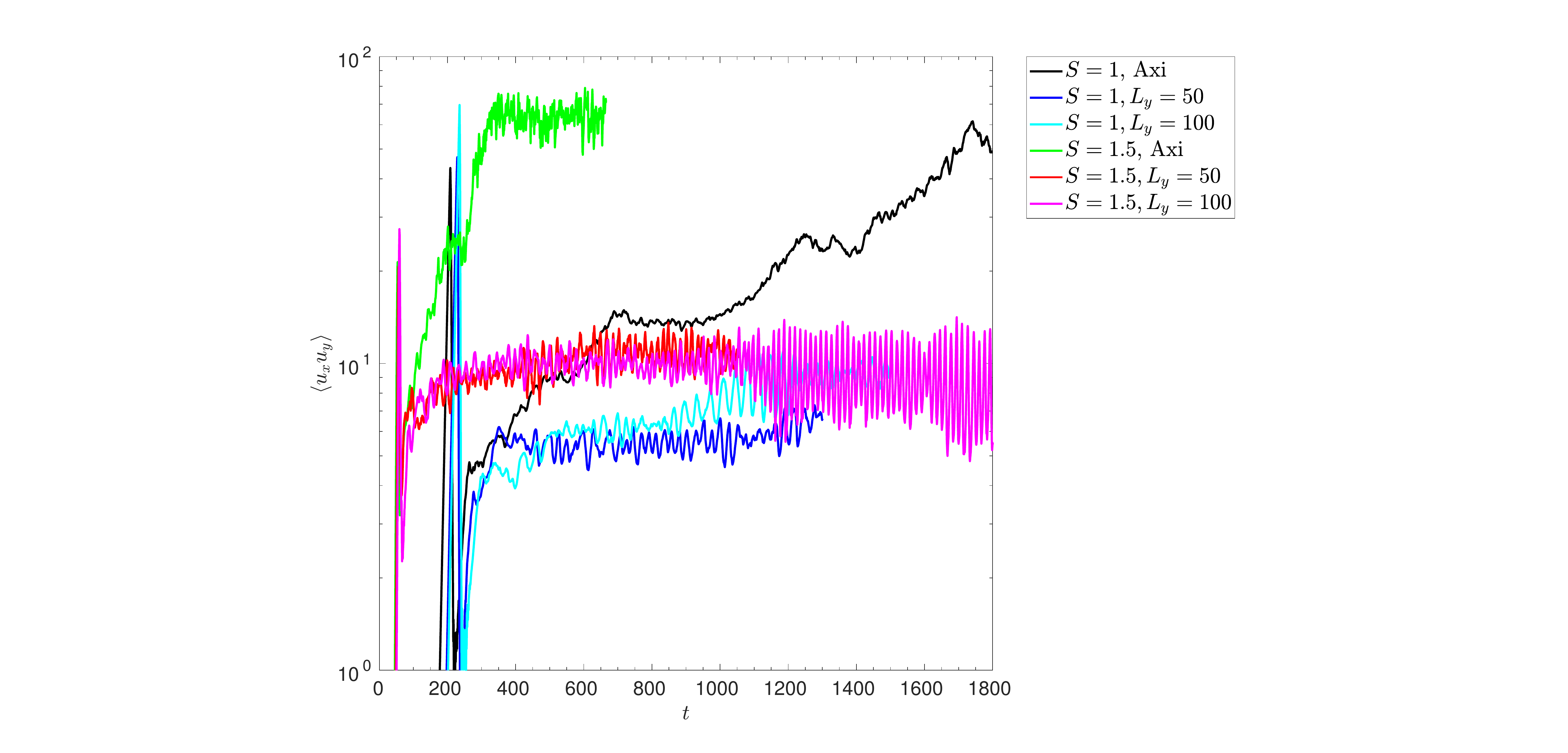}}
   \subfigure{\includegraphics[trim=9cm 1cm 9cm 1cm, clip=true,width=0.48\textwidth]{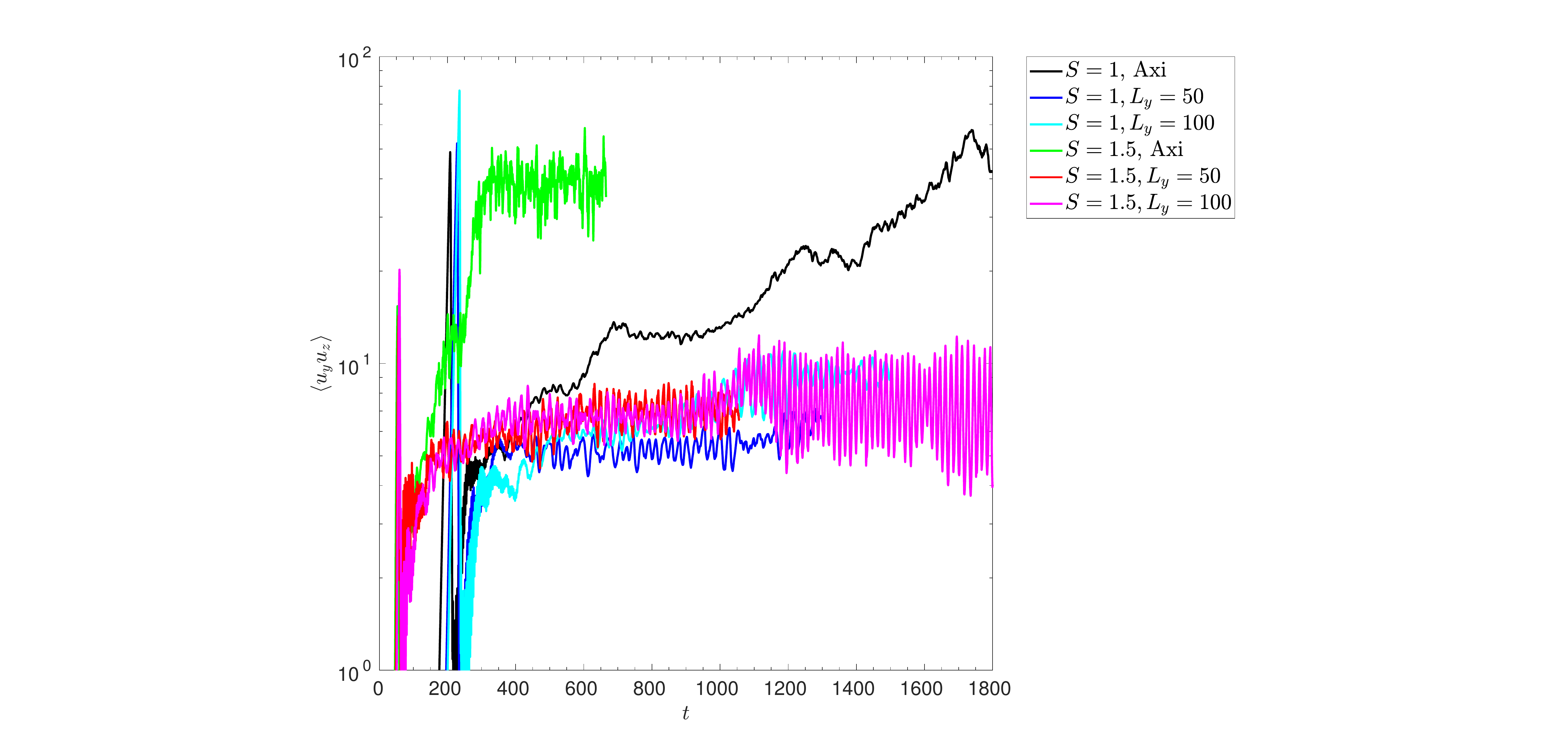}}
   \subfigure{\includegraphics[trim=9cm 1cm 9cm 1cm, clip=true,width=0.48\textwidth]{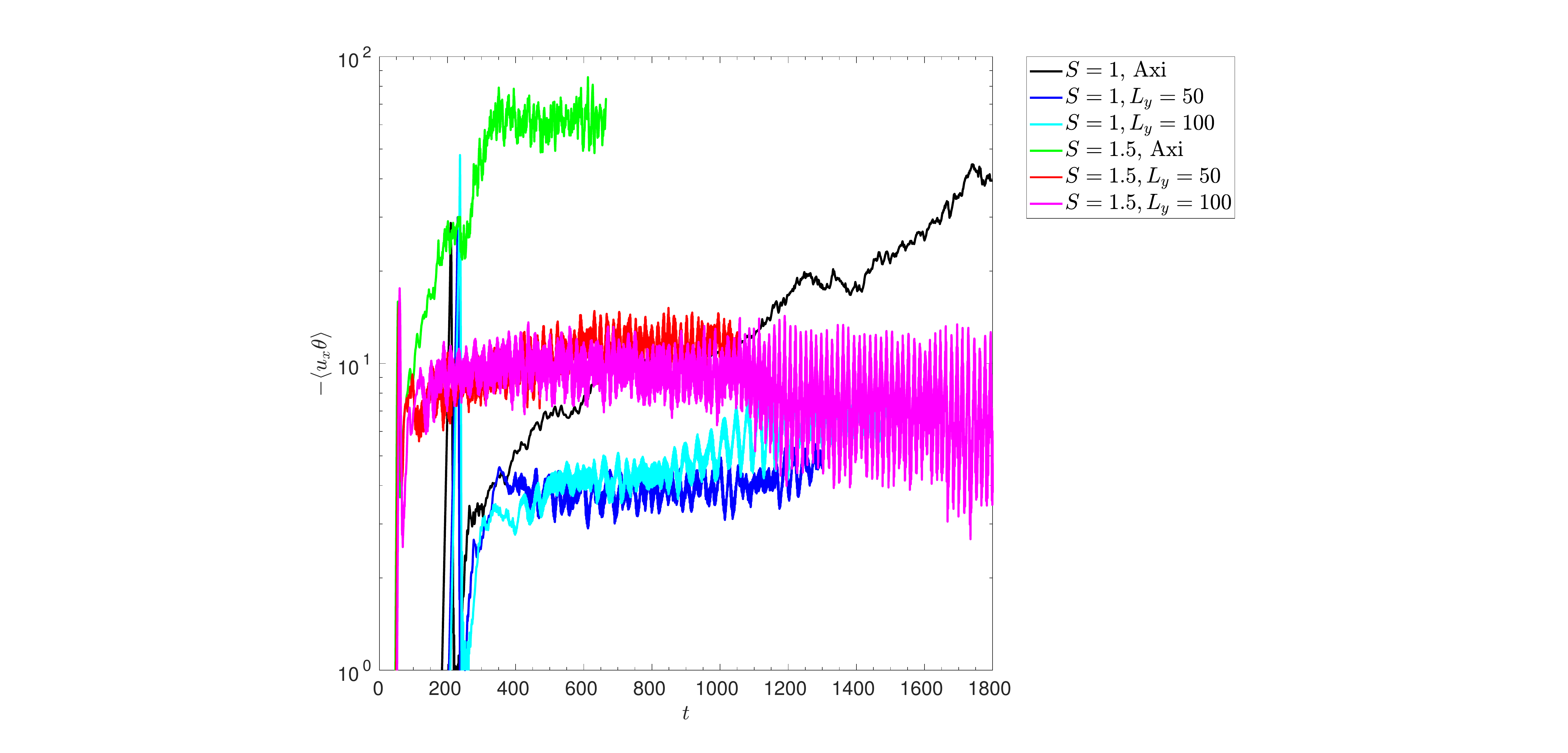}}
    \end{center}
  \caption{Same as Fig.~\ref{Seq2a} but for simulations with $S=1$ and $1.5$.}
  \label{Seq1a}
\end{figure}

\begin{figure}
  \begin{center}
     \subfigure[$t=160$]{\includegraphics[trim=4cm 0cm 5cm 0cm, clip=true,width=0.3\textwidth]{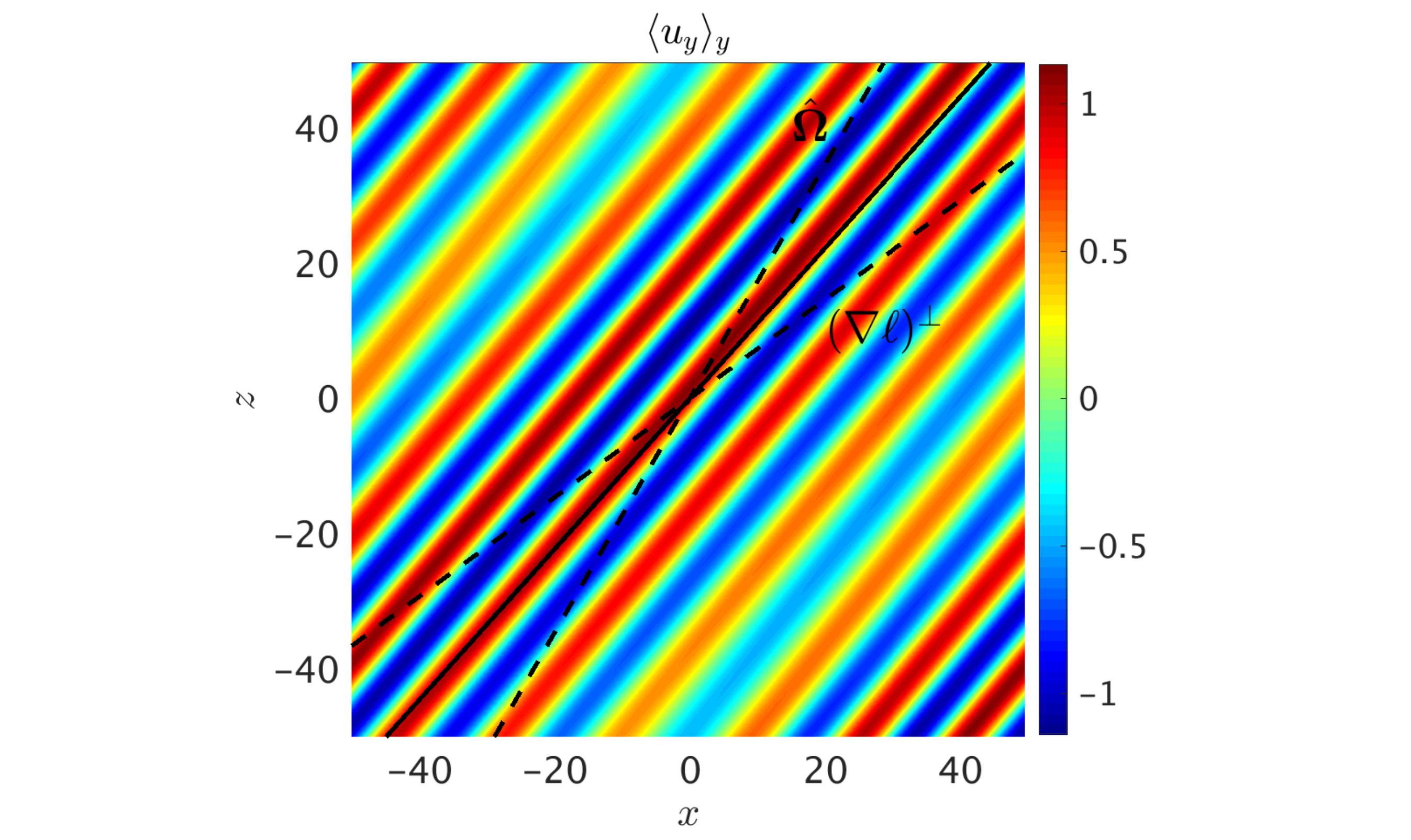}}
   \subfigure[$t=1500-1510$]{\includegraphics[trim=4cm 0cm 5cm 0cm, clip=true,width=0.3\textwidth]{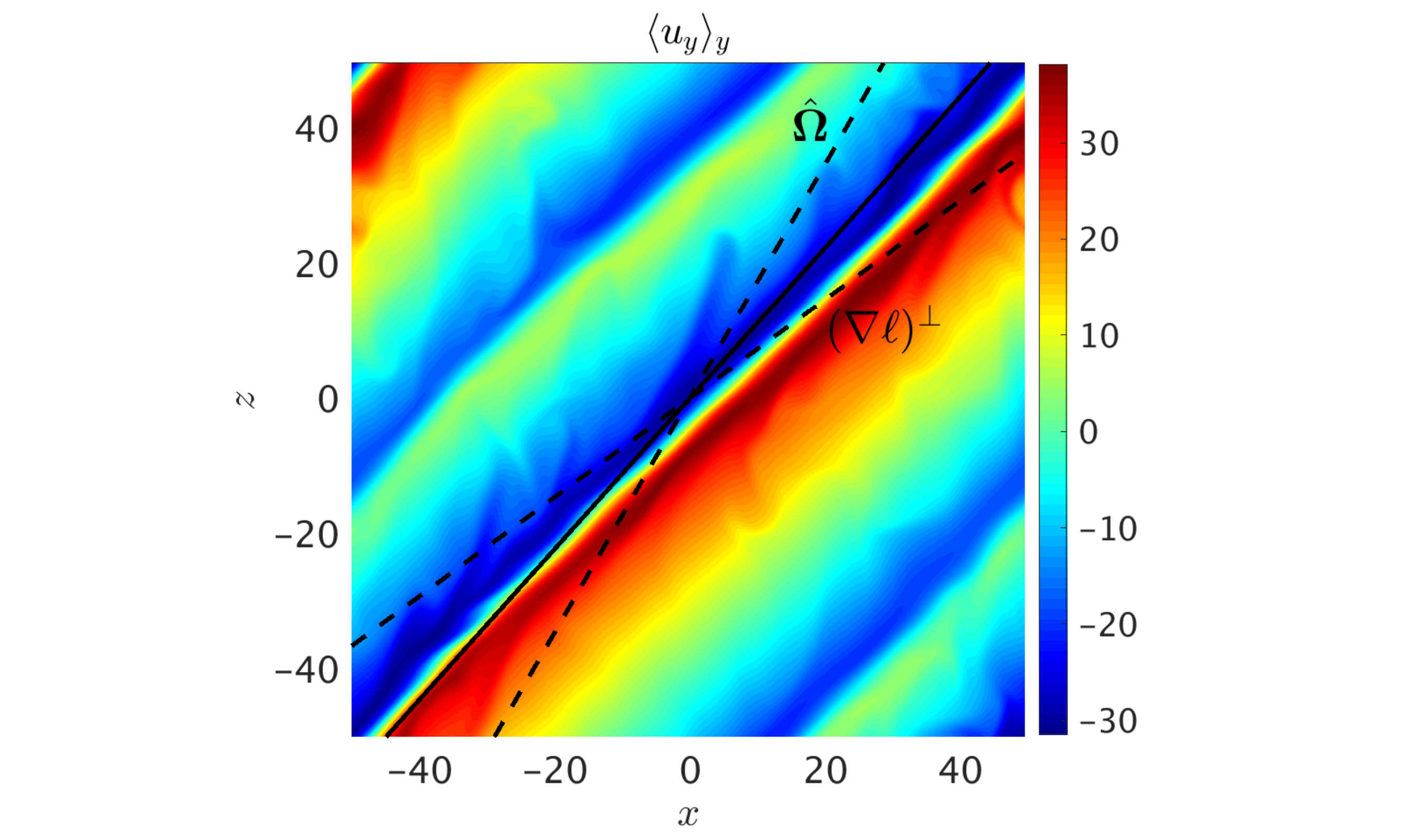}}
    \subfigure[$t=1500-1510$]{\includegraphics[trim=4cm 0cm 5cm 0cm, clip=true,width=0.3\textwidth]{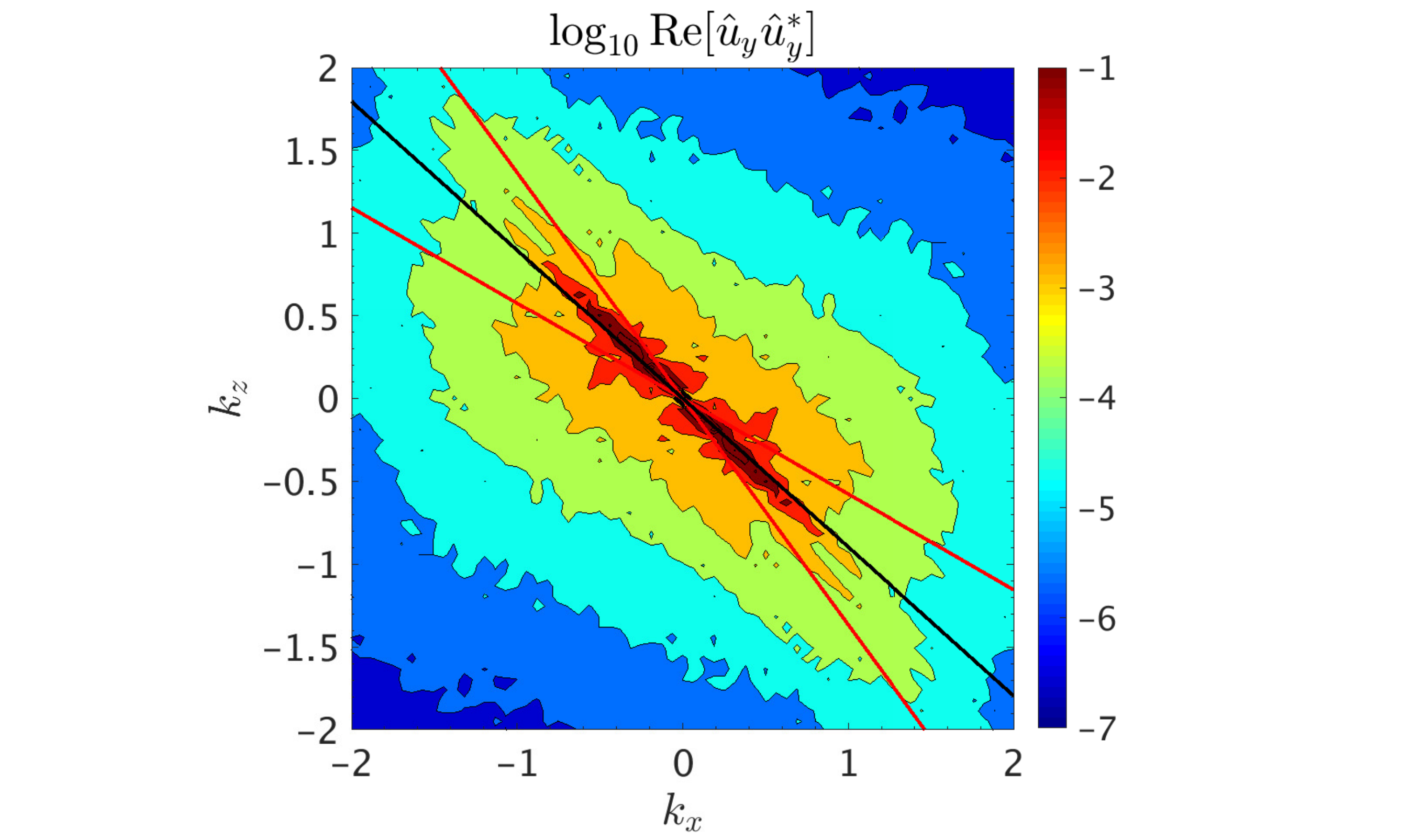}}
    \end{center}
  \caption{Snapshots of $u_y$ in the $(x, z)$-plane for an axisymmetric simulation with $S = 1$, $\Lambda=30^{\circ}$, $N^2 = 10$, and $\mathrm{Pr} = 10^{-2}$, at $t=160$ (top) and an average over 10 slices from $t=1500-1510$ (middle). Bottom: Fourier spectrum of $\log_{10} \mathrm{Re}[\hat{u}_y \hat{u}_y^*]$ on the ($k_x,k_z$)-plane using the same snapshots as the middle panel. The black and red lines are similar to those in previous figures but for the parameters of this simulation.}
  \label{Seq1flow}
\end{figure}

The differential rotation in stars is not always expected to be strong enough to violate Rayleigh's criterion, but the GSF instability can still operate on weaker shear. Here we explore two further weak shear cases, that would be Rayleigh-stable (if $N^2=0$), with $S=1$ and $S=1.5$, including axisymmetric and 3D simulations with various $L_y$. In Fig.~\ref{Seq1} and \ref{Seq1a} we show the time-evolution of various volume-averaged quantities, similarly to Figs.~\ref{Seq2} and \ref{Seq2a}. We immediately observe that axisymmetric simulations develop much stronger flows (Fig.~\ref{Seq1}) and lead to much more efficient transport compared with 3D cases (Fig.~\ref{Seq1a}), and that the 3D cases exhibit only a weak dependence on $L_y$. These results are consistent with those in \S~\ref{Seq2shearingperiodicAxi} and \ref{Seq2shearingperiodic3D}.

In Fig.~\ref{Seq1flow}, we present a snapshot of $u_y$ during the linear growth phase in the axisymmetric simulation with $S=1$ at $t=160$ (top panel), as well as $u_y$ during a subsequent nonlinear phase based on averaging over 10 time snapshots from $t=1500$ to $t=1510$ (middle panel). We also show $\log_{10} \mathrm{Re}[\hat{u}_y \hat{u}_y^*]$ on the $(k_x,k_z)$-plane in the bottom panel of the same figure. The flow in the linear growth phase consists of slanted finger-like jets along the direction expected from \S~\ref{lineargrowth} in each case. In the later nonlinear phases, these jets have merged to form strong larger-scale zonal jets approaching the size of the box, similar to those observed in \S~\ref{Seq2shearingperiodicAxi} and \S~\ref{Seq2shearingperiodic3D}. The middle and bottom panels of Fig.~\ref{Seq1flow} both indicate that the preferred direction of the flow is no longer aligned with the linear prediction, and is driven instead towards marginality, with the total flow being modified by the instability. 
As a result of the strong zonal jets in the axisymmetric simulations with $S=1$ and $S=1.5$, the transport is nearly as efficient as in the simulations with $S=2$ presented in Fig.~\ref{Seq2a}. This surprising result is a consequence of the strong zonal jets that develop. The 3D simulations exhibit very similar behaviour to the axisymmetric cases except that the zonal jets are considerably weaker and do not enhance the transport as efficiently. The flow is qualitatively similar with $S=1.5$, so we omit showing snapshots for this case. These examples indicate that the evolution described in \S~\ref{Seq2shearingperiodicAxi} and \S~\ref{Seq2shearingperiodic3D} may be generic for cases with weaker shear (which here correspond with Rayleigh stable cases).

\subsection{Strong shear cases ($S=2.5$ and $S=3$)}
\label{Gam30LargerS}

\begin{figure}
  \begin{center}
  \subfigure{\includegraphics[trim=9cm 1cm 9cm 1cm, clip=true,width=0.48\textwidth]{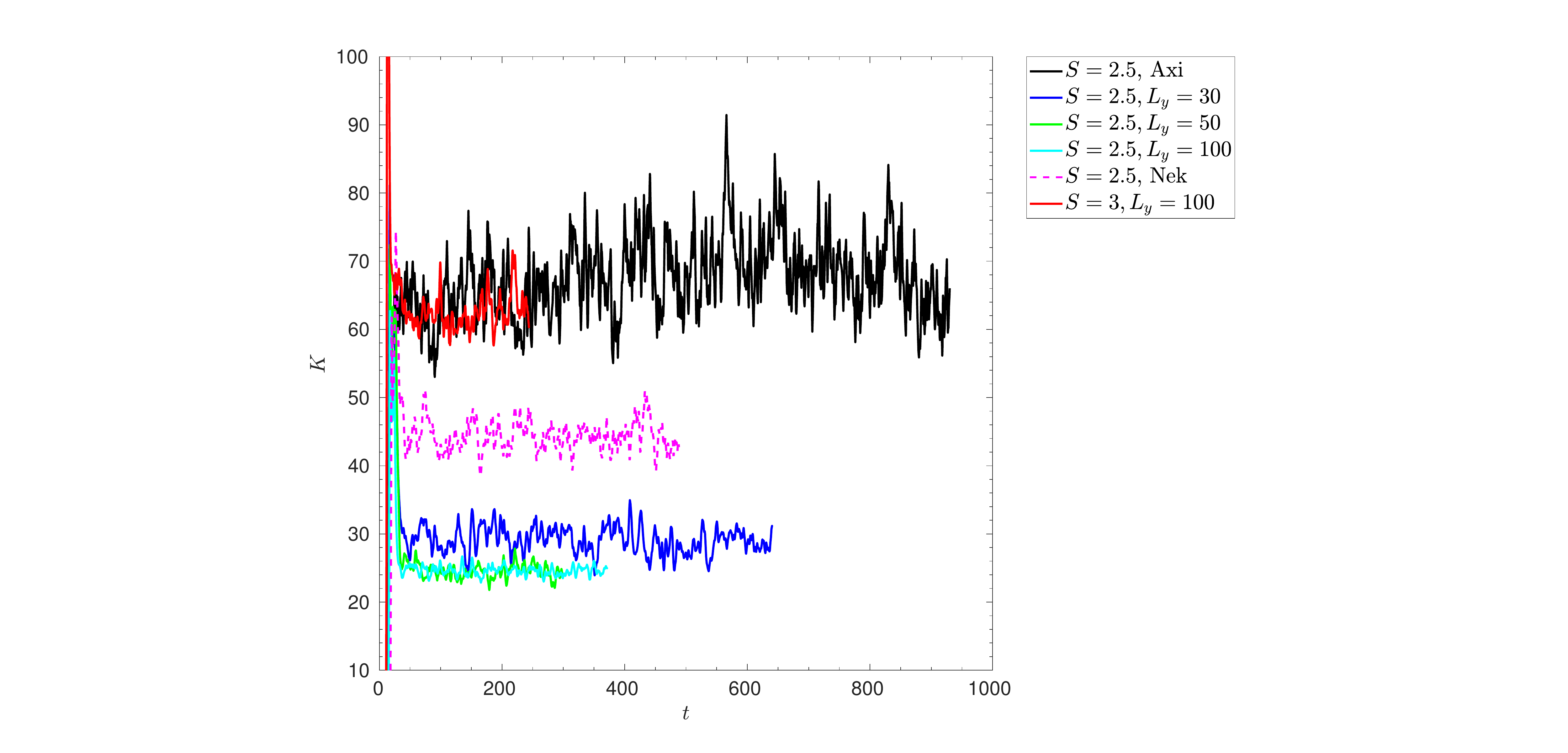}}
   \subfigure{\includegraphics[trim=9cm 1cm 9cm 1cm, clip=true,width=0.48\textwidth]{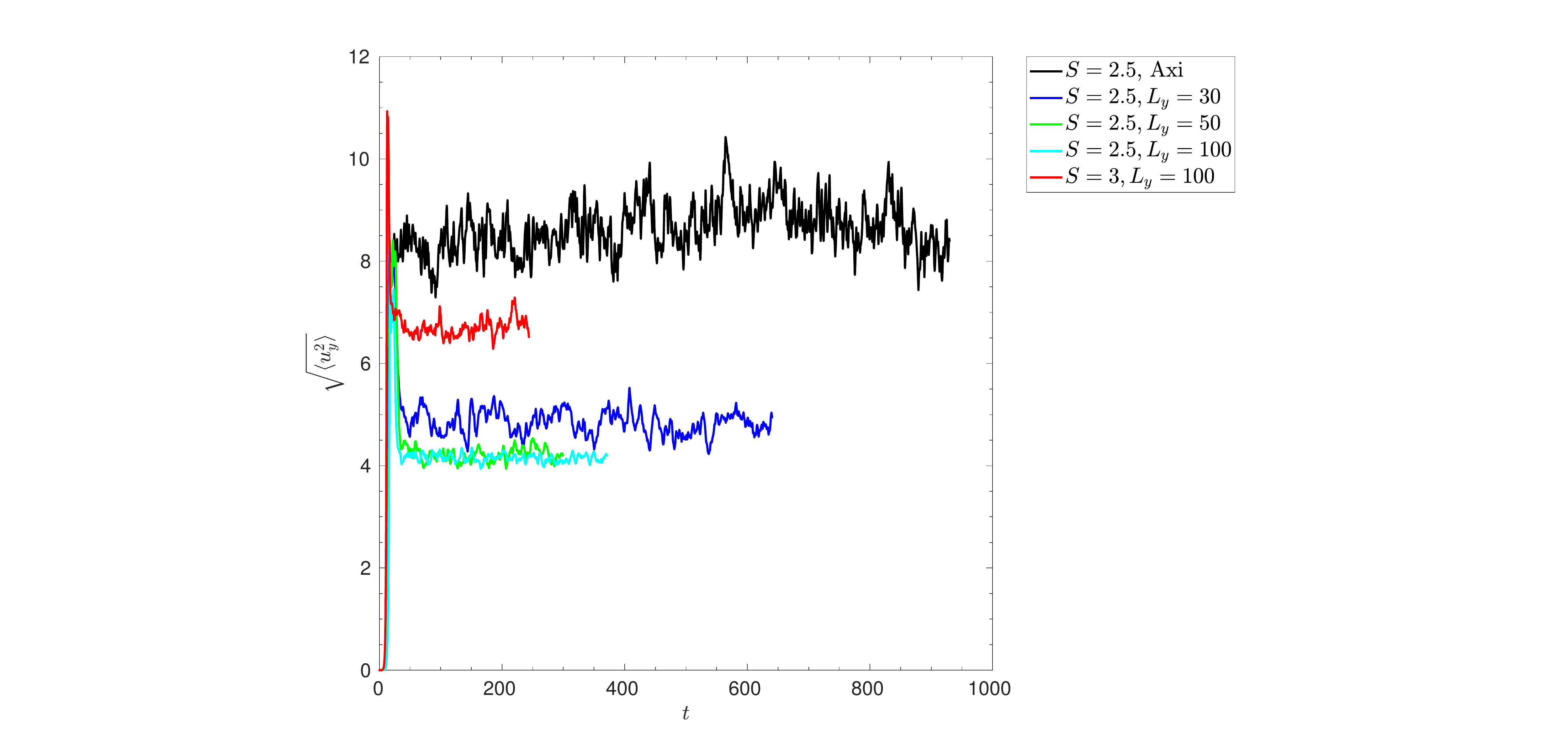}}
   \subfigure{\includegraphics[trim=9cm 1cm 9cm 1cm, clip=true,width=0.48\textwidth]{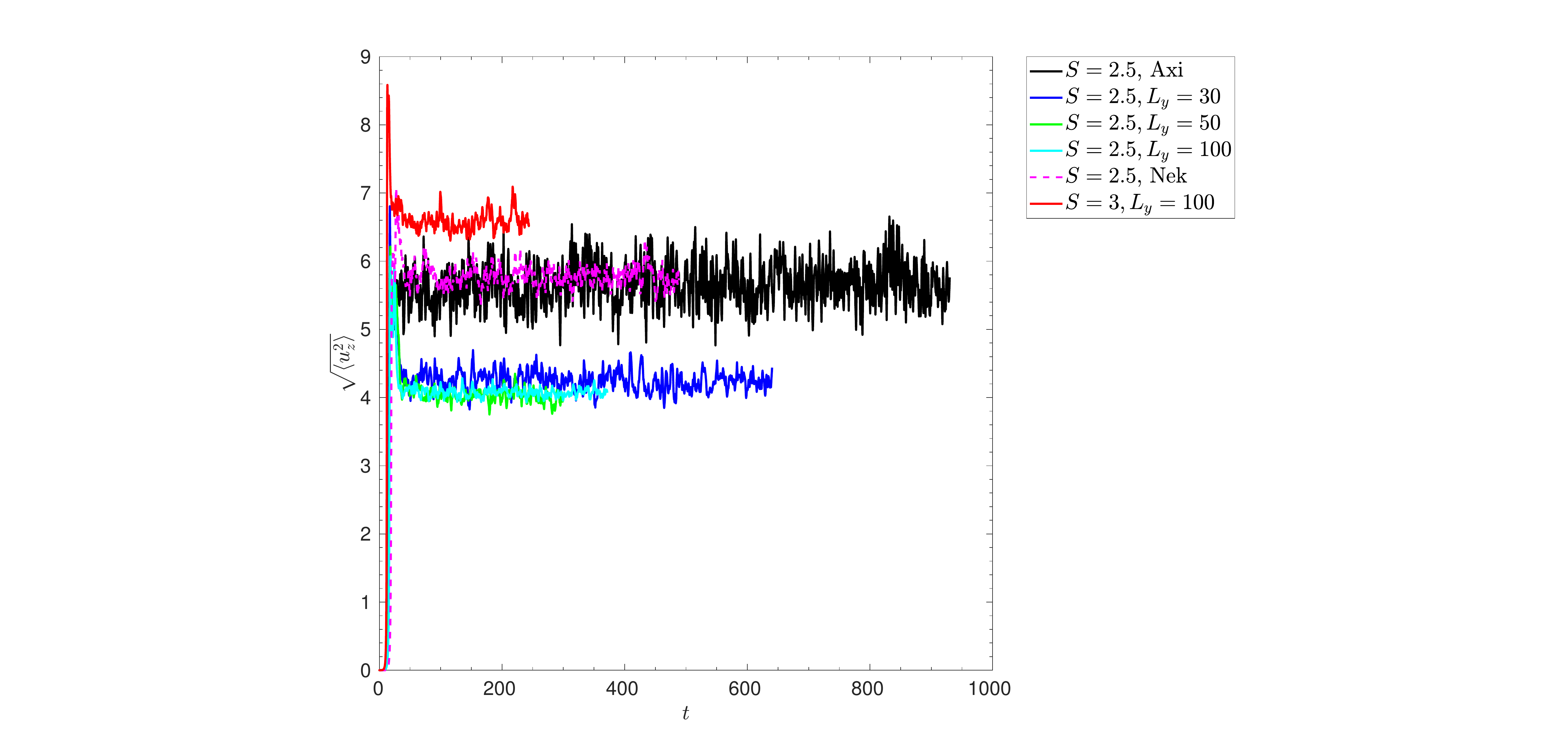}}
    \end{center}
  \caption{Same as Fig.~\ref{Seq2} but for simulations with $S=2.5$ and $S=3$.}
  \label{Seq2p5}
\end{figure}

\begin{figure}
  \begin{center}
  \subfigure{\includegraphics[trim=9cm 1cm 9cm 1cm, clip=true,width=0.48\textwidth]{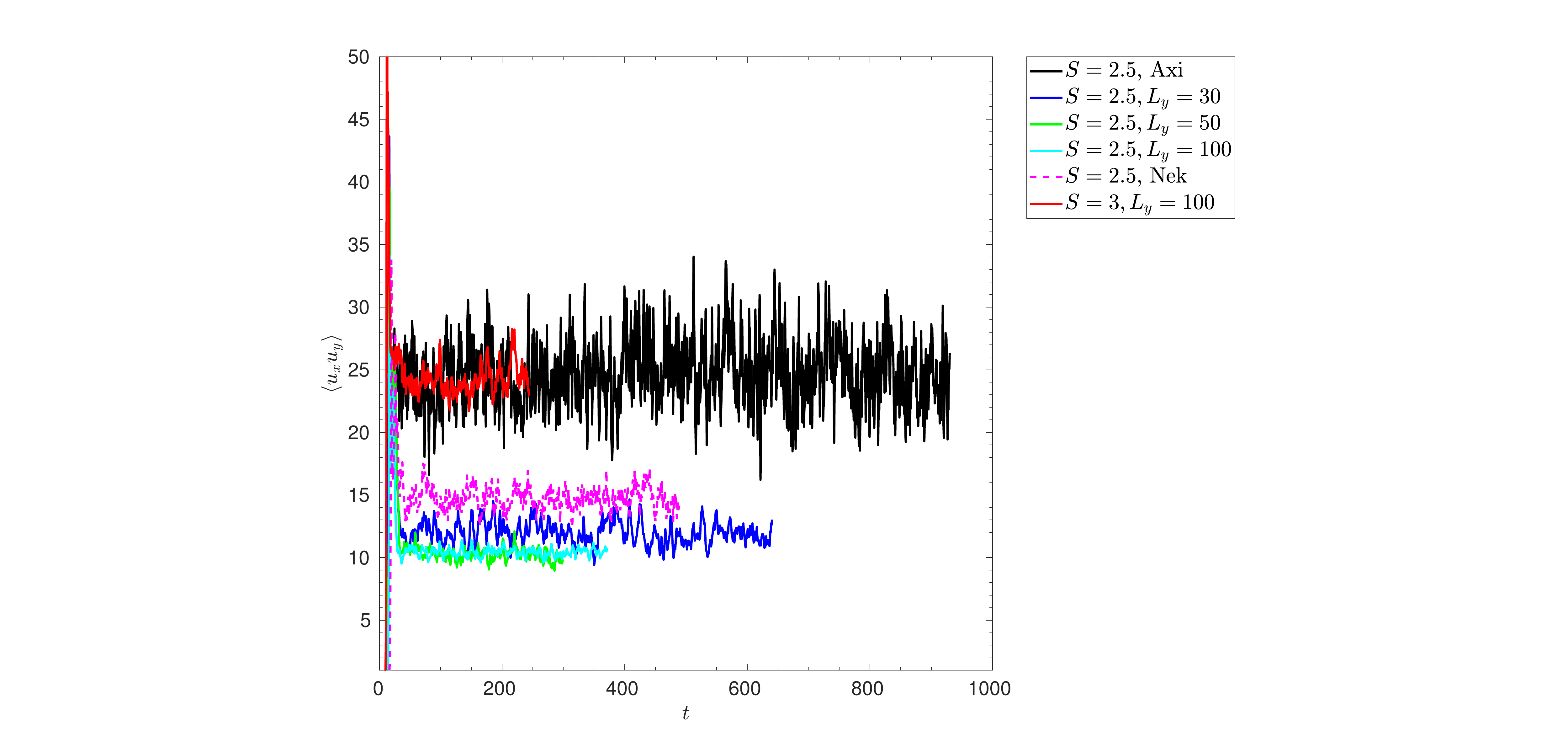}}
   \subfigure{\includegraphics[trim=9cm 1cm 9cm 1cm, clip=true,width=0.48\textwidth]{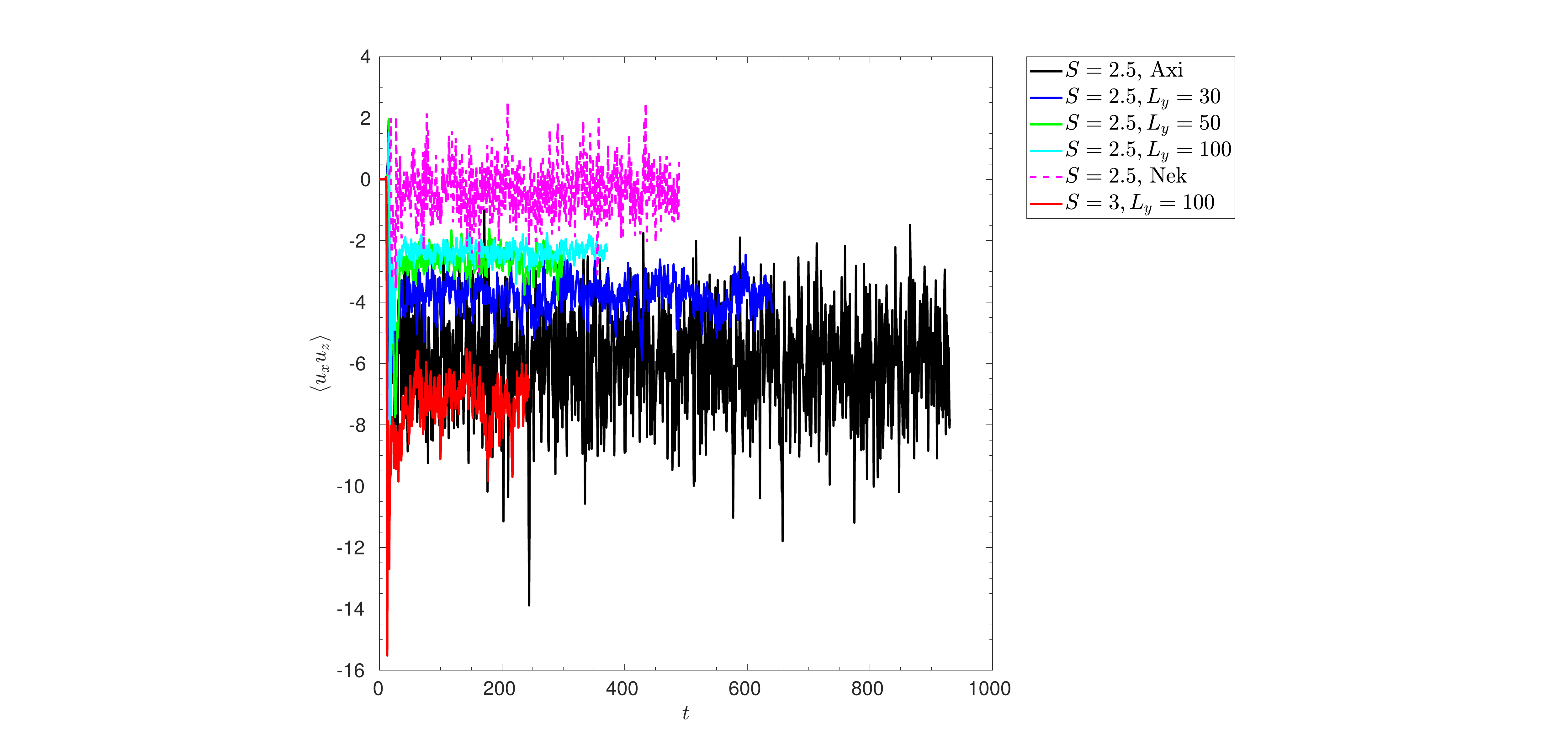}}
   \subfigure{\includegraphics[trim=9cm 1cm 9cm 1cm, clip=true,width=0.48\textwidth]{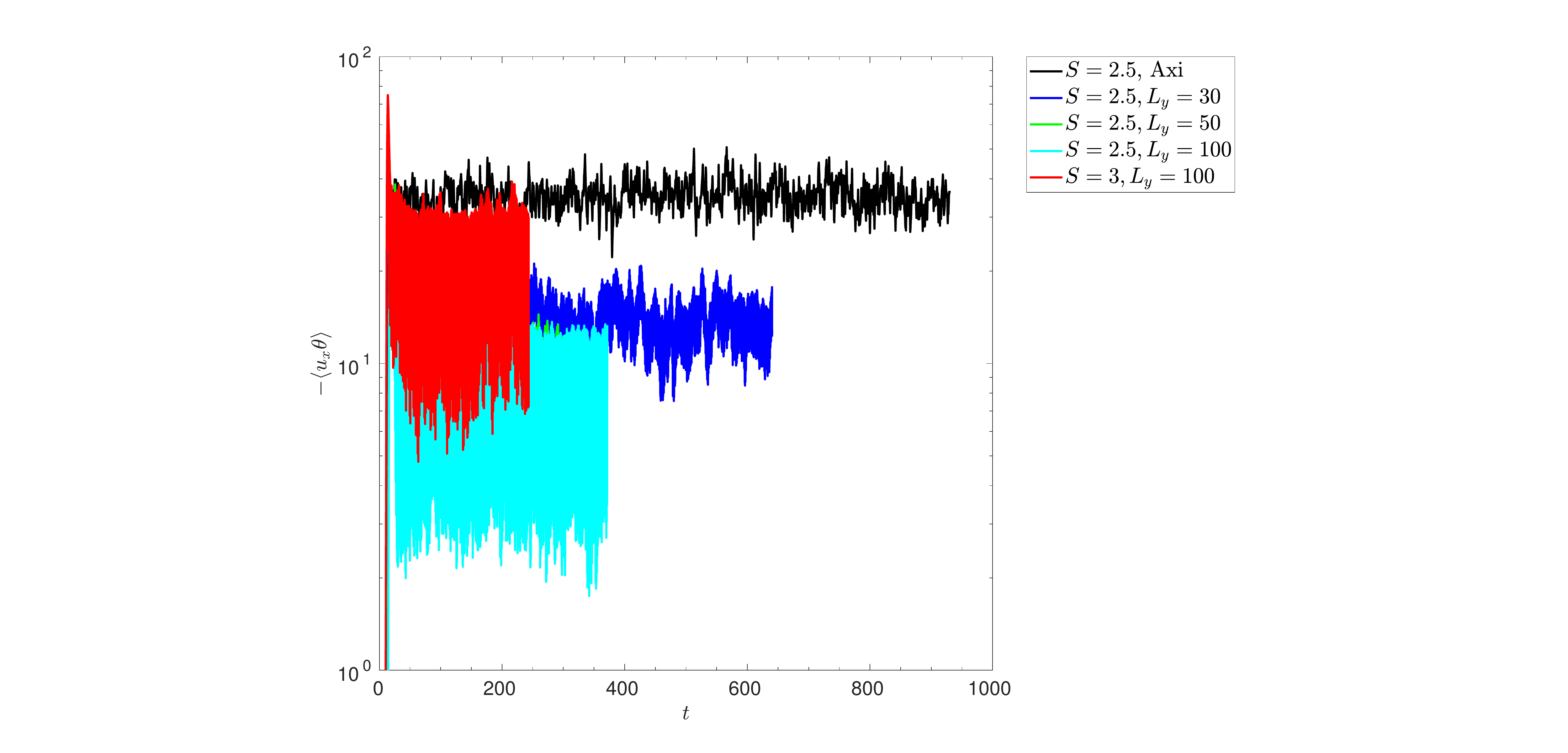}}
    \end{center}
  \caption{Same as Fig.~\ref{Seq2} but for simulations with $S=2.5$ and $S=3$.}
  \label{Seq2p5a}
\end{figure}

\begin{figure}
  \begin{center}
      \subfigure[$t=700$]{\includegraphics[trim=4cm 0cm 5cm 0cm, clip=true,width=0.3\textwidth]{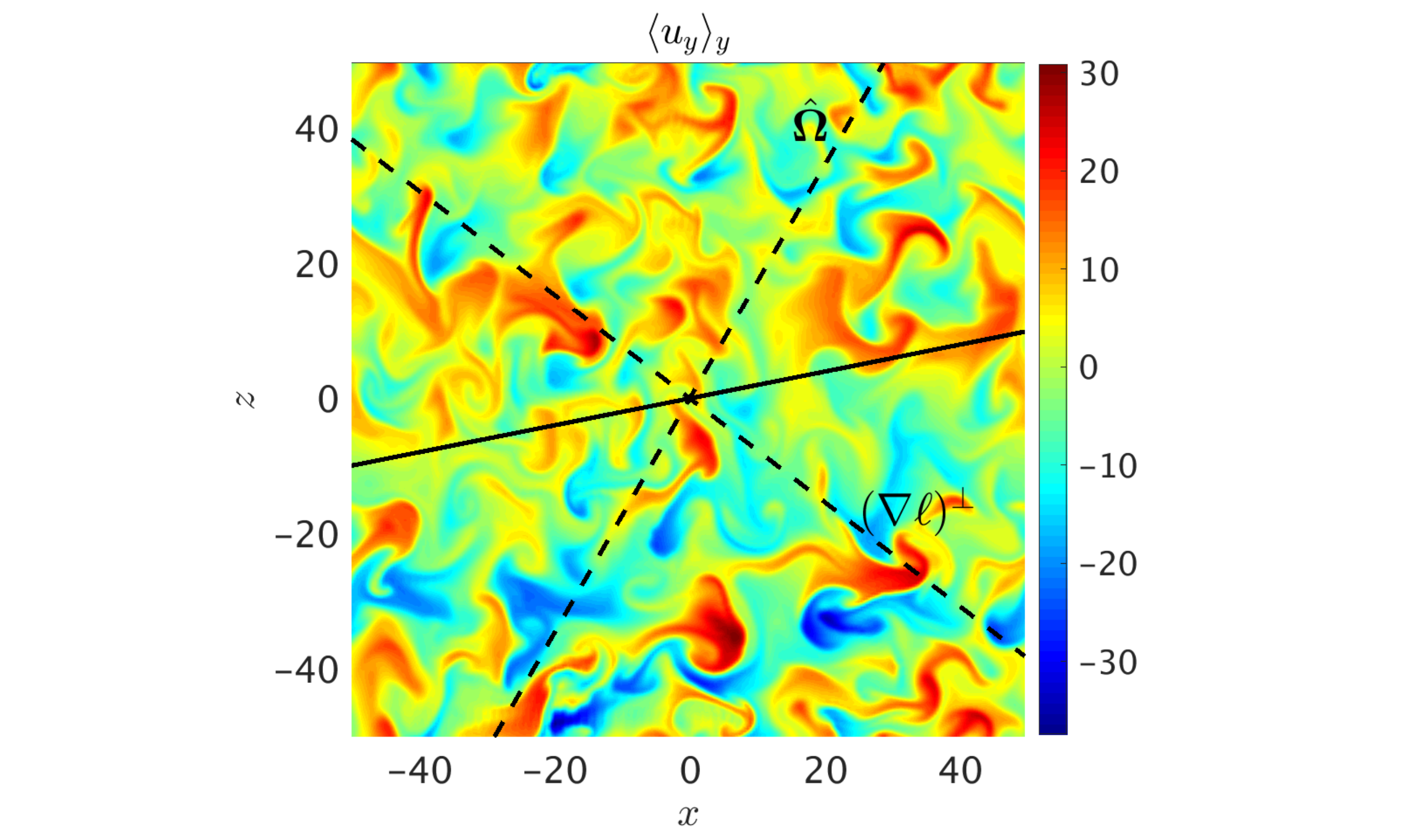}}
    \subfigure[$t=700-800$]{\includegraphics[trim=4cm 0cm 5cm 0cm, clip=true,width=0.3\textwidth]{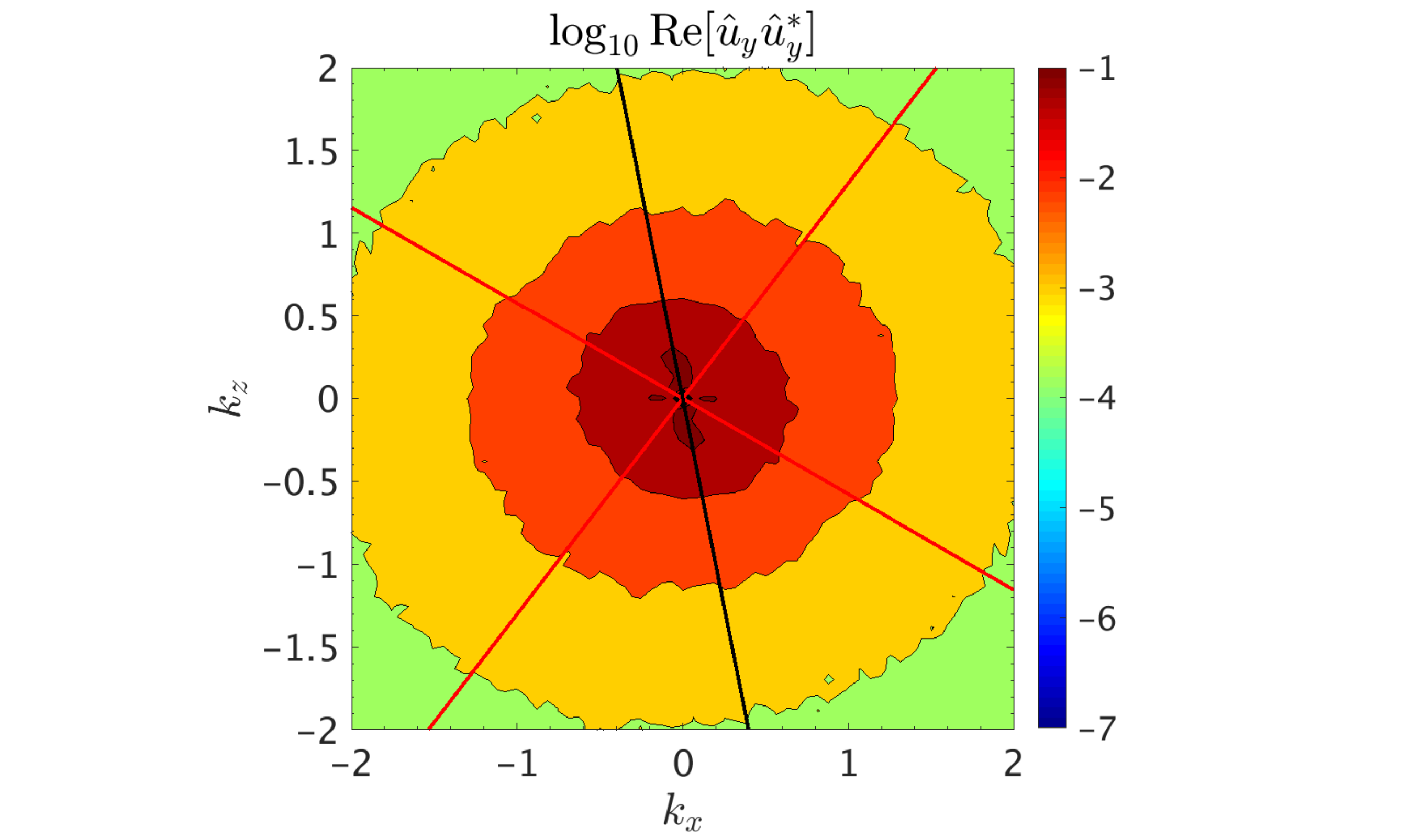}}
    \end{center}
  \caption{Top: snapshots of $u_y$ in the $(x, z)$-plane for an axisymmetric simulation with $S = 2.5$, $\Lambda=30^{\circ}$, $N^2 = 10$, and $\mathrm{Pr} = 10^{-2}$, at $t=700$ during the later nonlinear phases.
  Bottom: Fourier spectrum of $\log_{10} \mathrm{Re}[\hat{u}_y \hat{u}_y^*]$ on the ($k_x,k_z$)-plane from averaging 100 slices from $t=700-800$. The black and red lines are similar to those in previous figures but for the parameters of this simulation.
  }
  \label{Seq2p5flow}
\end{figure}

Our next set of simulations with $\Lambda=30^{\circ}$ explores stronger shear cases with $S=2.5$ and $3$ that would be Rayleigh-unstable in the absence of stratification. These simulations differ significantly from those with weaker shears presented previously. The evolution of volume-averaged quantities is presented in Fig.~\ref{Seq2p5} and \ref{Seq2p5a}, and a snapshot of $u_y$ in the axisymmetric simulation with $S=2.5$ at $t=700$ is shown in the top panel of Fig.~\ref{Seq2p5flow}. The latter shows that the flow primarily consists of finger-like jets, which are comparable in scale with the linear modes, unlike the large-scale zonal jets that were produced in the weaker shear cases, and the flow remains closer to a homogeneous turbulent state. As a result, the flow remains statistically steady with sustained transport properties, exhibiting a weaker dependence on $L_y$ than the cases with smaller $S$ presented previously. These simulations are superficially similar to those at the equator in paper I except that the finger-like jets have a preferred direction that is tilted from the $x$-axis. The modes continue to exhibit a preferential tilt angle that is similar to the prediction from linear theory even during later nonlinear phases. This is shown in the bottom panel of Fig.~\ref{Seq2p5flow}, where the $u_y$ spectrum is presented, based on an average of 100 snapshots from $t=700$ to $t=800$ in the turbulent state from the axisymmetric simulation with $S=2.5$.

The axisymmetric and 3D simulations behave in a qualitatively similar way. The main quantitative difference is that the 3D cases saturate with energies and Reynolds stresses that are smaller by approximately a factor of 2. Results for both $S=2.5$ and $S=3$ are observed to become approximately independent of $L_y$ once this exceeds 30. Presumably these cases differ from those with weaker shears in that the unstable modes instead saturate due to the action of parasitic shear instabilities which limit their amplitudes. These shear instabilities are expected to be weaker in cases with smaller $S$, and may require sufficiently large amplitude to onset that jet mergers occur before they become important. This may be related to the stability of GSF-unstable modes in astrophysical discs as a function of Ro as studied by \cite{LatterPap2018}.

\subsection{Evolution in larger boxes and different aspect ratios for $S=1$ and $S=2$} \label{LargerBoxes}

\begin{figure}
  \begin{center}
      \subfigure{\includegraphics[trim=1cm 0cm 2.5cm 0cm, clip=true,width=0.48\textwidth]{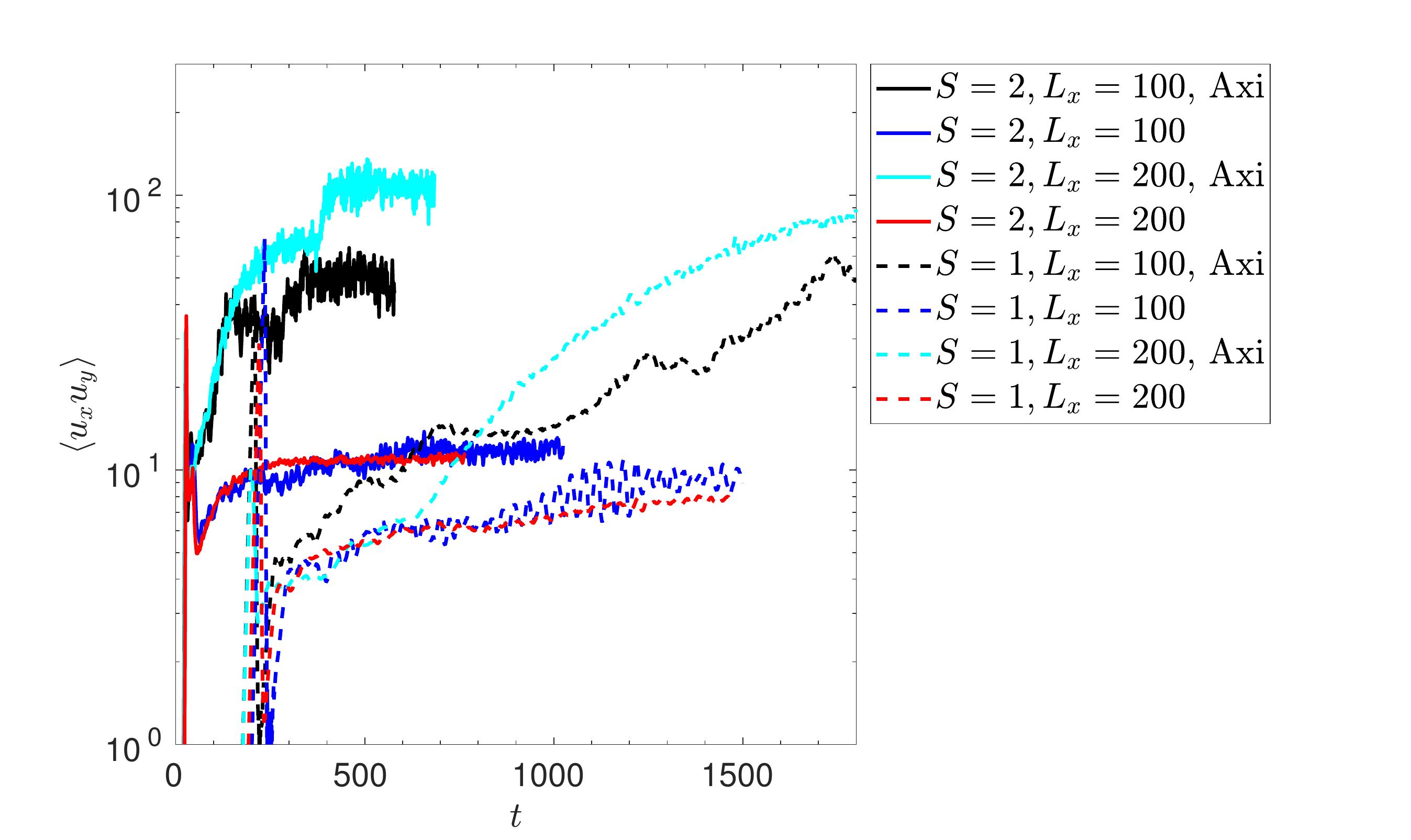}}
       \subfigure{\includegraphics[trim=1cm 0cm 2.5cm 0cm, clip=true,width=0.48\textwidth]{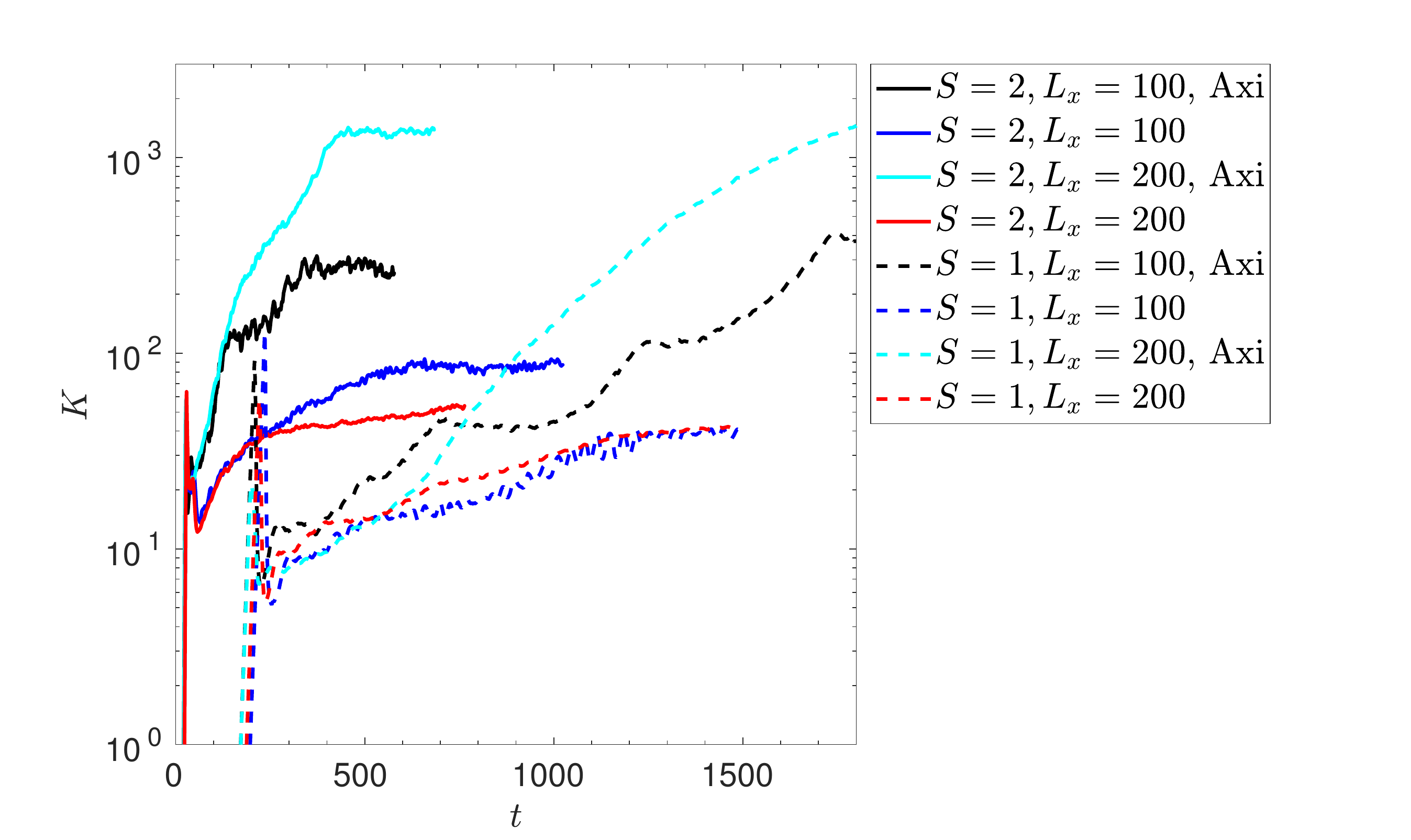}}
    \end{center}
  \caption{Temporal evolution of $\langle u_x u_y\rangle$ and $K$ in a set of simulations with $S=1$ or $S=2$, and $\Lambda=30^{\circ}$, $N^2 = 10$, $\mathrm{Pr} = 10^{-2}$, comparing cases with two different box sizes $L_x=L_z=100$ and $L_x=L_z=200$. Cases with $S=1$ are shown as dashed lines. This shows that 3D cases do not strongly depend on the box size, whereas axisymmetric cases are stronger in larger boxes.}
  \label{Seq2BIG}
\end{figure}

\begin{figure*}
  \begin{center}
   \subfigure[$S=1$, Axi, $t=2820$]{\includegraphics[trim=4cm 0cm 5cm 0cm, clip=true,width=0.35\textwidth]{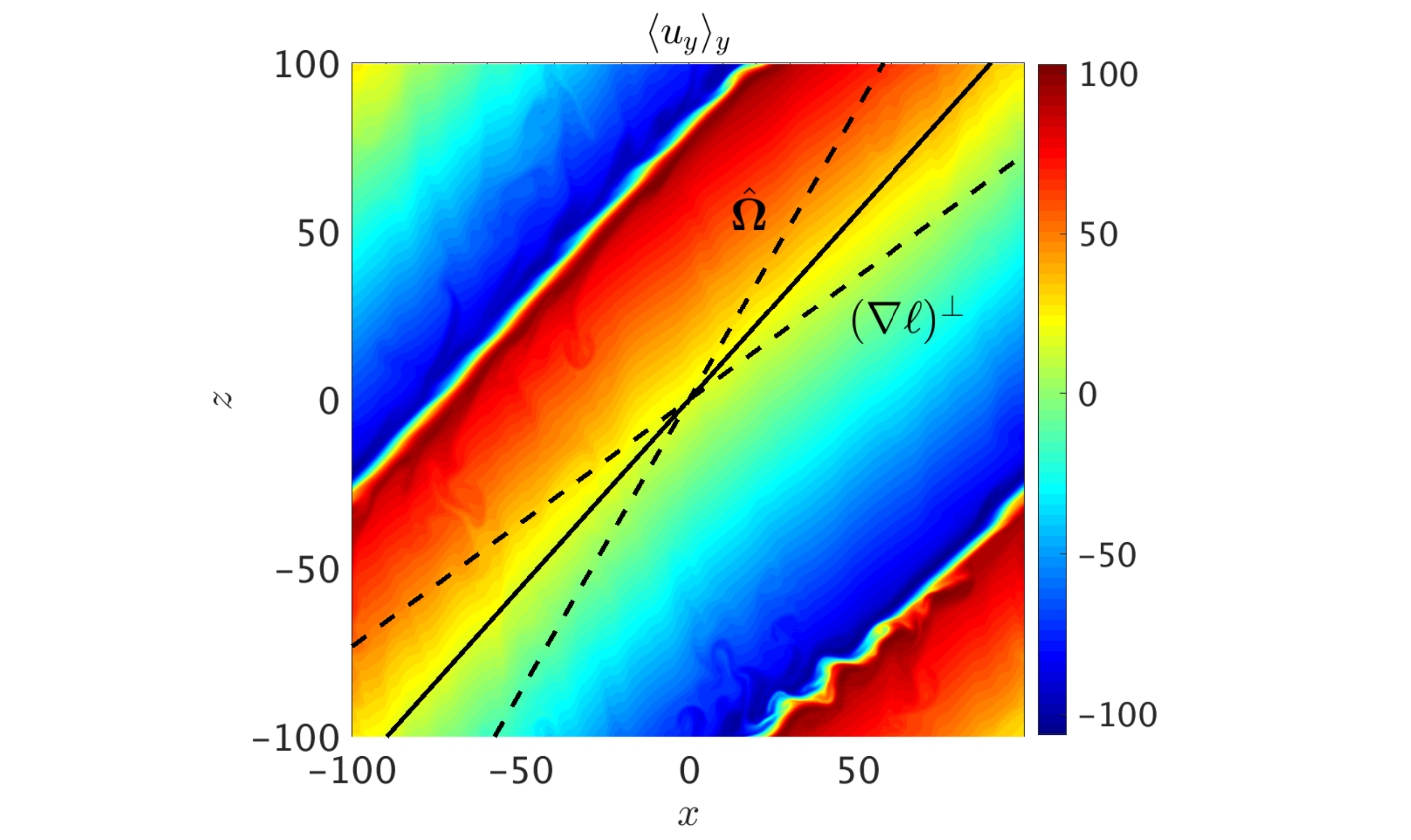}}
   \subfigure[$S=1$, 3D, $t=1460$]{\includegraphics[trim=4cm 0cm 5cm 0cm, clip=true,width=0.35\textwidth]{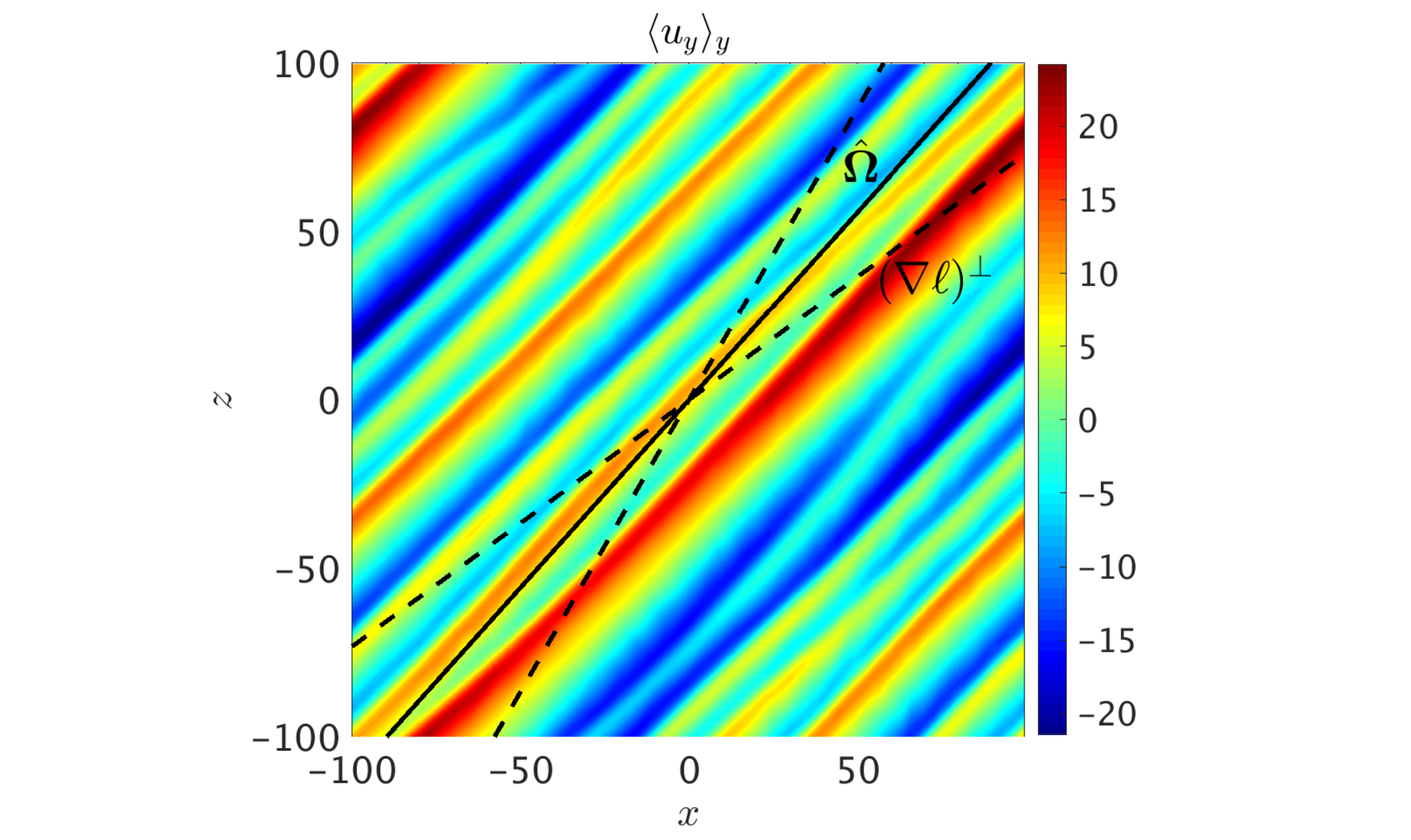}}
    \subfigure[$S=2$, Axi,$t=680$]{\includegraphics[trim=4cm 0cm 5cm 0cm, clip=true,width=0.35\textwidth]{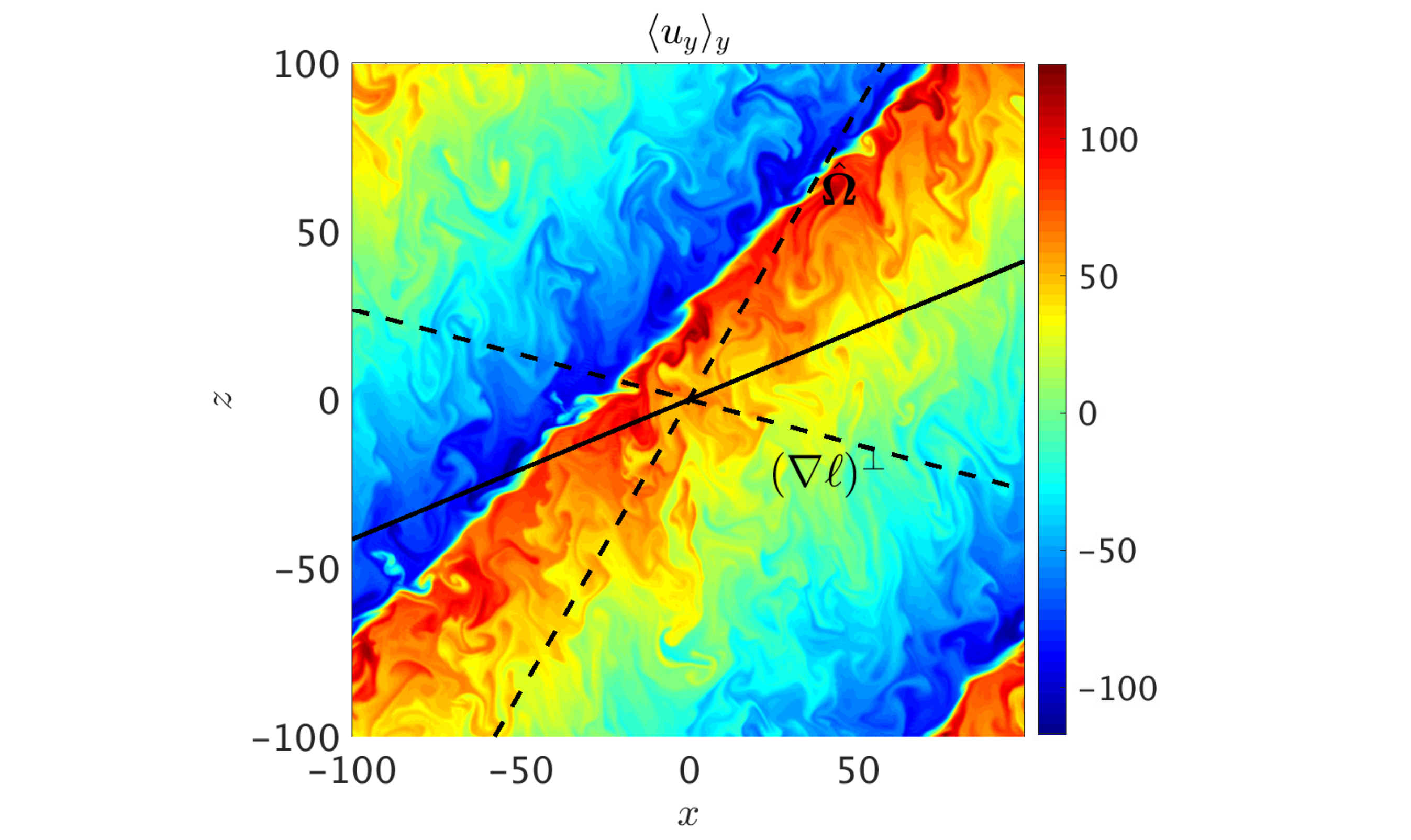}}
    \subfigure[$S=2$, 3D, $t=760$]{\includegraphics[trim=4cm 0cm 5cm 0cm, clip=true,width=0.35\textwidth]{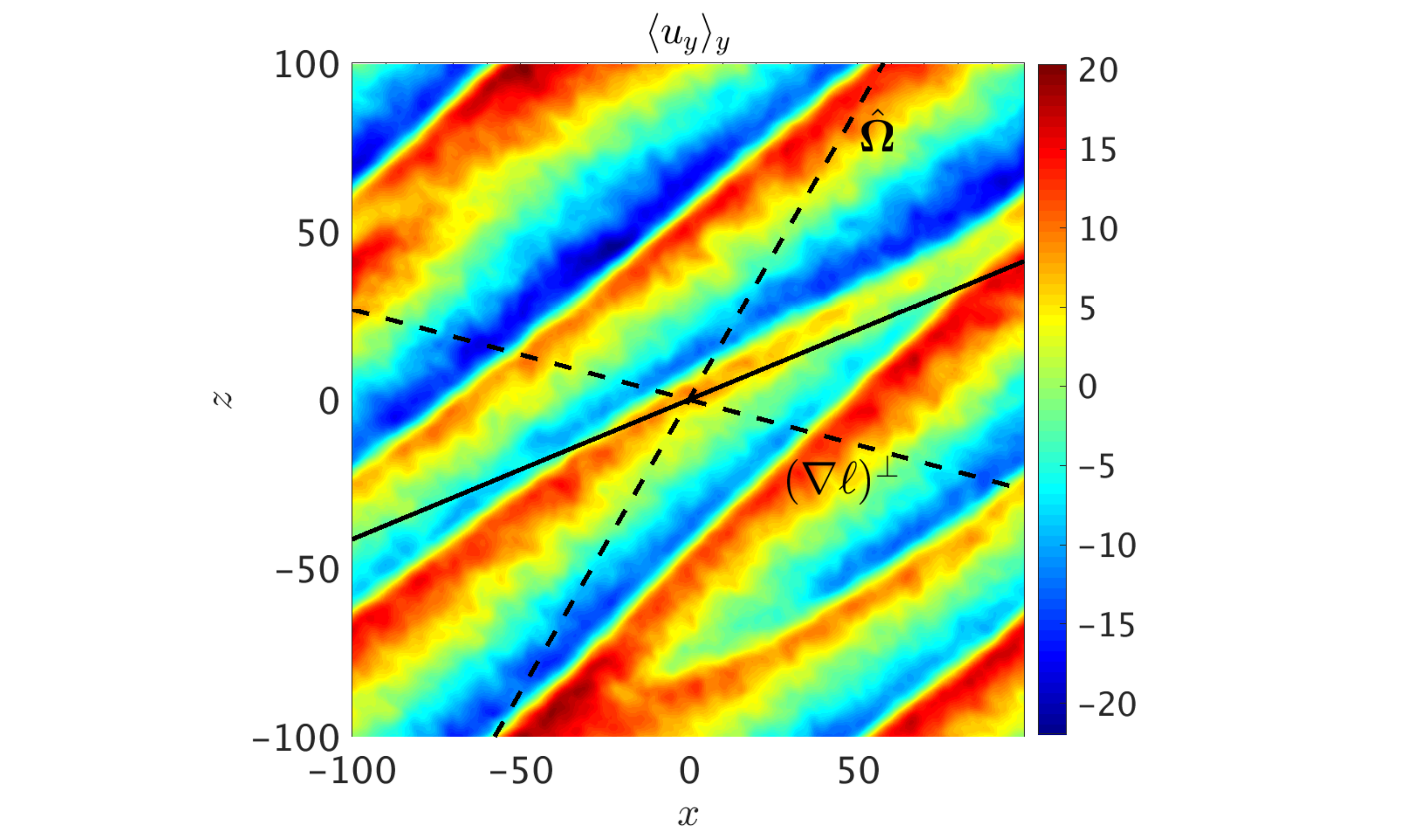}}
    \end{center}
  \caption{Snapshots of $u_y$ in the $(x, z)$-plane for a set of simulations with $L_x=L_z=200$, with either $S=1$ or $S=2$, and $\Lambda=30^{\circ}$, $N^2 = 10$, $\mathrm{Pr} = 10^{-2}$, at various times for both axisymmetric and 3D simulations (where the latter have $L_y=200$).}
  \label{Seq2BIG2}
\end{figure*}

The zonal jets in the weaker shear cases ($S=1, 1.5, 2$) with $L_x=L_z=100$ are observed to grow until they become comparable with the size of the box in $x$ and $z$ (as is most clearly seen in Fig.~\ref{Seq2_2D}). Does this behaviour continue as we increase $L_x$ and $L_z$, and how does the evolution differ in bigger boxes? To answer these questions, we have performed four additional simulations with $L_x=L_z=200$ that have either $S=1$ or $S=2$, and each for both an axisymmetric and a 3D case with $L_y=200$. 

We show the time history of $K$ and $\langle u_x u_y\rangle$ for these new simulations in Fig.~\ref{Seq2BIG}, where we have compared our results with the axisymmetric and 3D cases with $L_x=L_z=100$ (and $L_y=100$ in 3D). Snapshots of the $u_y$ flow in each of these simulations are presented in Fig.~\ref{Seq2BIG2}. We observe that the axisymmetric flow kinetic energy and corresponding transport grow to be substantially larger in the bigger box, with the final saturated value, after undergoing several ``jumps", being approximately a factor of 2 larger. Fig.~\ref{Seq2BIG2} shows that the zonal jets in both cases with a bigger box have grown to be comparable in size with the box in $x$ and $z$, having a wavelength that is twice as large compared with the smaller box snapshots in Figs.~\ref{Seq2_2D} and \ref{Seq1flow}. The flows in these bigger boxes are also much faster. These results suggest that the axisymmetric GSF instability behaves qualitatively like Boussinesq salt fingering (or double-diffusive convection), in which layers merge until they grow to the size of the box \citep{Garaud2018}.

The 3D cases behave in a strikingly different manner, at least for the run times considered here. Fig.~\ref{Seq2BIG} shows that the 3D cases in the biggest box saturate with a similar energy to the smaller box (in fact slightly smaller for the case with $S=2$). The mean value of the transport $\langle u_x u_y\rangle$ is almost identical between the two box sizes in 3D, though the turbulent fluctuations are smaller. Inspection of the flow in Fig.~\ref{Seq2BIG2} suggest the key difference with the axisymmetric cases: the zonal jets are not able to grow to the size of the box in 3D, at least for the run times explored here. This may be because the smaller-scale jets are subject to non-axisymmetric ``parasitic" shear instabilities that limit their amplitudes in 3D. Such non-axisymmetric modes are of course ruled out in axisymmetric simulations. The convergence with increasing $L_x=L_z$ in 3D is promising, and suggests that further simulations with larger boxes may not be necessary for our purposes. This can be confirmed conclusively only with much longer duration simulations however, since the largest scale may only emerge on a timescale proportional to $L_x^2/\nu$.

We speculate that the axisymmetric simulations behave qualitatively differently from the 3D cases because axisymmetric shear instabilities that act on the zonal jets are inhibited by rotation for small flow amplitudes (and presumably only set in if $u \gtrsim \Omega/k$, where $u$ is the velocity amplitude and $k$ is the wavenumber of the flow, by analogy with \citealt{LatterPap2018}), allowing them to reach much larger amplitudes than they could if non-axisymmetric modes were permitted. On the other hand, non-axisymmetric parasitic modes (which are likely to be more important than in the Keplerian case in \citealt{LatterPap2018}, at least for weaker $S$) are likely to operate in 3D for somewhat weaker flow amplitudes. As a result, we may expect the 3D cases to saturate with weaker flows than the axisymmetric cases.

\begin{figure}
  \begin{center}
      \subfigure{\includegraphics[trim=0cm 0cm 0cm 0cm, clip=true,width=0.48\textwidth]{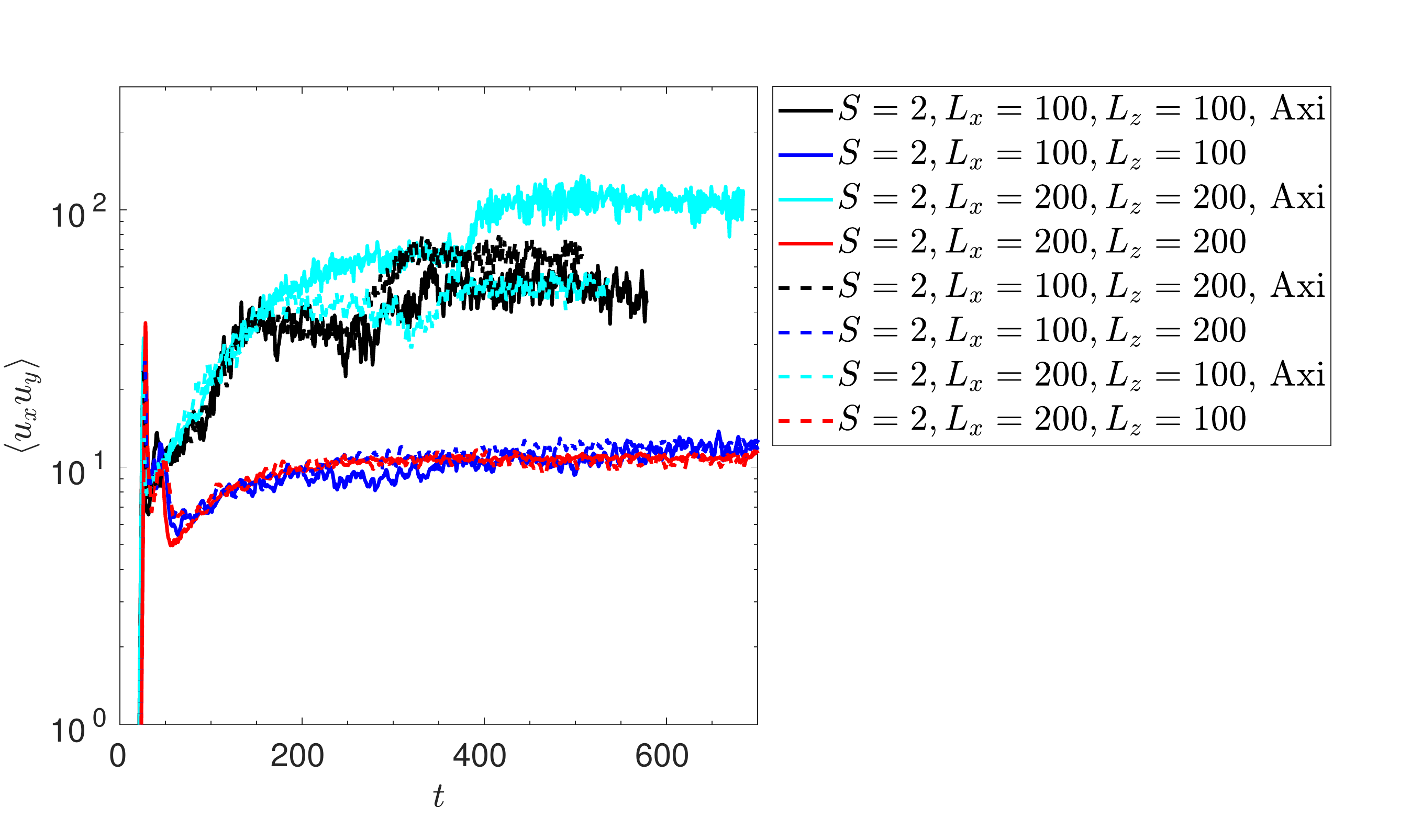}}
       \subfigure{\includegraphics[trim=0cm 0cm 0cm 0cm, clip=true,width=0.48\textwidth]{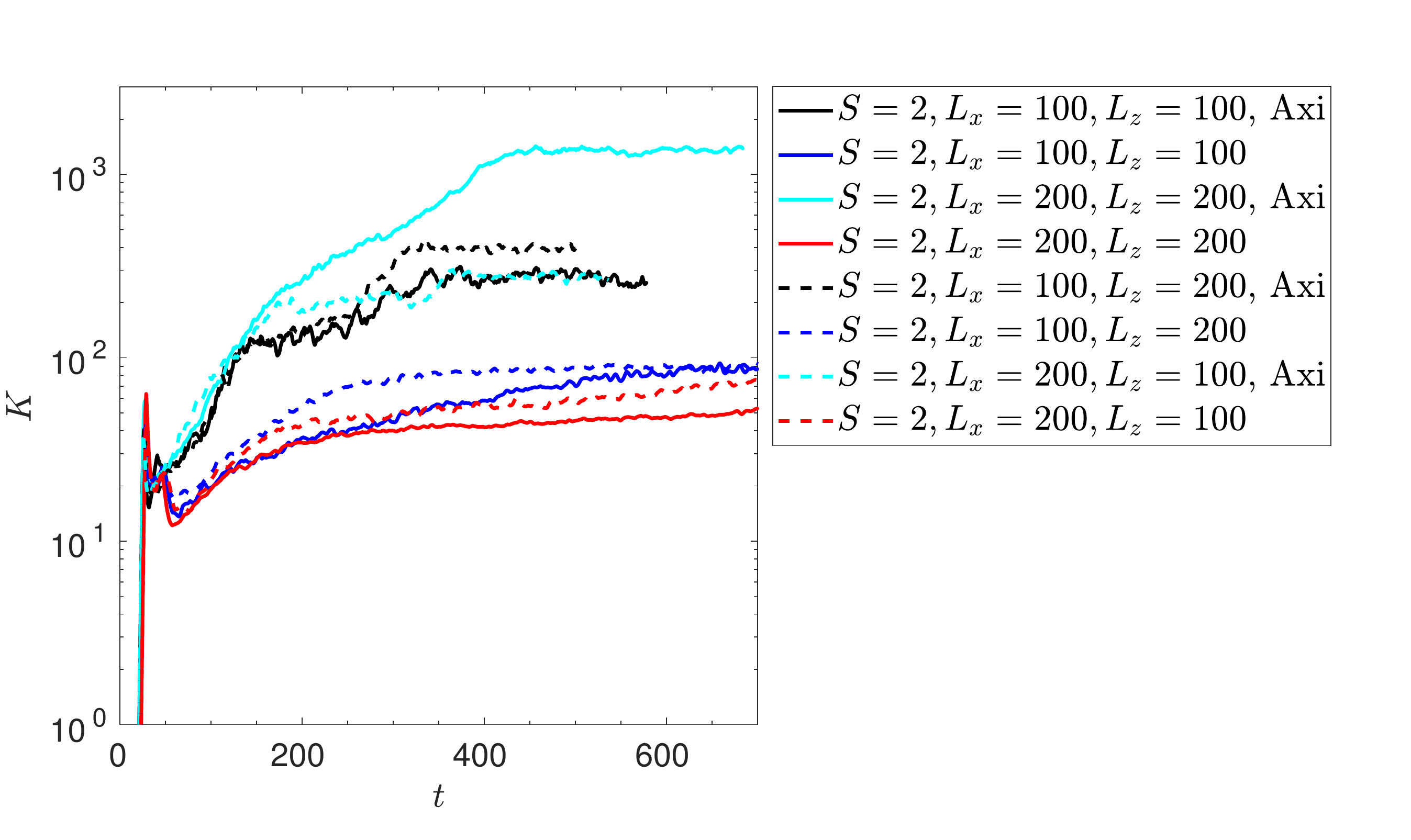}}
    \end{center}
  \caption{Temporal evolution of $\langle u_x u_y\rangle$ and $K$ in simulations with $S=2$, and $\Lambda=30^{\circ}$, $N^2 = 10$, $\mathrm{Pr} = 10^{-2}$, comparing cases with different aspect ratios. This shows that the transport in 3D is not significantly affected by varying the aspect ratio $L_x/L_z$, whereas axisymmetric cases are strongly affected.}
  \label{Seq2aspect}
\end{figure}

Finally, we briefly explore the effect of varying the aspect ratio $L_x/L_z$ in simulations with $S=2$. This quantity might be considered important because zonal jets grow to sizes comparable with the box, so that the dynamics of the jets could be affected by the periodic boundary conditions. For example, the dynamics of double-diffusive intrusions, in which similar (though not directly analogous) large-scale inclined structures are generated \citep{SimeonovStern2007,Medrano2014}, is affected by the degree of inclination of  the box relative to the intrusions. In Fig.~\ref{Seq2aspect} we show the time evolution of $\langle u_x u_y\rangle $ and $K$ in four additional simulations (both axisymmetric and 3D) with $L_x=100$, $L_z=200$ and $L_x=200$, $L_z=100$ together with those with $L_x=L_z=100$ and $L_x=L_z=200$ already presented. Axisymmetric simulations are affected by the aspect ratio, both in their kinetic energy and transport properties. On the other hand, while the kinetic energy in the 3D simulations can differ by $\sim 50\%$ as we vary the aspect ratio from 1/2 to 2, the Reynolds stress components such as $\langle u_x u_y\rangle $ are not significantly affected (other components not shown but behave similarly). This suggests that the aspect ratio (and hence the orientation of the box to the natural angle for the jet formation) does not significantly affect the transport properties that we have observed in 3D, and further indicates that they are less affected by the jets than the axisymmetric cases.

\subsection{Summary}

We have also performed an extensive suite of simulations in which $\Lambda$ is varied. Qualitatively similar behaviour was found at these other latitudes to the cases presented in this section. In particular: cases with weaker shears that are here Rayleigh-stable form strong zonal jets which enhance the transport in axisymmetric cases, and simulations with stronger shears remain closer to a statistically-steady and homogeneous turbulent state. There are significant differences between axisymmetric and 3D simulations, indicating that only the latter should be used to infer the transport properties for application to astrophysics. Since the stronger shear cases (that are here Rayleigh-unstable) saturate in a state of homogeneous turbulence, we might expect a generalisation of the simple single-mode theory in paper I to apply to these, whereas this may not be expected to work when large-scale zonal jets are important. In the next section, we turn to a comparison of the transport properties of the flow with the predictions of a simple single mode theory that is straightforward to compute (e.g.~in stellar evolution codes). 

\section{Theory for saturation of the GSF instability}
\label{theorycomparison}

For astrophysical applications we would like to quantify the angular momentum transport produced by the GSF instability in a simple way so that its effects can be incorporated in stellar evolution codes. For simplicity, we develop the theory introduced in paper I (based on analogy with salt fingering in \citealt{Brownetal2013}) so that it should apply to homogeneous turbulence driven by the instability. A quasi-linear theory may be required though to explain the transport in the presence of strong zonal jets.

Following paper I (see also \citealt{Brownetal2013} for salt fingering), we assume that the flow is dominated by the fastest growing linear mode, and that this mode saturates when its growth rate balances its nonlinear cascade rate. However, we must refine our previous arguments away from the equator, since the fastest growing modes have a preferential tilt in the $(k_x,k_z)$-plane, with an angle 
\begin{eqnarray}
\theta_k = -\mathrm{tan}^{-1}\left(\frac{k_z}{k_x}\right)
\end{eqnarray}
below the $x$-axis (e.g.~Fig.~\ref{lineargrowth}). This is readily calculable from linear theory once we have determined the fastest growing mode. To do this numerically, we solve Eq.~\ref{cubic}, in addition to the two equations obtained by differentiating Eq.~\ref{cubic} with respect to $k_x$ and $k_z$ and setting these equal to zero.
In the limit $\mathrm{Pr}\rightarrow 0$, the tilt angle can be obtained from Eq.~\ref{FGMdirection1}. The velocity vector of the fastest growing mode is tilted in the $(x,z)$-plane by an angle $\theta_u=\pm\pi/2 -\theta_k$. 

The fastest growing mode (with shearing-periodic BCs) is an ``elevator mode", which is a 1D shear flow ($u_\parallel$) along this preferred direction, with a perpendicular wavenumber $k_\perp=\sqrt{k_x^2+k_z^2}$. We expect parasitic instabilities to saturate these modes whenever $s\sim u_\parallel k_\perp$. As in paper I, we define a constant of proportionality $A$, which should only weakly depend on the parameters of the system if the theory is approximately correct, such that 
\begin{eqnarray} 
 u_\parallel \equiv \frac{A s}{k_\perp}.
\end{eqnarray}
We then relate this to $u_x$ by $|u_x| = |u_\parallel\cos\theta_u|=|u_\parallel \sin \theta_k|$. This model reduces to the theory in paper I at the equator, where $\theta_u =0$ and $k_x=0$.

For a single linear mode, the Fourier amplitudes of perturbations are related by
\begin{eqnarray}
\label{linear1}
u_y &=& \frac{\left(\mathcal{S}-2\Omega (\cos \Lambda + \frac{k_x}{k_z}\sin \Lambda)\right)}{s_\nu}u_x, \\
\label{linear2}
u_z &=& -\frac{k_x}{k_z}u_x, \\
\label{linear3}
\theta &=& \frac{-\mathcal{N}^2 (\cos \Gamma - \frac{k_x}{k_z}\sin \Gamma)}{s_\kappa} u_x,
\end{eqnarray}
in terms of the radial velocity $u_x$. Using Eqs.~\ref{linear1}--\ref{linear3} for a single mode, we can construct\footnote{Note that these relations are unchanged when we consider latitudinal differential rotation or moderate centrifugal effects in which $\boldsymbol{e}_g\ne\boldsymbol{e}_x$, and only $s,k_x,k_z$ and $\Lambda$ are modified in this case.}:
\begin{eqnarray}
\langle u_x u_y\rangle &=& \frac{1}{2 s_\nu}\left(\mathcal{S}-2\Omega (\cos\Lambda+\frac{k_x}{k_z}\sin\Lambda )\right)|u_x|^2, \\
\langle u_yu_z \rangle &=& -\frac{k_x}{2s_{\nu} k_z}\left(\mathcal{S}-2\Omega (\cos\Lambda+\frac{k_x}{k_z}\sin\Lambda )\right)|u_x|^2, \\
\langle u_xu_z \rangle &=& -\frac{k_x}{2k_z}|u_x|^2, \\
\langle u_x\theta \rangle &=& -\frac{\mathcal{N}^2(\cos \Gamma - \frac{k_x}{k_z}\sin \Gamma)}{2s_{\kappa}}|u_x|^2, \\
\langle u_z\theta \rangle &=& \frac{k_x\mathcal{N}^2(\cos \Gamma - \frac{k_x}{k_z}\sin \Gamma)}{2k_zs_{\kappa}}|u_x|^2.
\end{eqnarray}
We may now obtain simple predictions for the flow and its resulting transport (such as $\langle u_x u_y\rangle$) in terms of the linear mode properties and a single constant $A$, which we determine by comparison with numerical simulations. Our next task is to explore the validity of this simple theory.

\begin{figure}
  \begin{center}
     \subfigure[$\Lambda=30^\circ, A=5$]{\includegraphics[trim=0cm 0cm 0cm 0cm, clip=true,width=0.52\textwidth]{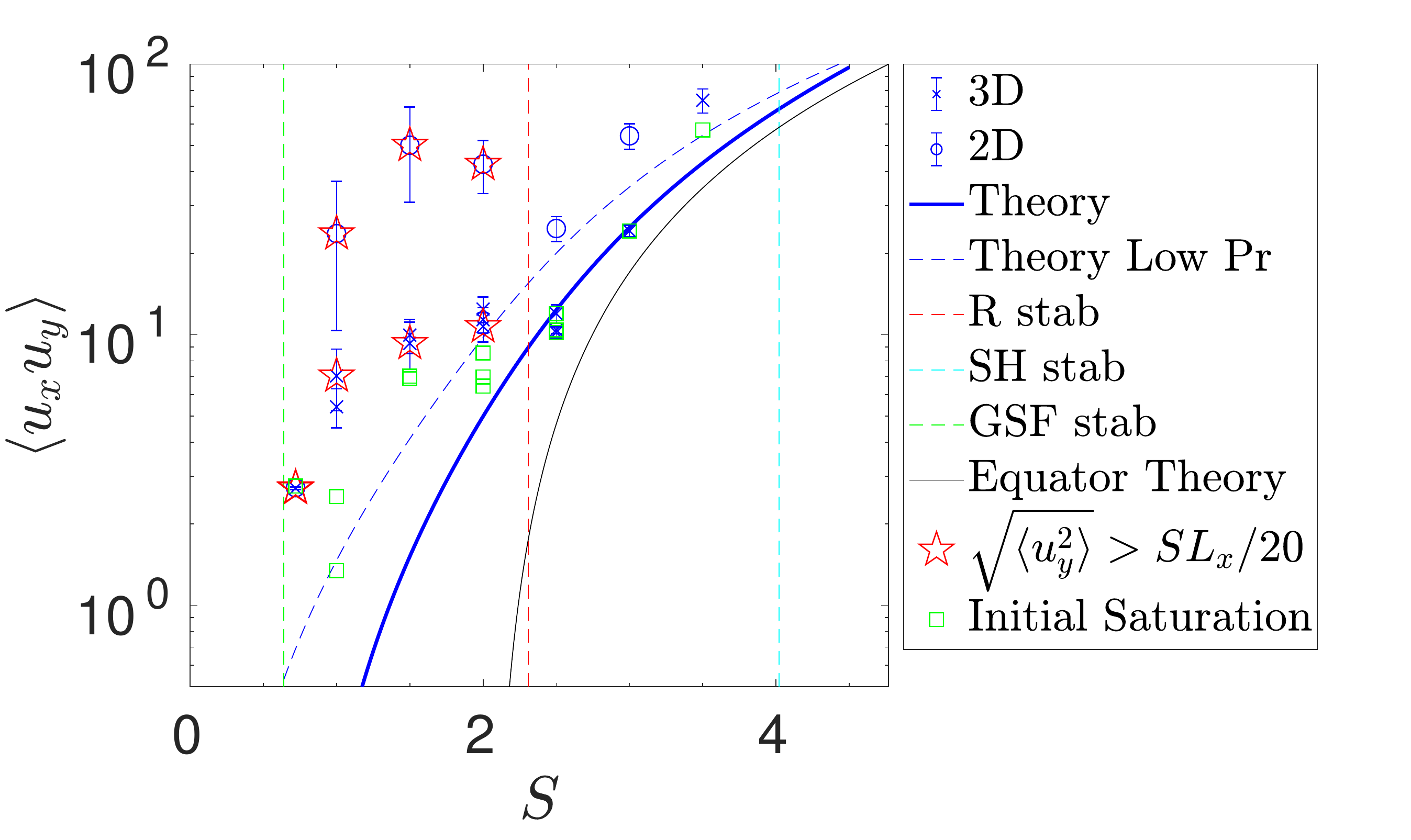}}
     \subfigure[$\Lambda=30^\circ, A=5$]{\includegraphics[trim=0cm 0cm 0cm 0cm, clip=true,width=0.52\textwidth]{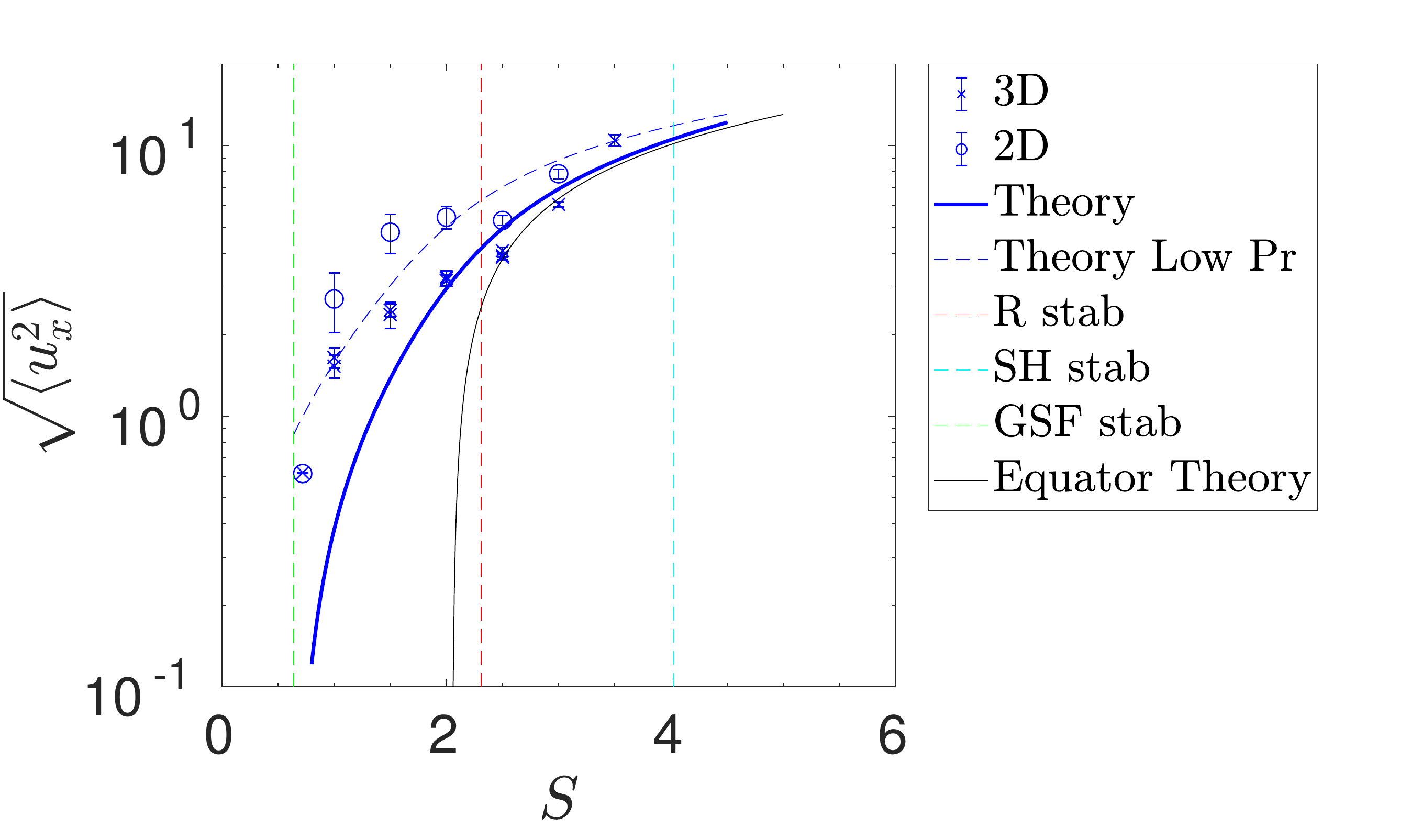}}
     \subfigure[$\Lambda=30^\circ, A=9$]{\includegraphics[trim=0cm 0cm 0cm 0cm, clip=true,width=0.52\textwidth]{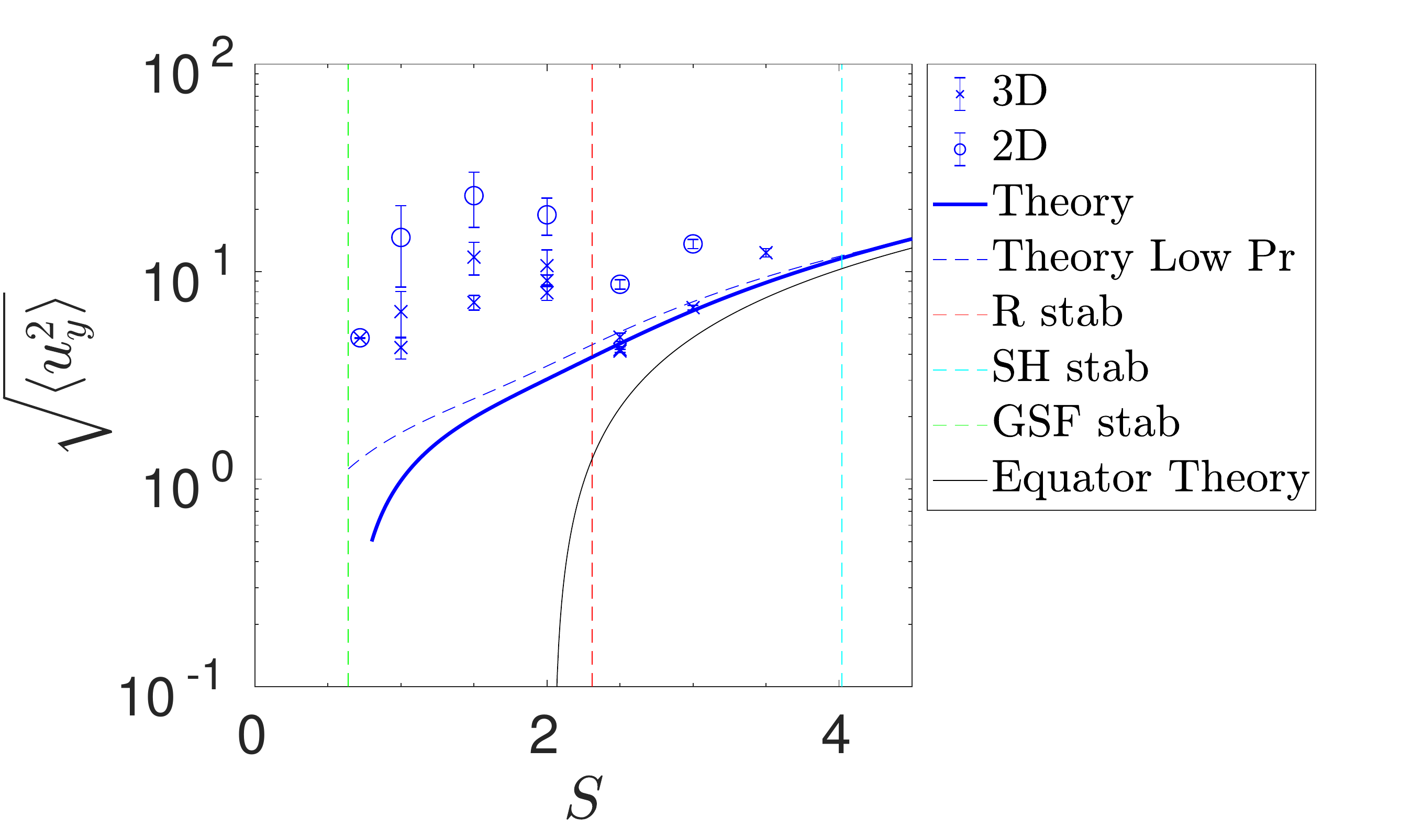}}
    \end{center}
  \caption{Comparison of $\langle u_xu_y\rangle$, $\sqrt{\langle u_x^2\rangle}$ and $\sqrt{\langle u_y^2\rangle}$ against the simple theory, as a function of $S$, showing a set of simulations with $\Lambda=30^{\circ}$, $N^2=10$, $Pr=10^{-2}$, for both axisymmetric (labelled 2D) and 3D cases. This shows that the weaker shear cases that are Rayleigh-stable have much larger flows and transport angular momentum more efficiently than predicted by the simplest homogeneous single-mode theory, by up to two orders of magnitude.}
  \label{TheoryComparison30}
\end{figure}

In Fig.~\ref{TheoryComparison30} (top panel), we show $\langle u_xu_y\rangle$ with error bars based on one standard deviation as a function of $S$ from a range\footnote{The values from simulations are taken as an average over the entire simulation after the linear growth phase, i.e.~they do not show the value during each layered state to avoid further cluttering the figure.} of axisymmetric (2D; blue circles) and 3D simulations (blue crosses, showing results with several different $L_y$) with $\Lambda=30^\circ$. In the bottom two panels we show $\sqrt{\langle u_y^2\rangle}$ and $\sqrt{\langle u_z^2\rangle}$. In each panel we indicate the line $\mathrm{RiPr}=\frac{1}{4}$ as the green-dashed line (Eq.~\ref{RiPrcrit}), Rayleigh stability (Eq.~\ref{Rayleigh}) as the red dashed line and Solberg-H\o iland stability (Eq.~\ref{adcrit2}) as the light blue dashed line. We also plot the theoretical prediction for these quantities according to the theory discussed above as the solid blue line, and a version based on the limit $\mathrm{Pr}\rightarrow 0$ (using Eqs.~\ref{FGMdirection} and \ref{FGMgrowthlowPr}) as the dashed blue line (which might be expected to provide the most efficient transport in the GSF unstable regime). Finally, the prediction according to the theory validated against simulations in paper I at the equator is plotted as the solid black line.

Firstly, we notice that the instability is much more efficient at transporting angular momentum, and drives much stronger flows, at non-equatorial latitudes compared with at the equator. The GSF instability at the equator requires $S>2$, whereas at other latitudes we only require $\mathrm{RiPr}<\frac{1}{4}$ (corresponding with $S>0.633$), which is much less restrictive. Secondly, we also observe here that the axisymmetric (2D) simulations typically produce stronger flows, and provide more efficient transport (by approximately a factor of 2), than the 3D simulations. This indicates that 3D simulations are probably required for understanding the instability in stellar interiors.

The simple single-mode theory with $A\approx 5$ does a reasonable job of capturing the transport in the stronger shear cases (that are here Rayleigh-unstable), albeit only for a narrow range of $S$ values. It does not work well for all $S$ however. Indeed, we might expect the theory to fail in the weaker shear cases in which strong zonal jets are generated. The top panel in Fig.~\ref{TheoryComparison30} indicates the value of $\langle u_xu_y\rangle$ for the 3D simulations after the initial saturation but before strong zonal jets have formed with green squares (note that simulations with various $L_y$ are plotted for certain $S$ values, as listed in Table~\ref{Table}). These values lie closer to the simple theoretical predictions, as we might expect. We have additionally indicated cases with strong zonal jets, defined as those simulations in which $\sqrt{\langle u_y^2\rangle}$ (based on a time-average of this quantity after the linear growth phase) exceeds $S L_x/20$ by over-plotting these points with red stars in the top panel of Fig.~\ref{TheoryComparison30}. This clearly demonstrates that the cases where the theory under-predicts the transport are those in which strong zonal jets have developed. Presumably a quasi-linear theory is required to explain the transport in these cases, which is a topic worthy of exploration in future work.

The main result in Fig.~\ref{TheoryComparison30} is that the transport is enhanced over the simple single-mode theory, by up to several orders of magnitude in the weakest shear cases dominated by zonal jets. Note that the largest values of $S$ considered are such that $\mathrm{Ri}=O(1)$, where we also expect the simple theory to no longer apply based on our observations in paper I.

\begin{figure}
  \begin{center}
    \subfigure[$\Lambda=60^\circ, A=5$]{\includegraphics[trim=0cm 0cm 0cm 0cm, clip=true,width=0.52\textwidth]{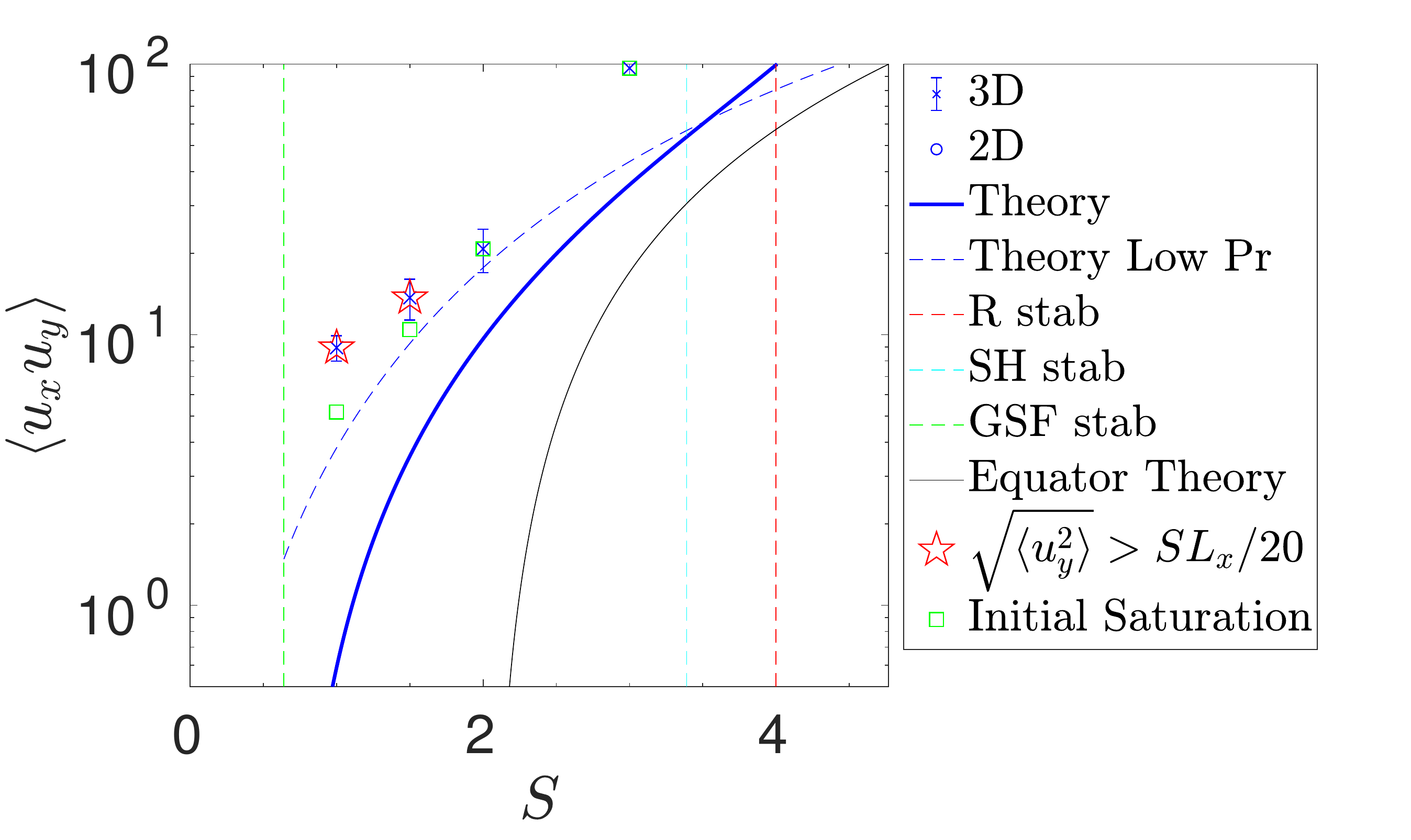}}
    \subfigure[$\Lambda=90^\circ, A=5$]{\includegraphics[trim=0cm 0cm 0cm 0cm, clip=true,width=0.52\textwidth]{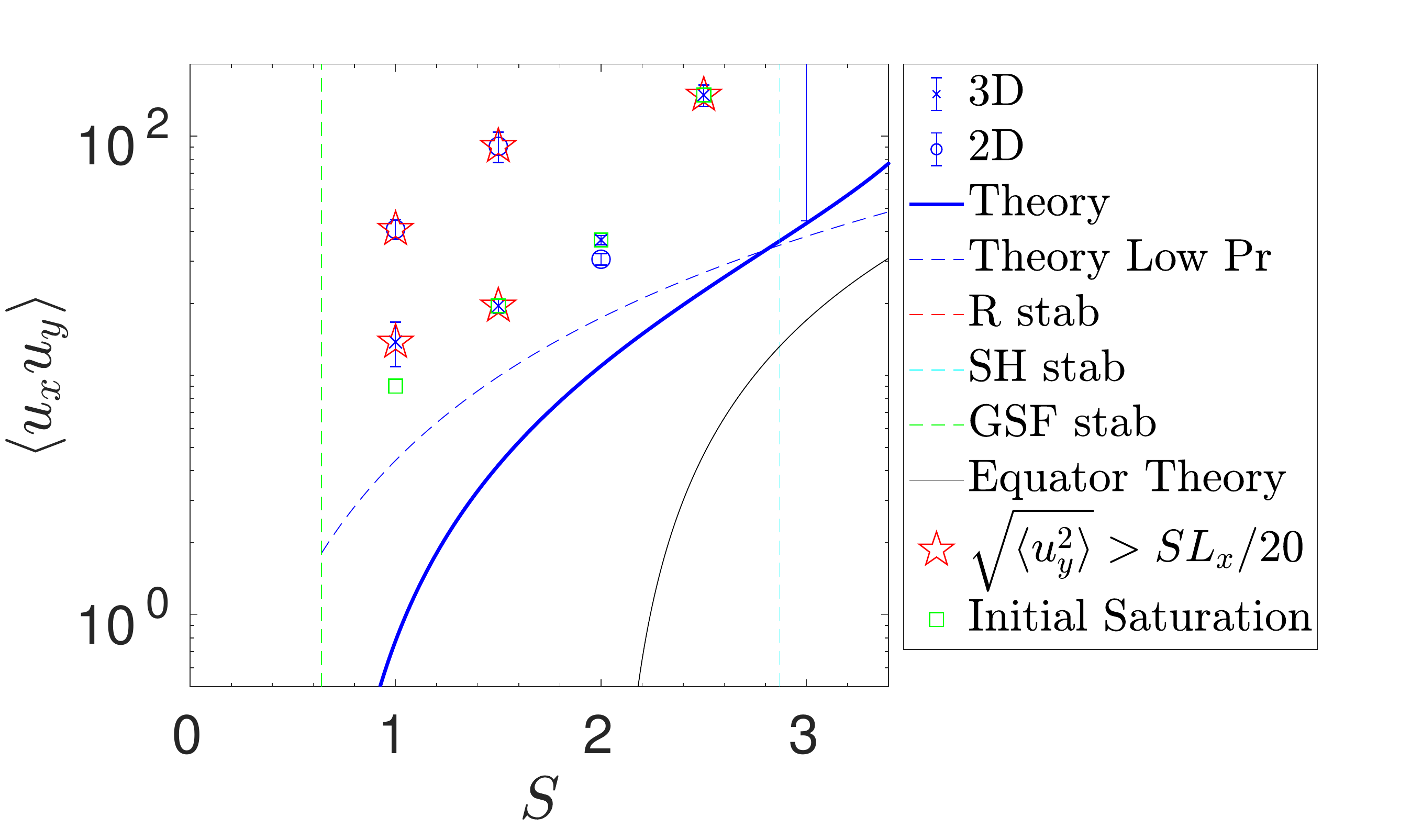}}
    \end{center}
  \caption{Comparison of $\langle u_xu_y\rangle$ against the simple theory, as a function of $S$, showing a set of simulations with $\Lambda=60^{\circ}$ and $\Lambda=90^\circ$, with $N^2=10$, $Pr=10^{-2}$, for both axisymmetric (labelled 2D) and 3D cases.}
  \label{TheoryComparison6090}
\end{figure}

We show a similar comparison for $\langle u_xu_y \rangle$ using simulations (not previously presented) at $\Lambda=60^\circ$ and $90^\circ$ (north pole) in Fig.~\ref{TheoryComparison6090}. These also show that the transport is significantly enhanced over the simple single-mode theory, due to the presence of strong zonal jets. The transport is also observed to be more efficient at higher latitudes. These figures indicate that $A$ may depend weakly on $\Lambda$, and so does not appear to be a universal constant for the non-equatorial GSF instability. One possibility to improve the match between simulations and theory would be to incorporate multiple-modes (rather than just the single fastest growing mode i.e. by instead fully accounting for the shape of the growth rate contours in $k$-space). For example, such an approach is required to apply rotating mixing length theory to explain the bulk properties of convection \citep{CBLB2020}. However, the strong zonal jets that form in the current problem may prevent this approach from removing the discrepancy. Further work is required to understand theoretically the transport by the GSF instability with weaker shears. We believe that such a theory may involve examining the quasilinear response of the shear and temperature fields to the Reynolds stresses and heat fluxes --- or the generalised quasilinear versions of the same theories \citep{TDM2011,MQT2014,mct2016}.

\section{Astrophysical Implications}
\label{astrophysical}

We now turn to estimate the astrophysical relevance of the GSF instability. However, we should note that uncertainties remain, particularly regarding the lack of a theory to describe the turbulent transport in weak shear cases with strong zonal jets. As in paper I, we must convert quantities from our dimensionless units to obtain the physical rates of angular momentum transport. We note that 
\begin{eqnarray}
\langle u_xu_y\rangle_{\mathrm{real}}= \Omega^2d^2 \langle u_xu_y\rangle_{\mathrm{code}}
\end{eqnarray}
which relates the Reynolds stress in physical units (subscript ``real") with the output from our simulations (subscript ``code"). For a crude estimate, we assume that the GSF instability transports angular momentum radially in the form of an eddy diffusion with a diffusivity $\nu_E$. An appropriate effective viscosity is given by 
\begin{eqnarray}
\label{nuEexp}
\nu_E = \frac{\langle u_xu_y\rangle_{\mathrm{real}}}{\mathcal{S}} = \nu 
S^{-1} N^{-1}  \mathrm{Pr}^{-1/2}\langle u_xu_y\rangle_{\mathrm{code}},
\end{eqnarray}
and an effective viscous timescale for angular momentum transport over a distance $L$ is
\begin{eqnarray}
\nonumber
t_{\nu}&=&\frac{1}{\langle u_xu_y\rangle_{\mathrm{code}}} \frac{L^2}{d^2} S \Omega^{-1}.
 \label{tnuExp}
\end{eqnarray}
Our simulations indicate that at non-equatorial latitudes, $\langle u_xu_y\rangle_{\mathrm{code}}\approx 10-100$, at least for $\mathrm{Pr}=10^{-2}$ for the $S$ and $N^2$ values considered in this work. This is typically much more efficient than instability at the equator. In the absence of strong zonal jets, we speculate that the transport will not strongly depend on Pr (for a partial justification, see \S~\ref{theorycomparison} and the discussion in paper I). However, the dynamics of the zonal jets, how the resulting transport depends on Pr, and whether they attain an ultimate size and strength, remain to be established. 

\subsection{Red giant and subgiant stars}
\label{redgiant}
Our first example is the application of our results to red giant stars, for which the models of \cite{Eggenberger2017} suggest an additional viscosity of $\nu=10^3-10^4 \mathrm{cm}^2\mathrm{s}^{-1}$ is required to explain their observed weak core-envelope differential rotations. We adopt the numbers from paper I (following e.g.~\citealt{Caleo2016a,Eggenberger2017}), to estimate an effective viscosity due to the non-equatorial instability,
\begin{eqnarray}
\nu_E\approx 500 \,\mathrm{cm}^2 \mathrm{s}^{-1} \frac{\langle u_xu_y\rangle_{\mathrm{code}}}{100},
\end{eqnarray}
which is slightly smaller than the required value. This crude estimate nevertheless indicates that the GSF instability could provide an important contribution to the ``additional viscosity" required in red giant stars. We advocate further work to explore the implementation of the theory in \S~\ref{theorycomparison} in stellar evolution codes incorporating rotation to explore whether this mechanism can work in practice.

\label{subgiant}
In a similar study to the above mentioned work on red giant stars, \cite{Eggenberger2019} suggest an additional viscosity of $\nu=10^3-10^4 \mathrm{cm}^2\mathrm{s}^{-1}$ is required to explain the observed weak core-envelope differential rotations of subgiant stars. The above crude estimate indicates that it would be worthwhile to explore further whether the GSF instability could also be important in the evolution of these stars.

\subsection{Solar Tachocline}
\label{Tachocline}
As already discussed, the stably-stratified layers in the lower parts of the solar tachocline may be GSF-unstable (away from the equator) -- even if this instability is not expected in the bulk of the radiation zone of the current Sun \citep{Rashid2008,Caleo2016a}. Using the numbers from paper I, we estimate
\begin{eqnarray}
\nu_E\approx 5\times 10^5 \mathrm{cm}^2 \mathrm{s}^{-1} \frac{\langle u_xu_y\rangle_{\mathrm{code}}}{100},
\end{eqnarray}
giving an effective viscous timescale 
\begin{eqnarray}
t_{\nu}\approx 0.03\mathrm{Myr} \left(\frac{L}{0.01 R_{\odot}}\right)^2\frac{1}{\langle u_xu_y\rangle_{\mathrm{code}}/100},
\end{eqnarray}
to transport angular momentum over the radial extent of the tachocline region (assumed to have $L=0.01R_{\odot}$). This estimate supports the suggestion in paper I that the GSF instability could be important for the long-term angular momentum transport in the tachocline. This mechanism may also be important in providing turbulent diffusion at mid-latitudes, which could play a crucial role in models of the tachocline (e.g.~\citealt{GM1998,McIntyre2007,Wood2011}).
This mechanism is also expected to have been even more important in the past, when the Sun was rotating more rapidly, so it may have played a role in the evolution of the internal rotation of Sun \citep{Menou2006}.

\subsection{Hot Jupiter atmospheric jets}
\label{HJ}

The atmospheric jets that advect heat from dayside to nightside on hot Jupiters occur in stably-stratified surface layers. The jets that are observed in simulations are often transonic (or possibly supersonic), with strong radial and latitudinal shear. Their atmospheres are also likely to have very small Pr and have effective thermal diffusion. These are conditions in which the GSF instability could operate, as first speculated by \cite{Goodman2009}. For a crude estimate, adopting numbers from \cite{Menou2019}, we find $\mathcal{N}\approx 2 \times 10^{-3}\mathrm{s}^{-1}$, the local rotation period is of order 1 day assuming synchronous rotation, i.e. $\Omega\approx 7 \times 10^{-5}\mathrm{s}^{-1}$, and we adopt a jet of shear strength $\mathcal{S}/\Omega\approx 140$. At $P\approx0.01$ bar, $\kappa\approx 10^{11}\mathrm{cm}^2 \mathrm{s}^{-1}$ \citep{Menou2019}, and we estimate $\nu\approx 10^4 \mathrm{cm}^2 \mathrm{s}^{-1}$ \citep{LiGoodman2010}. We therefore obtain
\begin{eqnarray}
d\approx 1 \;\mathrm{km},
\end{eqnarray}
indicating that this instability occurs on short length-scales. This is impossible to resolve in global simulations  \citep[e.g.][]{Showman2009,DobbsDixon2010,Fromang2016,Mayne2017}, so the effects of this instability on limiting jet strengths and modifying their profiles would not previously have been captured. The resulting effective viscosity is estimated to be
\begin{eqnarray}
\nu_E\approx 3\times 10^6 \mathrm{cm}^2 \mathrm{s}^{-1} \frac{\langle u_xu_y\rangle_{\mathrm{code}}}{100}.
\end{eqnarray}
This crude estimate suggests that this mechanism may be weaker than the related one discussed using order-of-magnitude estimates by \cite{Menou2019}, presumably because the GSF instability preferentially excites short-wavelength modes. Nevertheless, the consequences of this instability for the dynamics of hot Jupiter atmospheres should be explored further. The resulting vertical mixing could also be important for their atmospheric chemistry.

\section{Conclusions}
\label{Conclusions}
We have presented the first exploration into the nonlinear evolution of the Goldreich-Schubert-Fricke (GSF) instability at a general latitude in a star (or planet), building upon our initial study at the equator in paper I \citep{BJT2019}. This instability can provide an important contribution to angular momentum transport in the stably-stratified radiation zones of differentially-rotating stars (or giant planets), but its nonlinear evolution has not been explored in this general case previously (except for the weakly nonlinear analysis in \citealt{Knobloch1982}). We first revisited the linear instability (see also \citealt{Acheson1978,KnoblochSpruit1982}), discussed its properties in detail, and derived several new results. In particular, we derived the following simple criterion for onset of (diffusive) axisymmetric instability: $\mathrm{RiPr}<\frac{1}{4}$, where Ri is the local (gradient) Richardson number and Pr is the (thermal) Prandtl number. At the equator the flow must instead violate Rayleigh's criterion for centrifugal instability, which is typically much more restrictive. 

We presented the results from a suite of hydrodynamical simulations using a local Cartesian model (with both shearing-periodic and impenetrable, stress-free, radial boundaries) to explore the nonlinear evolution of this instability at a range of latitudes ($\Lambda=30^\circ$, $60^\circ$ and $90^\circ$) for various shear strengths, spanning the range from cases that would be Rayleigh-stable to those that would be Rayleigh-unstable in the absence of stable stratification. The GSF instability exhibits interesting dynamics at a general latitude, particularly in the weaker shear cases, where strong zonal jets were observed to develop. These jets propagate with a preferred direction in the meridional plane, which initially corresponds with that of the fastest growing linearly unstable modes. They subsequently merge and strengthen until they occupy a large fraction of our simulation domain, after which the tilt angle of these flows can depart from the linear prediction if they are sufficiently strong. When these jets form, they are observed to significantly enhance the turbulent transport, particularly in axisymmetric simulations. On the other hand, the strong shear cases exhibit a state that is closer to homogeneous turbulence, consisting of smaller-scale jets closer to the length-scale (and with the preferred direction) of the fastest growing linear modes.

The large-scale zonal jets can be thought of as angular momentum ``layering", by analogy with the layering in the density field observed in other stably-stratified flows such as salt fingering \citep[e.g.][]{Garaud2018}. Similarly with other double-diffusive problems, these jets are observed to merge until they occupy the full-extent of the box in our axisymmetric simulations. As with other double-diffusive problems in which layers are observed to form, their long-term evolution and ``ultimate" scale and strength are not currently well understood theoretically. However, the 3D simulations behave qualitatively differently, leading to zonal jets of finite size and strength that do not appear to continue to merge in larger boxes. Further work should explore the origin and dynamics of these jets to confirm whether they do indeed attain an ultimate scale and strength in 3D.

The GSF instability transports angular momentum much more efficiently at non-equatorial latitudes than it does at the equator, often by several orders of magnitude. We have compared the transport produced by our non-equatorial simulations with the predictions from a generalisation of the simple single-mode theory that we validated against equatorial simulations in paper I. We found that this theory significantly under-predicts the transport in the weak shear cases in which strong zonal jets are produced, potentially by more than an order of magnitude, though it may approximately apply in strong shear cases. The strong zonal jets in cases with weak differential rotation enhance the prospect that the GSF instability could provide efficient turbulent transport in stellar and planetary interiors.

We estimate that the GSF instability could play an important role in transporting angular momentum in red giant (e.g.~\citealt{Beck2012,Eggenberger2016,Eggenberger2017}) and subgiant stars (e.g.~\citealt{Eggenberger2019}), which could contribute to the ``additional viscosity" required to explain their observed core rotation rates. It could also play a role in the formation and evolution of the solar tachocline, and in the dynamics of atmospheric jets on hot Jupiters. It would be worth exploring the astrophysical consequences of the GSF instability further with stellar evolution codes incorporating rotation.

We have also found axisymmetric simulations to over-predict the transport and flow kinetic energy, compared with three-dimensional simulations. This indicates that three-dimensional simulations are probably required to determine the transport properties for astrophysical applications. However, astrophysically relevant values of Pr are currently impossible to achieve in simulations, which requires us to extrapolate our results, as with many other problems involving astrophysical fluids.

Topics worthy of exploration in future work include the incorporation of gradients in heavy elements (e.g.~\citealt{KnoblochSpruit1983}), the influence of magnetic fields (e.g.~\citealt{Menouetal2004}), and the investigation of smaller Pr fluids. It would also be worthwhile to perform global simulations to explore the evolution of the GSF instability in spherical geometry, and in particular the dynamics of the resulting zonal jets, though this will be a very challenging numerical problem. Finally, the derivation and analysis of an asymptotically-reduced model of the GSF instability (along the lines of e.g.~\citealt{Xie2019}) may shed some light on the low Pr limit, and potentially also on the origin and evolution of the zonal jets.

\section*{Acknowledgements}
We would like to thank the referee for a prompt and constructive report that helped us to improve the paper. AJB was supported by STFC grants ST/R00059X/1 and ST/S000275/1, and initially by the Leverhulme Trust through the award of an Early Career Fellowship. CAJ was supported by STFC grant ST/S00047X/1. SMT was supported by funding from the European Research Council (ERC) under the EU's Horizon 2020 research and innovation programme (grant agreement D5S-DLV-786780). This work was undertaken on ARC1, ARC2, ARC3 and ARC4, part of the High Performance Computing facilities at the University of Leeds, UK. Some simulations were also performed using the UKMHD1 allocation on the DiRAC Data Intensive service at Leicester, operated by the University of Leicester IT Services, which forms part of the STFC DiRAC HPC Facility (www.dirac.ac.uk). The equipment was funded by BEIS capital funding via STFC capital grants ST/K000373/1 and ST/R002363/1 and STFC DiRAC Operations grant ST/R001014/1. DiRAC is part of the National e-Infrastructure.

\setlength{\bibsep}{0pt}
\bibliography{GSF}
\bibliographystyle{mnras}

\appendix
\section{The GSF instability in the limit of small Prandtl number with the product of Richardson and Prandtl Numbers $O(1)$}
\label{RiPrO1}

In this appendix, we extend \S~\ref{lineargrowth} by presenting a complementary asymptotic linear analysis of the GSF instability in the limit as $\mathrm{Pr}\rightarrow 0$, with $\mathrm{RiPr} \sim O(1)$. For slow rotators, the Richardson number can be large, so that although Pr is small, Ri can be so large that RiPr remains finite in the limit Pr $\to 0$. This limit was considered at the poles
by \cite{Rashid2008}, but here we consider general latitudes. The appropriate scaling in this case is now $a \sim O(\Omega^2)$,
$b \sim O(\Omega^2/\mathrm{Pr})$, $s \sim {O(\Omega)}$ and $k^2 \sim O(\Omega / \nu)$, where $\mathcal{S} \sim O(\Omega)$ throughout. Note that the scaling for $k^2$, as well as the scaling for $b$, is different from that required to derive the results in \S~\ref{limprto0}. The cubic dispersion relation Eq.~\ref{cubic} here reduces to
\begin{eqnarray}
\label{dispRiPr1_eqn}
\kappa k^2 s^2 + s (2 \nu \kappa k^4 + b) 
+ \nu^2 \kappa k^6 + a \kappa k^2 + b \nu k^2 = 0.
\end{eqnarray}
Let the wavenumber
\begin{eqnarray}
\boldsymbol{k} = (k \cos \theta_k , 0, -k \sin \theta_k),
\end{eqnarray}
with magnitude $k$ and angle $\theta_k$ below the $x$-axis. Now 
\begin{eqnarray}
a=\frac{2 \Omega \vert \nabla \ell \vert}{\varpi} \sin(\Lambda-\theta_k) \sin(\gamma - \theta_k),
\label{a_eqn}
\end{eqnarray}
which is negative in the unstable case, so $\theta_k$ lies between $\gamma$ and $\Lambda$. At large Ri, $\Gamma$ is small, so $b$ simplifies to
\begin{eqnarray}
b=\mathcal{N}^2 \sin^2 \theta_k.
\label{b_eqn}
\end{eqnarray}
We define
\begin{eqnarray}
\lambda= \frac{\kappa \nu k^4}{b},
\label{lambda_def_eqn}
\end{eqnarray}
and maximise $s$ over $k^2$ by applying $k^2 \partial / \partial k^2$ to Eq.~\ref{dispRiPr1_eqn}, noting $\partial s / \partial k^2 = 0$, and subtracting  Eq.~\ref{dispRiPr1_eqn} to obtain
\begin{eqnarray}
s = \frac{2 \nu k^2 \lambda}{1-2 \lambda}
\label{s_eqn}
\end{eqnarray}
showing that for instability, $s>0$, $0 < \lambda < 1/2$.
Substituting this into Eq.~\ref{dispRiPr1_eqn} to eliminate $s$ gives 
\begin{eqnarray}
\label{lambda_eqn}
4 a \lambda^2 - (4 a + \mathrm{Pr} b) \lambda
+a + \mathrm{Pr}b = 0, 
 \end{eqnarray}
which can be written
\begin{eqnarray}
 (1-2\lambda)^2 a=(\lambda - 1) \mathrm{Pr} b .
\label{a_b_eqn}
\end{eqnarray}
We now maximise $s$ over $\theta_k$ . Eq.~\ref{s_eqn} can be written 
\begin{eqnarray}
(1-2\lambda) s = 2 \mathrm{Pr}^{1/2} \lambda^{3/2} b^{1/2}.
\label{s_b_eqn}
\end{eqnarray} 
Taking the $\log$ of this, differentiating with respect to $\theta_k$ and setting $\partial s / \partial \theta_k=0$ gives
\begin{eqnarray}
(2\lambda - 3) b \frac{\partial \lambda}{\partial \theta_k} = \lambda (1 - 2 \lambda) \frac{\partial b}{\partial \theta_k}.
\label{dlambda_dtheta_eqn}
\end{eqnarray}
Taking the $\log$ of Eq.~\ref{a_b_eqn} and differentiating with respect to $\theta_k$, using Eq.~\ref{dlambda_dtheta_eqn} to eliminate
$\partial \lambda / \partial \theta_k$, gives
\begin{eqnarray}
(2\lambda - 1)  \frac{\partial a}{\partial \theta_k} = \mathrm{Pr} \frac{\partial b}{\partial \theta_k}.
\label{da_dtheta_eqn}
\end{eqnarray} 
Using Eqs.~\ref{a_eqn}, \ref{b_eqn}, this can be written
\begin{eqnarray}
\label{theta_eqn}
(1-2 \lambda) \sin(\Lambda + \gamma - 2 \theta_k) =  \mathrm{RPr} \sin 2 \theta_k,
\end{eqnarray}
In general, the two equations  Eqs.~\ref{lambda_eqn}, \ref{theta_eqn}  for $\lambda$ and $\theta_k$  must be solved numerically, since $a$, $b$ and $\lambda$ depend on $\theta_k$. The growth rate $s$
can then be found using Eq.~\ref{s_eqn}.
 However, there are two limits within this scaling which shed light on the nature of the solutions.

\subsection{Limit $\mathrm{RiPr}\to 0$, $\lambda \to 1/2$}

First we consider
\begin{eqnarray}
1 \gg \mathrm{Ri Pr}  \gg \mathrm{Pr} >0.
\end{eqnarray}
Since $\mathcal{S} \sim O(\Omega)$, RPr is also small, and since $b/a \sim O(\mathrm{R})$, $b\mathrm{Pr} \ll a$. So in this limit Eq.~\ref{a_b_eqn} reduces to
$(1 - 2\lambda)^2 \to 0$, i.e. $\lambda \to 1/2$. Then from Eq.~\ref{s_eqn}, $s \gg \nu k^2$ and from Eq.~\ref{lambda_def_eqn}, 
$b \to 2 \nu \kappa k^4$ which means  Eq.~\ref{dispRiPr1_eqn} reduces to Eq.~\ref{smallRiPrdisp_eqn}, so $s$~$\to$~$\sqrt{-a}$ and the
limit $\lambda \to 1/2$ is the same limit as discussed in \S~\ref{limprto0}. Now expanding in powers of the small parameter $ (\mathrm{RPr})^{1/2}$, and using 
 Eqs.~\ref{a_eqn}, \ref{b_eqn} and \ref{a_b_eqn} 
\begin{eqnarray}
\frac{b \mathrm{Pr}}{a} = \frac{\mathrm{RPr} \sin^2 \theta_k}{\sin(\Lambda- \theta_k) \sin ( \gamma- \theta_k)} \to - 2 (1-2 \lambda)^2,
\label{bProvera_eqn}
\end{eqnarray}
so $(1-2 \lambda)$ is $O((\mathrm{RPr})^{1/2})$. Then  Eq.~\ref{theta_eqn} gives $\sin(\Lambda + \gamma - 2 \theta_k)$~$\sim O((\mathrm{RPr}
)^{1/2})$, so
\begin{eqnarray}
\theta_k \to \frac{\Lambda + \gamma}{2}
\end{eqnarray}
as $\mathrm{RPr} \to 0$,
which is the same result as in Eq.~\ref{FGMdirection} in \S~\ref{limprto0}.
 Eq.~\ref{bProvera_eqn} then becomes
\begin{eqnarray}
(1- 2 \lambda)^2 \to \frac{\mathrm{R Pr} \sin^2\left( \frac{\gamma+ \Lambda}{2}\right)}{2 \sin^2 \left(\frac{\gamma - \Lambda}{2}\right)}, 
\end{eqnarray}
giving $\lambda$ in terms of the small parameter $(\mathrm{RPr})^{1/2}$ accurate to first order in the small parameter.

\subsection{Limit $\mathrm{RiPr}\to 1/4$, $\lambda \to 0$}

The second limit of interest is $\lambda \to 0$ (recall that $0$~$<$~$\lambda$~$<$~$1/2$). We will see below that this limit corresponds to 
$\mathrm{RiPr} \to 1/4$. Taking $\lambda \to 0$, Eq.~\ref{lambda_eqn}  becomes
\begin{eqnarray}
a + \mathrm{Pr} b \to 0,
\label{lim2_lambda_eqn}
\end{eqnarray}
and Eq.~\ref{theta_eqn} becomes
\begin{eqnarray}
\label{lim2_theta_eqn}
 \sin(\Lambda + \gamma - 2 \theta_k) \to  \mathrm{RPr} \sin 2 \theta_k.
\end{eqnarray}
Using Eqs.~\ref{a_eqn}, \ref{b_eqn} and \ref{modifiedRi} these give
\begin{eqnarray*}
 \sin \Lambda \sin \gamma \cot^2 \theta_k  - \sin (\Lambda+\gamma) \cot \theta_k + 
 \cos \Lambda \cos \gamma  \to -  \mathrm{RPr} 
 \end{eqnarray*}
\vspace{-8mm}
\begin{eqnarray}
\end{eqnarray}

\vspace{-2mm}

\noindent and 
\begin{eqnarray}
 \frac{1}{2} \sin ( \Lambda + \gamma) ( \cot \theta_k - \tan \theta_k) - \cos(\Lambda + \gamma) \to \mathrm{RPr}.
\end{eqnarray}
Eliminating RPr between these leads to 
\begin{eqnarray*}
  \sin  \Lambda  \sin \gamma ( 1 + \cot^2 \theta_k ) \to \frac{1}{2} \sin (\Lambda + \gamma) (\cot \theta_k  + \tan \theta_k).
\end{eqnarray*}
Dividing by $ 1 + \cot^2 \theta_k$ gives
\begin{eqnarray}
\cot \theta_k \to \frac{1}{2} (\cot \gamma + \cot \Lambda),
\end{eqnarray}
providing $\theta_k$ in this limit. Since the $\cot$ function is monotonic between $0 < \theta_k < \pi$ this implies that $\theta_k$ again must lie in the wedge of instability between $\gamma$ and $\Lambda$, but it is no longer exactly half way between them. 
Inserting this into Eq.~\ref{a_eqn},
\begin{eqnarray}
a \to - \frac{\Omega | \nabla  \ell |}{2 \varpi} \frac{\sin^2 ( \gamma - \Lambda)  }{\sin \gamma \sin \Lambda} \sin^2 \theta_k.
\label{lim2_a_eqn}
\end{eqnarray}
From Eqs.~\ref{Ridef} and \ref{modifiedRi}, $\mathrm{R}/\mathrm{Ri} = \mathcal{S}^2 \varpi / 2 \Omega | \nabla \ell |$, and using Eq.~\ref{gradldef}
\begin{eqnarray} 
\frac{\mathrm{R}}{\mathrm{Ri}} = \frac{ \sin^2 ( \gamma - \Lambda)}{\sin \gamma \sin \Lambda}
\end{eqnarray}
and inserting this into Eq.~\ref{b_eqn}
\begin{eqnarray}
b =  4 \mathrm{Ri} \frac{\Omega | \nabla  \ell |}{2 \varpi} \frac{\sin^2 ( \gamma - \Lambda) }{\sin \gamma \sin \Lambda} \sin^2 \theta_k.
\label{lim2_b_eqn}
\end{eqnarray}
From Eqs.~\ref{lim2_lambda_eqn}, \ref{lim2_a_eqn} and \ref{lim2_b_eqn},
\begin{eqnarray}
\epsilon = \frac{a + \mathrm{Pr} b}{a} \to  1 - 4 \mathrm{RiPr} \to 0,
\label{lim2_RiPr}
\end{eqnarray}
defining the small parameter $\epsilon$ and justifying the earlier statement that the limit $\lambda \to 0$ is the same limit as $\mathrm{RiPr}$~$\to$~$1/4$. So we see that within the $\mathrm{Pr}\ll 1$ but $\mathrm{RiPr} \sim O(1)$ 
scaling, the two limits at the ends of the  available range of $  1/2 > \lambda > 0$ correspond to the two limits $\mathrm{RiPr} \to 0$ and $\mathrm{RiPr} \to 1/4$ respectively.
Intermediate values of RiPr correspond to intermediate values of $\lambda$. Ignoring squares of the small parameter $\epsilon$, Eq.~\ref{lambda_eqn} gives
\begin{eqnarray}
\lambda \to \frac{a + \mathrm{Pr} b}{3 a} = \frac{\epsilon}{3},
\end{eqnarray}  
so from Eqs.~\ref{lambda_def_eqn} and \ref{d_def_eqn}
 \begin{eqnarray}
k^4 \to \frac{\epsilon \, \mathcal{N}^2 \sin^2 \theta_k}{3 \nu \kappa} = \frac{\epsilon \sin^2 \theta_k}{3 d^4},
\end{eqnarray}
so $k^4$ is now small compared to the value given by Eq.~\ref{kvalue_1} (i.e. the instability in this case prefers larger wavelengths).
Using this, and Eq.~\ref{Ridef}, to eliminate $k^2$ in Eq.~\ref{s_eqn} gives the growth rate,
\begin{eqnarray}
s^2 \to \frac{1}{27} \epsilon^3  \mathcal{S}^2  \sin^2 \theta_k.
\end{eqnarray}
As expected, as RiPr approaches 1/4 from below, the growth rate decreases from $O(\Omega)$ to zero, since for $\mathrm{RiPr} > 1/4$ the system is stable to axisymmetric diffusive modes. 
This extends the study of instability at the pole by \cite{Rashid2008} to general latitudes. We have also confirmed each of the analytical results in this section by solving numerically Eq.~\ref{cubic} for appropriate parameter choices.

\section{Table of simulations}
\begin{table*}
\begin{center}
\begin{tabular}{ccccccccc|cc}
\hline
$\Lambda$ & $\Gamma$ & $S$ & Ri & RiPr & $L_x$ & $L_y$ & $N_x$ & $N_y$ & $\langle u_xu_y\rangle$ & $\sqrt{\langle u_y^2\rangle}$
 \\
\hline 
$30^{\circ}$ & $4.13^{\circ}$ & 0.72 & 19.3 & 0.19 & 100 & 0 & 256 & 1 & $2.72\pm0.0001$ & $4.80\pm0.0001$ \\
$30^{\circ}$ & $4.13^{\circ}$ & 0.72 & 19.3 & 0.19 & 100 & 100 & 256 & 256 & $2.71\pm0.03$ & $4.79\pm0.01$ \\
$30^{\circ}$ & $5.74^{\circ}$ & 1 & 10 & 0.1 & 100 & 0 & 256 & 1 & $23.6\pm13.3$ & $14.6\pm6.2$ \\
$30^{\circ}$ & $5.74^{\circ}$ & 1 & 10 & 0.1 & 100 & 50 & 256 & 256 & $5.41\pm0.89$ & $4.31\pm0.52$ \\
$30^{\circ}$ & $5.74^{\circ}$ & 1 & 10 & 0.1 & 100 & 100 & 256 & 256 & $7.04\pm1.81$ & $6.42\pm1.61$ \\
$30^{\circ}$ & $8.63^{\circ}$ & 1.5 & 4.44 & 0.044 & 100 & 0 & 256 & 1 & $50.17\pm19.3$ & $23.2\pm6.9$ \\
$30^{\circ}$ & $8.63^{\circ}$ & 1.5 & 4.44 & 0.044 & 100 & 50 & 256 & 256 & $9.97\pm1.44$ & $7.11\pm0.57$ \\
$30^{\circ}$ & $8.63^{\circ}$ & 1.5 & 4.44 & 0.044 & 100 & 100 & 256 & 256 & $9.29\pm1.83$ & $11.75\pm2.10$ \\
$30^{\circ}$ & $11.54^{\circ}$ & 2 & 2.5 & 0.025 & 100 & 0 & 256 & 1 & $42.68\pm9.47$ & $18.8\pm3.82$ \\
$30^{\circ}$ & $11.54^{\circ}$ & 2 & 2.5 & 0.025 & $100^\star$ & 0 & 256 & 1 & $43.38\pm7.89$ & $19.27\pm2.94$ \\
$30^{\circ}$ & $11.54^{\circ}$ & 2 & 2.5 & 0.025 & $200^\dagger$ & 0 & 512 & 1 & $51.16\pm16.9$ & $21.73\pm6.01$ \\
$30^{\circ}$ & $11.54^{\circ}$ & 2 & 2.5 & 0.025 & 200 & 0 & 512 & 1 & $81.17\pm28.65$ & $39.6\pm13.4$ \\
$30^{\circ}$ & $11.54^{\circ}$ & 2 & 2.5 & 0.025 & 100 & 30 & 256 & 128 & $12.44\pm1.35$ & $9.06\pm0.57$ \\
$30^{\circ}$ & $11.54^{\circ}$ & 2 & 2.5 & 0.025 & 100 & 50 & 256 & 256 & $11.36\pm1.23$ & $7.91\pm0.64$ \\
$30^{\circ}$ & $11.54^{\circ}$ & 2 & 2.5 & 0.025 & 100 & 100 & 256 & 256 & $10.73\pm1.34$ & $10.69\pm2.06$ \\
$30^{\circ}$ & $11.54^{\circ}$ & 2 & 2.5 & 0.025 & $100^\star$ & 100 & 256 & 256 & $11.1\pm1.24$ & $11.3\pm1.74$ \\
$30^{\circ}$ & $11.54^{\circ}$ & 2 & 2.5 & 0.025 & $200^\dagger$ & 100 & 512 & 256 & $10.64\pm0.88$ & $9.97\pm1.68$ \\
$30^{\circ}$ & $11.54^{\circ}$ & 2 & 2.5 & 0.025 & 200 & 200 & 512 & 512 & $10.49\pm0.82$ & $7.93\pm0.94$ \\
$30^{\circ}$ & $11.54^{\circ}$ & 2 & 2.5 & 0.025 & 100 & 30 & 200N & 60N & $6.21\pm1.12$ & $6.84\pm0.97$ \\
$30^{\circ}$ & $11.54^{\circ}$ & 2.5 & 1.6 & 0.016 & 100 & 0 & 256 & 1 & $24.69\pm2.60$ & $8.70\pm0.46$ \\
$30^{\circ}$ & $14.48^{\circ}$ & 2.5 & 1.6 & 0.016 & 100 & 30 & 256 & 128 & $11.97\pm0.92$ & $4.83\pm0.23$ \\
$30^{\circ}$ & $14.48^{\circ}$ & 2.5 & 1.6 & 0.016 & 100 & 50 & 256 & 256 & $10.19\pm0.48$ & $4.22\pm0.13$ \\
$30^{\circ}$ & $14.48^{\circ}$ & 2.5 & 1.6 & 0.016 & 100 & 30 & 200N & 60N & $14.83\pm1.24$ & $9.45\pm0.34$ \\
$30^{\circ}$ & $17.46^{\circ}$ & 3 & 1.11 & 0.011 & 100 & 0 & 256 & 1 & $54.32\pm5.90$ & $13.6\pm0.70$ \\
$30^{\circ}$ & $17.46^{\circ}$ & 3 & 1.11 & 0.011 & 100 & 100 & 256 & 256 & $24.18\pm1.17$ & $6.71\pm0.18$ \\
$30^{\circ}$ & $20.49^{\circ}$ & 3.5 & 0.82 & 0.0082 & 100 & 100 & 256 & 256 & $73.51\pm7.51$ & $12.34\pm0.57$ \\
\hline
$90^{\circ}$ & $11.54^{\circ}$ & 1 & 10 & 0.1 & 100 & 0 & 256 & 1 & $40.83\pm3.90$ & $15.34\pm1.21$ \\
$90^{\circ}$ & $11.54^{\circ}$ & 1 & 10 & 0.1 & 100 & 100 & 256 & 256 & $13.79\pm2.92$ & $7.94\pm1.90$ \\
$90^{\circ}$ & $17.46^{\circ}$ & 1.5 & 4.44 & 0.044 & 100 & 0 & 256 & 1 & $90.7\pm13.1$ & $16.08\pm1.48$ \\
$90^{\circ}$ & $17.46^{\circ}$ & 1.5 & 4.44 & 0.044 & 100 & 100 & 256 & 256 & $19.54\pm1.37$ & $7.89\pm0.86$ \\
$90^{\circ}$ & $23.58^{\circ}$ & 2 & 2.5 & 0.025 & 100 & 0 & 256 & 1 & $30.59\pm1.72$ & $7.93\pm0.23$ \\
$90^{\circ}$ & $23.58^{\circ}$ & 2 & 2.5 & 0.025 & 100 & 100 & 256 & 256 & $36.83\pm1.62$ & $9.34\pm0.15$ \\
$90^{\circ}$ & $30^{\circ}$ & 2.5 & 1.6 & 0.016 & 100 & 100 & 256 & 256 & $148.6\pm15.1$ & $20.2\pm1.71$ \\
$90^{\circ}$ & $36.87^{\circ}$ & 3 & 1.1 & 0.011 & 100 & 100 & 256 & 256 & $317.4\pm273.1$ & $33.77\pm9.87$ \\
\hline
$60^{\circ}$ & $9.97^{\circ}$ & 1 & 10 & 0.1 & 100 & 100 & 256 & 256 & $8.95\pm0.97$ & $6.21\pm1.05$ \\
$60^{\circ}$ & $15.06^{\circ}$ & 1.5 & 4.44 & 0.044 & 100 & 100 & 256 & 256 & $13.69\pm2.36$ & $8.20\pm2.08$ \\
$60^{\circ}$ & $20.27^{\circ}$ & 2 & 2.5 & 0.025 & 100 & 100 & 256 & 256 & $20.76\pm3.79$ & $8.14\pm0.98$ \\
$60^{\circ}$ & $31.31^{\circ}$ & 3 & 1.1 & 0.011 & 100 & 100 & 256 & 256 & $96.65\pm5.46$ & $13.08\pm0.29$ \\
\end{tabular}
\caption{
Table of simulation parameters. All simulations have $\mathrm{Pr}=10^{-2}$, $N^2=10$, $L_z=L_x$, and $N_x=N_z$, unless otherwise specified. Time-averages are based on the entire simulation after the initial linear growth. The eighth and ninth column give the number of Fourier modes in each direction. Simulations with Nek5000 have `N' in their $N_x$ and $N_y$ column entries and these numbers give the total number of grid points in each direction for $N_x$ and $N_y$, computed using an element distribution with $\mathcal{N}_p=10$ points in each element (15 fully de-aliased). Simulation parameters not listed in this table are given in \S~\ref{model}. The data listed to the right of the vertical lines are derived from the simulation results. Our simulation units are determined by setting $\Omega=d=1$. The two cases labelled with a $\star$ have $L_z=200$ (i.e.~$L_x/L_z=1/2$) and $N_z=512$, and those labelled with a $\dagger$ have $L_z=100$ (i.e.~$L_x/L_z=2$) and $N_z=256$.}
\end{center}
\label{Table}
\end{table*}

\label{lastpage}
\end{document}